\documentclass[
 reprint,
 amssymb, amsmath,
 aip,jcp
]{revtex4-1}

\usepackage{comment}
\usepackage{dcolumn}
\usepackage[nolist]{acronym}

\usepackage[colorlinks=true,linkcolor=blue]{hyperref}%
\expandafter\ifx\csname package@font\endcsname\relax\else
 \expandafter\expandafter
 \expandafter\usepackage
 \expandafter\expandafter
 \expandafter{\csname package@font\endcsname}%
\fi
\usepackage{xspace}

\usepackage{graphicx}

\usepackage[version=3]{mhchem}
\usepackage[normalem]{ulem}
\usepackage[utf8]{inputenc}

\usepackage{tikz}
\usepackage{multirow}
\usepackage{physics}
\usetikzlibrary{positioning}

\usepackage{bm}
\usepackage{blkarray}

\newcommand{\elemm}[3]{\mel{#1}{#2}{#3}}
\newcommand{\basis}[0]{\mathcal{B}}

\newcommand{\tcbio}[0]{{\ac{TC}\text{-}\ac{BiO}}}
\newcommand{\varj}[0]{{\text{TCVar/J}}}
\newcommand{\hfj}[0]{{\text{HF/J}}}
\newcommand{\tcbioj}[0]{{\text{TC-BiO/J}}}

\newcommand{\phiiub}[0]{\ket{{\Phi_i^\basis}[u]}}
\newcommand{\chiiub}[0]{\ket{{\chi_i^\basis}[u]}}

\newcommand{\eib}[0]{\tilde{E}_i^\basis[u]}
\newcommand{\ei}[0]{\tilde{E}_i}

\newcommand{\Ne}[0]{N_{\text{e}}}

\newcommand{\ban}[1]{\hat{c}_{#1}}
\newcommand{\bac}[1]{\hat{b}^\dagger_{#1}}

\newcommand{\etcbio}[0]{E_{\text{TC-BiO}}}

\newcommand{\pphii}[0]{{\Phi_i}}
\newcommand{\cchii}[0]{{\chi_i}}

\newcommand{\grid}[0]{{\text{grid}}}
\newcommand{\br}[1]{{\mathbf{r}_{#1}}}
\newcommand{\brbig}[1]{{\mathbf{R}_{#1}}}

\newcommand{\brb}[1]{{\bf r}_{#1}}
\newcommand{\rab}[0]{|\brb{1} - \brb{2}|}

\newcommand{\ai}[1]{\hat{a}_{#1}}
\newcommand{\aic}[1]{\hat{a}^{\dagger}_{#1}}

\newcommand{\nuclnum}{N_{\text{nucl}}}
\newcommand{\ku}[2]{\hat{K}[u](\br{#1},\br{#2})}

\newcommand{\lu}[3]{\hat{L}[u](\br{#1},\br{#2},\br{#3})}

\newcommand{\uu}[2]{u(\br{#1},\br{#2})}
\newcommand{\umu}[2]{u_\mu(r_{#1#2})}
\newcommand{\unew}[2]{\mathcal{U}(\br{#1},\br{#2})}
\newcommand{\unewa}[2]{\mathcal{U}(\br{#1},\br{#2})}
\newcommand{\envlop}[1]{\bar{g}(\br{#1})}
\newcommand{\tu}{\hat{\tau}_u}

\newcommand{\deriv}[3]{\frac{\partial^{#3} #1}{\partial {#2}^{#3}}}

\newcommand{\wmuijkl}[4]{{w}_{ij}^{kl}}
\newcommand{\kijkl}[0]{{K}_{ij}^{kl}}
\newcommand{\linearkijkl}[0]{{\gamma}_{ij}^{kl}}
\newcommand{\quadrkijkl}[0]{{\theta}_{ij}^{kl}}

\newcommand{\lmuijmkln}[0]{{L}_{ijm}^{kln}}

\newcommand{\htc}[0]{\tilde{H}}

\newcommand{\hu}[0]{\tilde{H}[u]}
\newcommand{\hub}[0]{\tilde{H}^\basis[u]}

\newcommand{\adi}[1]{a^{\dagger}_{#1}}

\newcommand{\fderiv}[2]{\frac{\delta #1}{\delta{#2}}}

\newcommand{\etchiphi}[0]{\tilde{E}[\chi,\Phi]}

\newcommand{\etchi}[1]{\tilde{E}[\chi,#1]}

\newcommand{\etphi}[1]{\tilde{E}[#1,\Phi]}
\newcommand{\Lset}[1]{{\mathcal{L}_{#1}}^{\nuclnum}}

\usepackage{csquotes}

\begin{document}

\author{Abdallah Ammar}
\email{aammar@irsamc.ups-tlse.fr}
\newcommand{\LCPQ}{Laboratoire de Chimie et Physique Quantiques (UMR 5626), Universit\'e de Toulouse, CNRS, UPS, France}
\affiliation{\LCPQ}
\author{Anthony Scemama}
\affiliation{\LCPQ}
\author{Emmanuel Giner}%
\email{emmanuel.giner@lct.jussieu.fr}
\affiliation{Laboratoire de Chimie Théorique, Sorbonne Université and CNRS, F-75005 Paris, France}

\title{Bi-orthonormal orbital optimization with a cheap core-electron free three-body correlation factor
for Quantum Monte Carlo and Transcorrelation}

\begin{abstract}
We introduce a novel three-body correlation factor that is designed to vanish in the core region around each nucleus and approach a universal two-body correlation factor for valence electrons. 
The Transcorrelated Hamiltonian is used to optimize the orbitals of a single Slater determinant within a biorthonormal framework.
The Slater-Jastrow wave function is optimized on a set of atomic and molecular systems containing both second-row elements and $3d$ transition metal elements. 
The optimization of the correlation factor and the orbitals, along with increasing the basis set, results in a systematic lowering of the Variational Monte Carlo energy for all systems tested.
Importantly, the optimal parameters of the correlation factor obtained for atomic systems are transferable to molecules.
Additionally, the present correlation factor is computationally efficient, using a mixed analytical-numerical integration scheme that reduces the costly numerical integration from $\mathbb{R}^6$ to $\mathbb{R}^3$.
\end{abstract}

\maketitle

\begin{acronym}
  \acro{FCI}{Full Configuration Interaction}
  \acro{CI}{Configuration Interaction}
  \acro{QMC}{Quantum Monte Carlo}
  \acro{AO}{Atomic Orbital}
  \acro{MO}{Molecular Orbital}
  \acro{HF}{Hartree-Fock}
  \acro{CAS}{Complete Active Space}
  \acro{VMC}{Variational Monte Carlo}
  \acro{DMC}{Diffusion Monte Carlo}
  \acro{TC}{Transcorrelated}
  \acro{FCIQMC}{Full Configuration Interaction Quantum Monte Carlo}
  \acro{SCF}{Self Consistent Field}
  \acro{RHF}{Restricted   Hartree-Fock}
  \acro{UHF}{Unrestricted Hartree-Fock}
  \acro{ROHF}{Restricted Open-shell Hartree-Fock}
  \acro{DIIS}{Direct Inversion in the Iterative Subspace}
  \acro{LS}{Level-Shifting}
  \acro{MP}{M{\o}ller-Plesset}
  \acro{MPPT2}[MP2]{M{\o}ller-Plesset Perturbation Theory at second order}
  \acro{BiO}{Bi-Orthogonal}
  \acro{MBPT2}[MBPT(2)]{Many Body Perturbation Theory to the second order}
  \acro{LCCSD}{Linearized Coupled Cluster Singles and Doubles}
  \acro{DMRG}{Density Matrix Renormalization Group}
  \acro{CC}{Coupled Cluster}
  \acro{DFT}{Density Functional Theory}
  \acro{WFT}{Wave Function Theory}
  \acro{Var}{Variational}
  \acro{CBS}{Complete Basis Set}
\end{acronym}

\section{Introduction}

The \ac{TC} approach\cite{Boys_Handy_1969_determination}, has proven to be a valuable tool for describing electronic correlation in a variety of chemical systems.
This approach involves a similarity transformation of the many-electron wavefunction by a Jastrow factor (or Gutzwiller factor for lattice model systems), which simplifies the computational treatment of strong electron-electron interactions and enables accurate results for complex systems using reasonable computational resources.
The resulting \ac{TC} effective Schrödinger equation can be solved with conventional methods based on single-particle basis functions, with faster convergence due to the removal of electron-electron singularities present in the standard Schrödinger equation.
The \ac{TC} method has been successfully combined with many post-\ac{HF} theories, including
\ac{MBPT2}\cite{Hino_etal_2001,Hino_etal_2001,Ochi_Tsuneyuki_2015},
\ac{DMC}\cite{Prasad_etal_2007},
\ac{CI}\cite{Zweistra_etal_2003,Tsuneyuki_2008,Luo_2011,Ochi_Tsuneyuki_2014,Gin-JCP-21,Ammar_etal_2022_extension,Ammar_etal_2022_optimization},
\ac{FCIQMC}\cite{Luo_Alavi_2018,Dobrautz_etal_2019,Cohen_etal_2019,Dobrautz_etal_2022,Jeszenszki_etal_2020,Guther_etal_2021,Haupt_etal_2023},
\ac{CC}\cite{Hino_etal_2002,TenNo_Hino_2002,Liao_etal_2021,Schraivogel_etal_2021},
\ac{DMRG}\cite{Baiardi_Reither_2020,Baiardi_etal_2022},
and
\ac{DFT}\cite{Imamura_Scuseria_2003,Umezawa_2017}.

The \ac{TC} approach has also been employed to mitigate the repulsive nature of nucleon-nucleon interactions at short ranges \cite{Suzuki_Matsumura_2005}, enhance the convergence properties of quantum gases with contact interactions in one dimension \cite{Jeszenszki_etal_2018}, and enhance the precision of quantum simulations of molecular ground and excited states as well as solid-state systems \cite{McArdle_Tew_2020,Sokolov_etal_2022}.

Optimizing the Jastrow parameters and the coefficients of both the orbitals and Slater determinants in the \ac{TC} framework is a challenging task for several reasons.
Firstly, the non-Hermitian nature of the \ac{TC} Hamiltonian makes it impossible to utilize the standard variational method, which relies on the exact eigenvalue being a lower bound to the approximated one.
Secondly, the additional two-body terms resulting from the similarity transformation require intermediate computations that scale as $M^2 \times N_{\grid}$, where $M$ is the number of basis functions and $N_{\grid}$ is the number of grid points used for integrating over $\mathbb{R}^3$.
These intermediates are themselves integrals over $\mathbb{R}^3$ and, given that closed-form evaluations are not possible for most of the correlation factors proposed in the literature, each integral must be approximated numerically. This leads to a typical scaling of $M^2 \times (N_{\grid})^2$, which can be prohibitively expensive.
Finally, the three-body operator that arises in the \ac{TC} Hamiltonian requires significant computational resources to handle.

The concept of a similarity transformation of the Hamiltonian by a correlation factor was originally proposed by Hirschfelder \cite{Hirschfelder-JCP-63} and further developed by Boys and Handy.
Three types of equations were developed for optimizing the parameters of a Slater-Jastrow wavefunction ($\Phi e^{\tau}$). The wavefunction includes a Slater determinant represented by $\Phi$ and an exponential term $\tau$ that is symmetric with respect to permutation of the electrons.
The first set of equations is dedicated to optimizing the coefficients of the one-electron functions of the Jastrow \cite{Boys_Handy_1969_determination, Boys_Handy_1969_condition, Bernardi_Boys_1973}.
While these parameters are redundant and only affect the one-electron orbitals, they are introduced and optimized to minimize the non-Hermitian feature of the \ac{TC} Hamiltonian.
The electron-electron and electron-electron-nucleus parameters of the Jastrow and the orbital coefficients were determined through two sets of projection equations over derivatives of the wavefunction with respect to those parameters \cite{Boys_Handy_1969_determination, Boys_Handy_1969_calculation, Boys_Handy_1969_first, Handy_1969, Hall_Solomon_1969, Lim_1974}.
The optimization equations for the parameters of the correlation factor have been extensively investigated and applied to compute first- and second-order properties of small atoms and molecules\cite{Boys_1969,Boys_Bernardi_1970,Handy_Epstein_1970,Armour_1972_convergence,Armour_1973,Armour_1974, Handy_1975,Huggett_Armour_1976,Hall_Solomon_1969,Armour_1980}.
Despite the non-Hermitian nature of the \ac{TC} Hamiltonian, these equations can provide results that are similar to those obtained by a \ac{VMC} optimization if the Jastrow factor is sufficiently flexible\cite{Luo_etal_2010}. However, the use of these projection-type equations to optimize the molecular orbitals can result in a more significant violation of the variational bounds.

In Ref.~\onlinecite{Bernardi_1973}, the \ac{TC} equations were reformulated as stationary solutions of an energy functional of left- and right-wavefunction.
This approach can be seen as an application of the \ac{BiO} Quantum Mechanics\cite{Cloizeaux_1960,Brandow_1967,Moshinsky_Seligman_1971,Gouyet_1973_champ,Dahl_1978, Payne_1982,Brody_2014,Ashida_etal_2020} to the non-Hermitian \ac{TC} Hamiltonian.
In this article, we refer to this method as the \ac{TC}-\ac{BiO} theory.
Two (left and right) sets of \ac{BiO} one-electron orbitals can be employed to realize the \ac{TC}-\ac{BiO} method.
Another way to perform a \ac{TC}-\ac{BiO} calculation is by using one set of orbitals with two different left and right \ac{CI} expansions\cite{Zweistra_etal_2003,Gin-JCP-21,Ammar_etal_2022_extension}.
The \ac{TC}-\ac{BiO} formulation using \ac{BiO} orbitals was first introduced by Fimple and Unwin\cite{Fimple_Unwin_1976} and was applied to the Helium atom.
Ten-no \textit{et al} applied the \ac{BiO} formalism in the second quantization framework to optimize two sets of \ac{BiO} orbitals using a \ac{TC}-\ac{SCF} technique\cite{Hino_etal_2001}.
The procedure involves iteratively building and diagonalizing a non-Hermitian \ac{TC} Fock matrix for a fixed Jastrow factor until orbital convergence is achieved.
They also presented \ac{BiO} versions of the \ac{MPPT2}\cite{Hino_etal_2001} and \ac{LCCSD}\cite{Hino_etal_2002} for the \ac{TC} Hamiltonian.
This method has been applied to small atoms and molecules\cite{Hino_etal_2001,Hino_etal_2002,TenNo_Hino_2002, Lee_Thom_2023}, as well as to solid-state systems\cite{Ochi_Tsuneyuki_2015,WahlenStrothman_etal_2015,Ochi_etal_2017,Ochi-CPC-23}.

Tsuneyuki and collaborators introduced a \ac{TC}-\ac{SCF} approach\cite{Umezawa_Tsuneyuki_2003_tc_self,Umezawa_Tsuneyuki_2003_tc_method} for optimizing orbitals by minimizing the variance of the \ac{TC} Hamiltonian\cite{Handy_1971_minimization,Handy_1972,Handy_1973}, which uses a single set of orthogonal orbitals.
The orthogonality between the orbitals is enforced through a Gram-Schmidt process at each \ac{SCF} iteration, as the resulting \ac{TC} Fock matrix is non-Hermitian.
This method has been successfully applied to solid-state systems\cite{Tsuneyuki_2008,Ochi_etal_2012,Ochi_Tsuneyuki_2013,Ochi_Tsuneyuki_2014,Ochi_etal_2014,Ochi_etal_2017} and small atoms\cite{Umezawa_etal_2006,Prasad_etal_2007}.
The \ac{TC}-\ac{SCF} equations are similar to the \ac{TC}-\ac{BiO}-\ac{SCF} equations, where the former can be obtained from the latter when the left- and right-orbitals are the same.
However, the \ac{TC}-\ac{SCF} orbitals only satisfy Brillouin's theorem for the left block, while the \ac{TC}-\ac{BiO}-\ac{SCF} orbitals satisfy it for both blocks.
Despite being less stable due to the non-Hermiticity of the \ac{TC} Hamiltonian, the \ac{TC}-\ac{SCF} method has been successfully applied in solid-state calculations and small atoms by imposing orthogonality between orbitals through Gram-Schmidt process.
However, this method can lead to nonphysical complex eigenvalues with significant imaginary parts, while the \ac{BiO} formulation is more stable.
An example of the difference between the two methods is that the orthogonal \ac{TC}-\ac{SCF} approach fails to calculate the band structure of wurtzite ZnO, while the results obtained using the \ac{TC}-\ac{BiO}-\ac{SCF} are in good agreement with the experiment\cite{Ochi_etal_2017}.

The \ac{TC}-\ac{Var} method proposed by Luo\cite{Luo_2010,Luo_2011} offers an alternative to overcome the non-Hermitian nature of the \ac{TC} Hamiltonian by replacing it with a Hermitian operator to optimize the orbitals and \ac{CI} coefficients of a \ac{CI}-Jastrow wavefunction.
This allows for optimization within a traditional \ac{Var} framework and has demonstrated rapid convergence.
However, our work in this article reveals only a small improvement of the \ac{TC}-\ac{Var} compared to the \ac{TC}-\ac{BiO} approach (see Sec.~\ref{sec:tcbio_vs_vartc}).

In this article, we introduce a novel three-body correlation factor that considers the electron-electron-nucleus coordinates.
Unlike previous three-body correlation factors, our new factor is designed to vanish in the core regions and gradually approach a universal two-electron correlation factor outside the core regions.
The advantages of this correlation factor include its simplicity, which reduces the computational cost by using a $\mathbb{R}^3$ instead of a $\mathbb{R}^6$ numerical integration scheme, and its easy parameterization with only one parameter per nucleus.
Furthermore, its role in the core regions ensures that optimal parameters for a single atom or an atom in a molecule are similar, allowing for a straightforward \ac{VMC} optimization with a single parameter for each atomic system.
We tested this new correlation factor at both the \ac{TC} and \ac{VMC} levels using a robust \ac{TC}-\ac{BiO}-\ac{SCF} scheme for both open- and closed-shell systems.
Finally, we performed optimization of the nucleus parameter on various atomic and molecular systems, including second-row elements and $3d$ transition metal atoms.

The structure of this work is as follows.
Section \ref{sec:theo_tc} provides a brief overview of the \ac{TC} framework, followed by a detailed presentation of the new correlation factor in Section \ref{sec:new_j}.
The biorthonormal framework is recalled in Section \ref{sec:n_h_theo}, which includes the functional stationary principle in Section \ref{subsubsec:stationary}, its application to the \ac{TC} framework with a single Slater determinant in Section \ref{sec:lef-right_det}, the corresponding left- and right-Brillouin conditions in Section \ref{subsubsec:Brillouin_theorem}, and the iterative scheme used to achieve the latter in Section \ref{sec:fock-op}.
In Section \ref{sec:results}, we present the numerical results supporting our work.
Specifically, we compare the $\tcbio$ and \ac{TC}-\ac{Var} orbital optimization techniques in Section \ref{sec:tcbio_vs_vartc}, optimize the correlation factor on a series of atomic systems including $3d$ transition metal atoms in Section \ref{opt-atom}, and optimize the correlation factor for molecules in Section \ref{transfer-mol}, showing that the obtained optimal parameters are similar to those obtained for atomic systems.

\section{Theory}
\label{sec:theo}

\subsection{Basics of the transcorrelated formalism}
\label{sec:theo_tc}
In this paper, $\mathbf{r}_i$ refers to the electron coordinate of electron $i$, and $r_{ij} = |\mathbf{r}_i-\mathbf{r}_j|$.

The general form of the transcorrelated Hamiltonian for a symmetric correlation factor $\uu{1}{2}$ is given by
\begin{equation}
 \label{ht_def_g}
 \begin{aligned}
  \hu  &\equiv e^{-\tu} \hat{H} e^{\tu} \\
                & = \hat{H} + \qty[ \hat{H},\tu ] + \frac{1}{2}\qty[ \qty[ \hat{H},\tu ],\tu],
 \end{aligned}
\end{equation}
where $\tu = \sum_{i<j}u(\br{i},\br{j})$ and $\hat{H} = -\sum_i \frac{1}{2} \nabla^2_i + v(\br{}_i) + \sum_{i<j}   1/r_{ij}$.
Eq. \eqref{ht_def_g} leads to the following transcorrelated Hamiltonian
\begin{equation}
 \label{ht_def_g2}
 \hu = \hat{H} - \sum_{i<j} \ku{i}{j} - \sum_{i<j<k} \lu{i}{j}{k},
\end{equation}
where the effective two- and three-body operators $\ku{1}{2}$ and $\lu{1}{2}{3}$ are defined as
\begin{equation}
 \label{def:k_u}
 \begin{aligned}
  \ku{1}{2} = \frac{1}{2} \bigg( &\Delta_1 \uu{1}{2} + \Delta_2 \uu{1}{2} \\
                                         + &\big(\nabla_1 \uu{1}{2} \big) ^2 + \big(\nabla_2 \uu{1}{2}      \big) ^2 \bigg) \\
                                         + &\nabla_1 \uu{1}{2} \cdot \nabla_1 + \nabla_2 \uu{1}{2}\cdot     \nabla_2,
 \end{aligned}
\end{equation}
and
\begin{equation}
 \label{def:l_u}
 \begin{aligned}
  \lu{1}{2}{3} = &   \nabla_1 \uu{1}{2} \cdot \nabla_1 \uu{1}{3} \\
                                          + & \nabla_2 \uu{2}{1} \cdot \nabla_2 \uu{2}{3}   \\
                                          + & \nabla_3 \uu{3}{1} \cdot \nabla_3 \uu{3}{2}   .
 \end{aligned}
\end{equation}
The definition of Eq.~\eqref{ht_def_g} implies that $\hu$ is not Hermitian as
\begin{equation}
 \qty(\hu)^\dagger = e^{+\tu} \hat{H} e^{-\tu} \ne \hu.
\end{equation}
In practice, the \ac{TC} Hamiltonian is projected into a one-particle basis set $\basis$
\begin{equation}
 \label{eq:def_hub}
 \begin{aligned}
 &\hub   = P^{\basis}\, \hu\, P^\basis,
 \end{aligned}
\end{equation}
where $P^{\basis}$ is the projector onto the Hilbert space spanned by the one-particle basis set $\basis$.
Since $\hub$ is non-Hermitian, a given eigenvalue $\eib$ is associated with a couple of right- and left-eigenvectors
\begin{equation}
 \label{eq:tc_eigv}
 \begin{aligned}
 & \hub \phiiub = \eib \phiiub \\
 & \qty(\hub)^\dagger \chiiub = \eib \chiiub,\\
 \end{aligned}
\end{equation}
and the exact eigenvalue $E_i$ is recovered in the \ac{CBS} limit due to the properties of the similarity transformation
\begin{equation}
 \lim_{\basis \rightarrow \text{CBS}} \eib = E_i.
\end{equation}
Since the correlation factor accounts for a part of the correlation effects, one can anticipate a faster convergence of $\eib$ compared to the conventional \ac{WFT}-based method.

$\hub$ can be written in a second-quantized form using a basis $\basis =\{\phi_i(\br{})\}$ of standard real-valued orthonormal spatial \acp{MO}:
\begin{equation}
 \label{eq:def_hub_sec_q}
 \begin{aligned}
 &\hu  = \sum_{i,j \in \basis} \,\, \sum_{\sigma = \uparrow,\downarrow} h_{ij}\, \adi{j,\sigma}\ai{i,\sigma}\\
 & + \frac{1}{2}\sum_{i,j,k,l \in \basis}  \,\, \sum_{\sigma,\lambda = \uparrow,\downarrow}
 \big( V_{ij}^{kl} - \kijkl\big)\, \adi{k,\sigma} \adi{l,\lambda} \ai{j,\lambda} \ai{i,\sigma} \\
       & - \frac{1}{6} \sum_{i,j,m,k,l,n \in \basis} \,\, \sum_{\sigma,\lambda,\kappa = \uparrow,\downarrow}
\lmuijmkln\, \adi{k,\sigma} \adi{l,\lambda} \adi{n,\kappa} \ai{m,\kappa} \ai{j,\lambda} \ai{i,\sigma},
 \end{aligned}
\end{equation}
where $h_{ij}$ are the usual one-electron integrals, $V_{ij}^{kl}$ are the usual two-electron integrals,
$\kijkl$ are the two-electron integrals corresponding to the effective two-body operator $\ku{1}{2}$
\begin{equation}
 \kijkl = \int \text{d} \br{1} \text{d} \br{2}\, \phi_k(\br{1}) \phi_l(\br{2})\, \ku{1}{2}\, \phi_i(\br{1}) \phi_j(\br{2}),
\end{equation}
and $\lmuijmkln$ are the three-electron integrals corresponding to the effective three-body operator $\lu{1}{2}{3}$
\begin{equation}
 \begin{aligned}
  \lmuijmkln  = \int & \text{d} \br{1} \text{d} \br{2} \text{d} \br{3}\,  \phi_k(\br{1}) \phi_l(\br{2}) \phi_n(\br{3}) \\ & \lu{1}{2}{3}\, \phi_i(\br{1}) \phi_j(\br{2}) \phi_m(\br{3}).
 \end{aligned}
\end{equation}
From Eqs.~\eqref{def:k_u} and \eqref{def:l_u} one can remark that the \ac{TC} Hamiltonian
depends on the gradients and Laplacian of the correlation factor $\uu{1}{2}$.
One can nevertheless perform an integration by parts, as suggested in Ref.~\onlinecite{Cohen_etal_2019},
in order to express the \ac{TC} Hamiltonian $\hub$ only as a function of the gradient of $\uu{1}{2}$.
The integrals of the two-body operator $\ku{1}{2}$ can be then rewritten as
\begin{equation}
 \label{def:k_grad}
 \kijkl = \linearkijkl + \quadrkijkl,
\end{equation}
with
\begin{equation}
 \begin{aligned}
 \label{def:k_grad_lin}
 \linearkijkl = & \frac{1}{2} \, \int \text{d} \br{1} \text{d} \br{2}\, \phi_l(\br{2}) \phi_j(\br{2})\, \nabla_1 u(\br{1},\br{2}) \\
 & \cdot \big( \phi_k(\br{1})\, \nabla_1 \phi_i(\br{1}) - \phi_i(\br{1})\, \nabla_1 \phi_k(\br{1})\big) \\
 + &  \frac{1}{2} \, \int \text{d} \br{1} \text{d} \br{2}\, \phi_k(\br{1}) \phi_i(\br{1})\, \nabla_2 u(\br{1},\br{2}) \\
 & \cdot \qty[ \phi_l(\br{2})\, \nabla_2 \phi_j(\br{2}) - \phi_j(\br{2})\, \nabla_2 \phi_l(\br{2})], \\
 \end{aligned}
\end{equation}
and
\begin{equation}
 \begin{aligned}
 \label{def:k_grad_quad}
 \quadrkijkl=   \frac{1}{2} \, \int &\text{d} \br{1} \text{d} \br{2}\,  \phi_k(\br{2}) \phi_l(\br{1}) \phi_j(\br{2}) \phi_i(\br{1}) \\
  &\qty[\nabla_1 u(\br{1},\br{2})]^2 + \qty[\nabla_2 u(\br{1},\br{2})]^2.\\
 \end{aligned}
\end{equation}

\subsection{Introducing a core-removing three-body correlation factor}
\label{sec:new_j}
In previous works\cite{Dobrautz_etal_2022,Ammar_etal_2022_extension}, we used the one-parameter correlation factor introduced in Ref.~\onlinecite{Gin-JCP-21},
$\uu{1}{2} = u_{\mu}(\rab)$, which is defined by the analytical form
\begin{equation}
 \label{eq:def_j}
 \umu{1}{2} =  \frac{1}{2}\, r_{12}\, \bigg( 1 - \erf(\mu\, r_{12})  \bigg) 
                - \frac{1}{2\sqrt{\pi}\,\mu}\,e^{-(r_{12}\,\mu)^2} 
\end{equation}
This correlation factor is controlled by the parameter $\mu$, which determines the depth and range of the correlation hole induced by $\umu{1}{2}$ (see  Ref.~\onlinecite{Gin-JCP-21} for a pictorial representation), and its explicit form yields a non-divergent scalar effective two-body interaction by exactly restoring the $s$-cusp condition\cite{Gin-JCP-21}.
However, to obtain sensible values for total energies and energy differences, a system-specific value of $\mu$ must be chosen, which can be obtained through an average over the mean-field density of quantities related to the correlation hole of the uniform electron gas, as shown in Ref.~\onlinecite{Gin-JCP-21}.
Even though the results reported in previous studies\cite{Gin-JCP-21,Dobrautz_etal_2022,Ammar_etal_2022_extension}
were promising for atomic and small molecular systems, the approach of using a system-specific value of $\mu$ has at least three major disadvantages.
\begin{enumerate}
 \item
In the case of a dissociated heterogeneous system $A\cdots B$, the system-specific value of $\mu$ obtained is an average between the values of $\mu$ on subsystem $A$ and $B$.
This average value is necessarily different from the two distinct values of $\mu$ for each subsystem, which leads to non-size-consistent results, unless the \ac{CBS} limit is reached and the exact energy is obtained.
This is not ideal for practical calculations done in incomplete basis sets, where one of the goals of \ac{TC} methods is to use smaller basis sets.
 \item
To achieve strict size-consistency for the correlation factor, it would be preferable to use a universal value of $\mu$.
However, atomic or molecular systems can be qualitatively divided into regions with vastly different typical densities. While a universal value of $\mu$ may be reasonable for valence electrons, the value of $\mu$ for core electrons depends heavily on the nuclear charge of the corresponding atom. Therefore, a universal value of $\mu$ cannot be applied as long as the same correlation factor is applied to both valence and core electrons.
 \item
Finally, it should be noted that the value of $\mu$ obtained by averaging is necessarily too low for the core, which often leads to an underestimation of ground-state energies in moderate basis sets. Additionally, the value is too high for the valence, which does not compact the wavefunction enough for valence electrons.
\end{enumerate}
An approach that is preferable is to employ a value of $\mu$ that is commonly used for valence electrons and eliminate the influence of the correlation factor in the core regions. This method can be easily implemented in the $f_{12}$ framework since the correlation factor is formulated in the orbital space. However, the correlation factor in the \ac{TC} framework is defined in real space, making it challenging to achieve a clear distinction between the core and valence regions.
The issue was initially addressed in the \ac{VMC} method by introducing three-body correlation factors that utilize the electron-electron-nucleus distances to decrease the depth and range of the correlation hole in the proximity of the nuclei. This results in intricate expressions for the correlation factors, which, in the context of \ac{TC} calculations, necessitate numerical integration in six dimensions.

To address the problem of the core-valence splitting, we propose a new correlation factor:
\begin{equation}
 \label{def:unew}
 \unew{1}{2} = \uu{1}{2}\, \envlop{1}\, \envlop{2}
\end{equation}
with
\begin{equation}
 \label{def:envlop}
 \envlop{}= \prod_{m=1}^{\nuclnum} \bigg( 1 - \exp\big( - \alpha_m \big| \br{}-\brbig{m} \big|^2 \big)\bigg),
\end{equation}
where $\brbig{m}$ is the position of the $m-$th nucleus.
The envelope $\envlop{}$ acts as a damping function which cancels the effect of the correlation factor $\uu{1}{2}$ near each nucleus.
Since the two-body correlation function $\umu{1}{2}$ is used to correlate valence electrons, we set $\mu=0.87$ as suggested in Ref.~\onlinecite{Gin-JCP-21}, a value which is capable of reproducing the \ac{TC} results obtained with Ten-No's frozen Gaussian geminal (FROGG) for valence electrons~\cite{TenNo-CPL-00-a}.
The correlation factor defined in Eq.~\eqref{def:unew} can be considered as a simplified version of the three-body correlation factor, as it introduces an explicit dependence between the inter-electron distance and the distance of each electron to each nucleus.
The parameters $\alpha_m$ determine the typical range in which the correlation factor $\uu{1}{2}$ is suppressed by the envelope $\envlop{1}\envlop{2}$ around the nucleus located at $\brbig{m}$.

\begin{figure}
	\includegraphics[width=\linewidth]{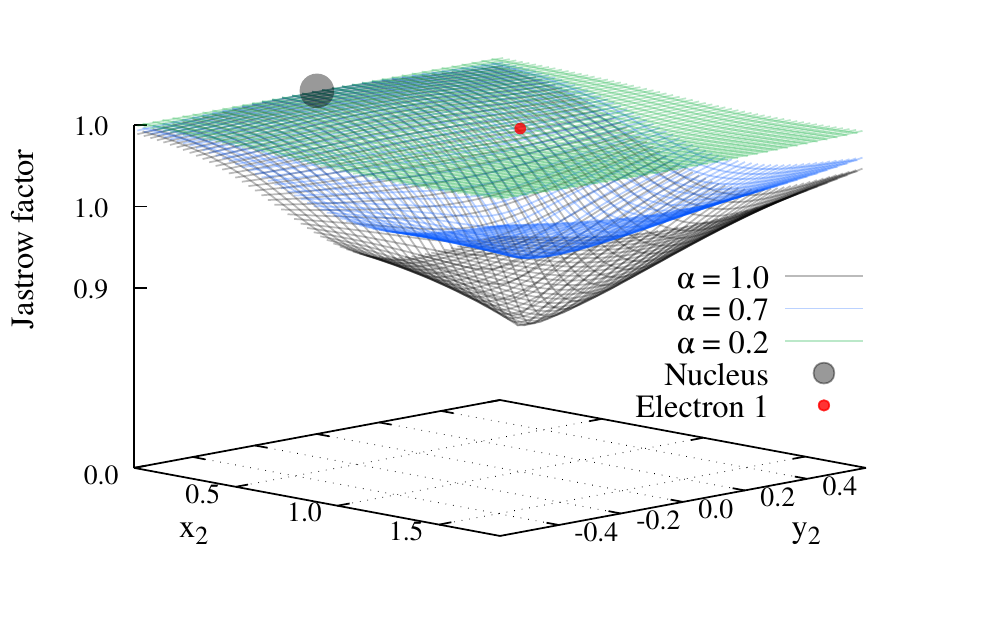}
	\caption{\label{fig:J3d_2e}
The Jastrow factor being the exponential form of Eq.~\eqref{def:unew} is visualized for a nucleus located at the origin with two electrons. The first electron (in red) is fixed at $(x_1=1,y_1=0)$ and the Jastrow factor is plotted as a function of the second electron's position $(x_2,y_2)$ for three values of $\alpha$: 0.2, 0.7, and 1.0.
	}
\end{figure}

\begin{figure}
	\includegraphics[width=\linewidth]{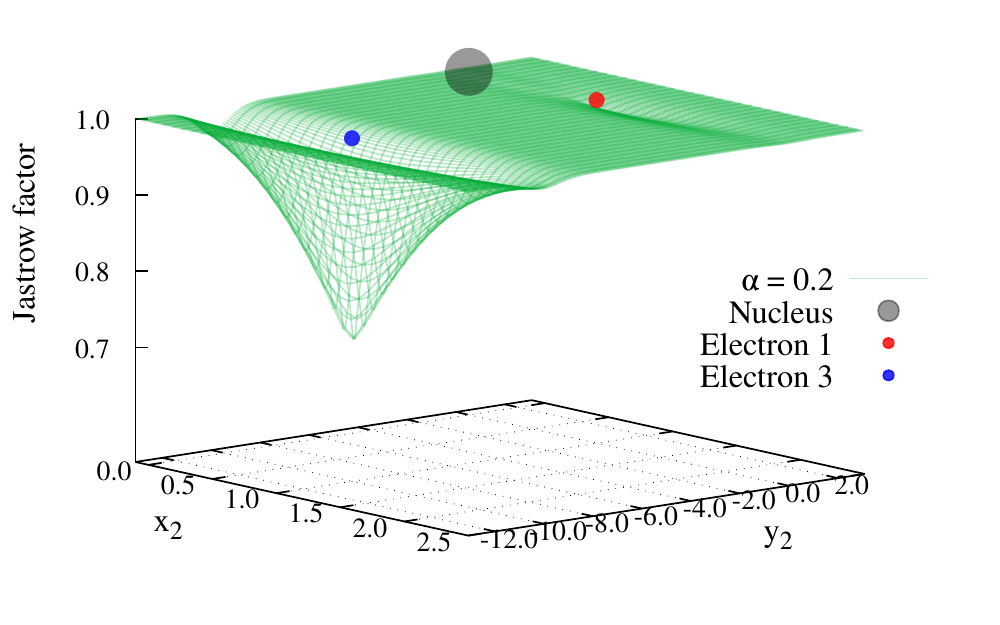}
	\caption{\label{fig:J3d_3e}
The Jastrow factor being the exponential form of Eq.~\eqref{def:unew} for a system of three electrons located around a nucleus at the origin is visualized in this figure. The first and third electrons are fixed at positions $(x_1=1,y_1=0)$ and $(x_3=1,y_3=-10)$, respectively, and are shown in red and blue. The Jastrow factor is plotted as a function of the position of the second electron $(x_2,y_2)$, for a chosen value of $\alpha=0.2$.
	}
\end{figure}

To demonstrate the impact of the parameter $\alpha$ on the correlation factor for the case of a core electron, we consider a nucleus positioned at the origin of the coordinate system, with an electron fixed at $(x_1=1,y_1=0,z_1=0)$, and a second electron whose position is allowed to vary in the two-dimensional plane $z_2=0$.
Figure~\ref{fig:J3d_2e} displays the value of the correlation factor
$e^{\qty(\sum_{i<j} \unewa{i}{j})}$
as a function of $(x_2,y_2)$ for three different values of $\alpha=0.2,0.7,1.0$.
The blue and black curves in Figure~\ref{fig:J3d_2e}, representing $\alpha=0.7$ and $\alpha=1$ respectively, indicate a significant distortion in the total electronic wavefunction. In contrast, a small $\alpha=0.2$ (green curve) results in a Jastrow factor that is close to one. When a third electron is added at $(x_3=1,y_3=-10,z_3=0)$, which is relatively far from the nucleus, the Jastrow factor for $\alpha=0.2$ is shown in Figure~\ref{fig:J3d_3e}.
A minor deviation from one is observed for the Jastrow near the electron close to the nucleus (in red), while the contribution from the electron far from the nucleus (in blue) is significant. This demonstrates why the proposed Jastrow factor allows for the \enquote{core electrons} to be \enquote{frozen} with respect to the dynamical correlation provided by the Jastrow factor.

The advantage of the form of the correlation factor given by Eqs.~\eqref{def:unew} and \eqref{def:envlop} is that the corresponding integrals can be evaluated in closed form (see Appendix for details).
\begin{equation}
 \label{def:w_munu}
 \mathbf{w}_{\mu \nu}(\br{1}) = \int \text{d}\br{2}\, \phi_{\mu}(\br{2}) \phi_{\nu}(\br{2})\, \nabla_1\unew{1}{2},
\end{equation}
\begin{equation}
 \label{def:w_munu_2}
 \mathbf{W}_{\mu \nu}(\br{1}) = \int \text{d}\br{2}\, \phi_{\mu}(\br{2}) \phi_{\nu}(\br{2})\, \big(\nabla_1\unew{1}{2} \big)^2.
\end{equation}
Using a typical DFT integration grid, the integrals $\linearkijkl$, $\quadrkijkl$, and $\lmuijmkln$ can be reconstructed from these intermediates. 
This is different from other correlation factors, where the equivalents of Eqs.~\eqref{def:w_munu} and \eqref{def:w_munu_2} need to be calculated numerically, resulting in a computational expense that scales as the sixth power of the number of integration points. In contrast, our approach has a more favorable cubic scaling.

\subsection{Orbital optimization using the TC Hamiltonian within a biorthonormal framework}
\label{sec:n_h_theo}

The biorthonormal approach in Quantum Mechanics\cite{Cloizeaux_1960,Brandow_1967,Moshinsky_Seligman_1971,Gouyet_1973_champ,Dahl_1978,Payne_1982,Brody_2014,Ashida_etal_2020}
has proven to be effective in various Quantum chemistry applications that involve non-orthogonal orbitals or non-Hermitian Hamiltonians. 
These applications include the method of moments\cite{Roeggen_Aashamar_1977},
the Valence Bond Method\cite{Norbeck_McWeeny_1975,McDouall_1992,McDouall_1993,Thorsteinsson_Cooper_1998}, effective Hamiltonians\cite{Fuchikami_Block_1982}, perturbation
techniques\cite{Gouyet_1973_use,Gouyet_1974,Mayer_1983,Surjan_etal_1985,Surjan_Poirier_1986,Surjan_Mayer_1991,Finley_Hirao_2000,Visentin_etal_2001},
and the \ac{TC} approach\cite{Fimple_Unwin_1976,Hino_etal_2001}.
In this section, we focus on the use of the biorthonormal framework within the \ac{TC} approach for optimizing orbitals. While this section does not aim to provide a comprehensive overview of the biorthogonal approach, interested readers can refer to
Refs.~\onlinecite{Moshinsky_Seligman_1971,Gouyet_1973_champ,Dahl_1978,Payne_1982,Brody_2014,
Cooper_2002,Surjan_2011}, and the references provided therein.


\subsubsection{Stationary points of the energy functional and non-Hermitian eigenvalue problems}
\label{subsubsec:stationary}

Due to the non-Hermitian nature of the \ac{TC} Hamiltonian, the standard energy minimization approach cannot be used to optimize wavefunctions since the variational principle does not apply. Instead, the search for an energy minimum over a wavefunction $\Psi$ can be replaced by the search for a stationary point of a functional $\etchiphi$ that depends on two wavefunctions $\chi$ and $\Phi$ known as the left- and right-wavefunctions, respectively. 
\begin{equation}
 \label{e_phichi}
 \etchiphi = \frac{\elemm{\chi}{\htc}{\Phi}}{\braket{\chi}{\Phi}}.
\end{equation}
A eigenvalue $\ei$ is obtained as soon as one of the left- or right-function is an eigenfunction
\begin{equation}
 \begin{aligned}
 \etchi{\pphii} = \frac{\elemm{\chi}{\htc}{\pphii}}{\braket{\chi}{\pphii}} = \ei\,\, \forall \chi, \\
 \etphi{\cchii} = \frac{\elemm{\chi_i}{\htc}{\Phi}}{\braket{\chi_i}{\Phi}} = \ei\,\, \forall \Phi. \\
 \end{aligned}
\end{equation}
Hence, finding $\ei$ and the corresponding left- and right-eigenvectors is equivalent to canceling the right- and left-functional derivative
\begin{equation}
 \begin{aligned}
   \fderiv{\etchi{\pphii}}{\chi} = 0 \,\, \forall  \,\, \chi, \\
   \fderiv{\etphi{\cchii}}{\Phi} = 0 \,\,  \forall  \,\, \Phi, \\
 \end{aligned}
\end{equation}
which have general forms given by
\begin{equation}
 \label{deriv_chi}
 \fderiv{\etchiphi}{\chi} =
 \frac{\left( \htc \Phi \right) \, \braket{\chi}{\Phi} - \elemm{\chi}{\htc}{\Phi} \Phi}
      {|\braket{\chi}{\Phi}|^2},
\end{equation}
\begin{equation}
 \label{deriv_phi}
 \fderiv{\etchiphi}{\Phi} =
 \frac{\left( \htc^\dagger \chi \right) \, \braket{\chi}{\Phi} - \elemm{\chi}{\htc}{\Phi} \chi}
      {|\braket{\chi}{\Phi}|^2}.
\end{equation}
It is worth highlighting that canceling the \emph{left}-functional derivative \emph{for all left-wavefunctions $\chi$} enables the determination of the \emph{optimal right-wavefunction} (and \textit{vice versa}).


\subsubsection{Left- and right-Slater determinants: biorthonormal framework}
\label{sec:lef-right_det}

To optimize the orbitals at the single Slater determinant level, one can search for a stationary point of $\etchiphi$, where ${\chi}$ and ${\Phi}$ are single-determinant wavefunctions.
These two Slater determinants are constructed from two different sets of real-valued orbitals $\basis^l=~\{~\chi_i(\br{}),i=1,n \}$ and $\basis^r=\{ \phi_i(\br{}),i=1,n \}$, which are chosen to satisfy the biorthonormal relation
\begin{equation}
 \label{eq:def_bi_ortho}
 \braket{\chi_i}{\phi_j} = \delta_{ij}.
\end{equation}
The use of two biorthonormal bases has the advantage of enabling the construction of creation operators $\bac{k,\sigma}$ and annihilation operators $\ban{l,\lambda}$
(where $k$ and $l$ are labels of spin-free orbitals, and $\sigma$ and $\lambda$ are labels of spins) that satisfy the usual anticommutation relation\cite{Moshinsky_Seligman_1971,Gouyet_1973_champ,Fimple_Unwin_1976,Dahl_1978,Payne_1982,Hino_etal_2001,Surjan_2011}
\begin{equation}
\begin{aligned}
        \label{eq:commut_b}
        [\bac{k,\sigma},\bac{l,\lambda}]_{+} &= 0 \\
        [\ban{k,\sigma},\ban{l,\lambda}]_{+} &= 0 \\
        [\bac{k,\sigma},\ban{l,\lambda}]_{+} &= \delta_{kl} \delta_{\sigma \lambda}
        \text{.}
\end{aligned}
\end{equation}
and avoid the complications introduced by overlap integrals between non-orthogonal basis
functions\cite{Takano_1959,Cantu_etal_1975,Kvasnicka_1977,Kojo_Hirose_2009}.
As a result, expressing an operator in second quantization using the biorthonormal bases $\basis^l$ and $\basis^r$ involves two straightforward rules: (i) replace the conventional creation and annihilation operators $\aic{k,\sigma}$ and $\ai{k,\sigma}$ with the biorthonormal basis's creation and annihilation operators $\bac{k,\sigma}$ and $\ban{k,\sigma}$, and (ii) write the integrals of the operator using the functions ${\chi_i}$ in the bra and ${\phi_j}$ in the ket.
For example, a two-electron operator $\hat{O} = \sum_{i>j}^{\Ne} o(\br{i},\br{j})$ can be expressed in second quantization using a biorthonormal basis as
\begin{equation}
 \hat{O} = \frac{1}{2}\sum_{\sigma,\lambda = \uparrow,\downarrow}\,\sum_{i,j \in \basis^r}\, \sum_{k,l \in \basis^l}
 \tilde{O}_{ij}^{kl} \, \bac{k,\sigma} \bac{l,\lambda} \ban{j,\lambda} \ban{i,\sigma},\\
\end{equation}
where $\tilde{O}_{ij}^{kl}$ represents the integrals in the biorthonormal basis
\begin{equation}
 \begin{aligned}
 \tilde{O}_{ij}^{kl} & = \elemm{\chi_l \chi_k }{\hat{o}}{\phi_j \phi_i} \\
                     & = \int \text{d} \br{1} \text{d} \br{2}\, \chi_l(\br{2}) \chi_k(\br{1})\,
                                      o(\br{1},\br{2})\, \phi_j(\br{2})\phi_i(\br{1}).
 \end{aligned}
\end{equation}
It should be noted that the integrals $\tilde{O}_{ij}^{kl}$ form a tensor that is not necessarily symmetric with respect to permutations between the bra and ket indices, even if the operator $\hat{O}$ is Hermitian.

The determinants $\ket{\chi}$ and $\ket{\Phi}$ can be expressed using orbital rotations
of reference Slater determinants $\ket{\chi_0}$ and $\ket{\Phi_0}$
\begin{equation}
 \begin{aligned}
  \ket{\Phi[\hat{\kappa}^R]} = e^{\hat{\kappa}^R} \ket{\Phi_0} \\
  \ket{\chi[\hat{\kappa}^L]} = e^{\hat{\kappa}^L} \ket{\chi_0},\\
 \end{aligned}
\end{equation}
where $\hat{\kappa}^L$ and $\hat{\kappa}^R$ are the anti-Hermitian orbital rotation operators given by
\begin{equation}
 \begin{aligned}
 &\hat{\kappa}^{R/L}  = \sum_{p>q} \kappa_{pq}^{R/L} \big(\hat{E}_{pq} - \hat{E}_{qp} \big) \\
 &\hat{E}_{pq} = \bac{p\uparrow} \ban{q\uparrow} + \bac{q\uparrow} \ban{p\uparrow} ,
 \end{aligned}
\end{equation}
where the coefficients $\kappa_{pq}^{L}$ and $\kappa_{pq}^{R}$ can differ from each other due to the non-Hermitian nature of the \ac{TC} Hamiltonian.

\subsubsection{Left- and right-Brillouin condition in the TC framework}
\label{subsubsec:Brillouin_theorem}
The bi-functional corresponding to the TC-SCF functional is defined as
\begin{equation}
 \label{eq:tc_bio}
 \tilde{E}\big[\chi_0,\hat{\kappa}^L,\Phi_0,\hat{\kappa}^R\big] =
\frac{\elemm{\chi_0}{e^{-\hat{\kappa}^L}\htc e^{\hat{\kappa}^R}}{\Phi_0}}
{ \elemm{\chi_0}{e^{-\hat{\kappa}^L}e^{\hat{\kappa}^R}}{\Phi_0} }.
\end{equation}
The equivalent of the usual minimization condition in the HF equations is to search for a stationary point $(\tilde{\chi}_0,\tilde{\Phi}_0)$ of the functional in Eq.~\eqref{eq:tc_bio} evaluated at $\kappa^L=\kappa^R=0$, which translates into the two following equations
\begin{equation}
 \label{eq_stat_1}
 \deriv{}{\kappa^L_{pq}}{} \tilde{E}\big[\tilde{\chi}_0,\hat{\kappa}^L=0,\tilde{\Phi}_0,\hat{\kappa}^R=0\big] = 0 \quad \forall \,\,p,q,
\end{equation}
\begin{equation}
 \label{eq_stat_2}
 \deriv{}{\kappa^R_{pq}}{} \tilde{E}\big[\tilde{\chi}_0,\hat{\kappa}^L=0,\tilde{\Phi}_0,\hat{\kappa}^R=0\big] = 0 \quad \forall \,\,p,q.
\end{equation}
Since the derivatives are taken at $\kappa^L=\kappa^R=0$, one can Taylor expand 
the numerator and denominator of $\tilde{E}\big[\chi_0,\hat{\kappa}^L,\Phi_0,\hat{\kappa}^R\big]$ up to first-order,
\begin{equation}
 \begin{aligned}
 \elemm{\tilde{\chi}_0}{e^{-\hat{\kappa}^L}\htc e^{\hat{\kappa}^R}}{\tilde{\Phi}_0} \approx
\elemm{\tilde{\chi}_0}{\htc}{\tilde{\Phi}_0} & - \elemm{\tilde{\chi}_0}{\kappa^L\htc}{\tilde{\Phi}_0} \\
                                             &+ \elemm{\tilde{\chi}_0}{\htc \kappa^R}{\tilde{\Phi}_0},
 \end{aligned}
\end{equation}
\begin{equation}
 \begin{aligned}
\elemm{\tilde{\chi}_0}{e^{-\hat{\kappa}^L} e^{\hat{\kappa}^R}}{\tilde{\Phi}_0} \approx 
\braket{\tilde{\chi}_0}{\tilde{\Phi}_0} & - \elemm{\tilde{\chi}_0}{\kappa^L}{\tilde{\Phi}_0} \\
					& + \elemm{\tilde{\chi}_0}{\kappa^R}{\tilde{\Phi}_0}.
 \end{aligned}
\end{equation}
The use of a biorthonormal orbital basis implies that $\braket{\tilde{\chi}_0}{\tilde{\Phi}_0}=1$ and
$\elemm{\tilde{\chi}_0}{\kappa^L}{\tilde{\Phi}_0} = \elemm{\tilde{\chi}_0}{\kappa^R}{\tilde{\Phi}_0} = 0$.
Taking the derivative with respect to $\kappa^L_{pq}$ evaluated at $\kappa^R=\kappa^L=0$ yields then
\begin{equation}
  \deriv{}{\kappa^L_{pq}}{} \tilde{E}\big[\tilde{\chi}_0,\hat{\kappa}^L=0,\tilde{\Phi}_0,\hat{\kappa}^R=0\big] =  \elemm{\tilde{\chi}_0}{\hat{E}_{qp}\htc}{\tilde{\Phi}_0},
\end{equation}
and similarly for the derivative with respect to $\kappa^R_{pq}$ evaluated at $\kappa^R=\kappa^L=0$
\begin{equation}
 \deriv{}{\kappa^R_{pq}}{} \tilde{E}\big[\tilde{\chi}_0,\hat{\kappa}^L=0,\tilde{\Phi}_0,\hat{\kappa}^R=0\big] =  \elemm{\tilde{\chi}_0}{\htc\hat{E}_{pq}}{\tilde{\Phi}_0},
\end{equation}
which give the left- and right-Brillouin condition for the set of orbitals composing $\ket{\tilde{\chi}_0}$ and $\ket{\tilde{\Phi}_0}$.
The functional of Eq. \eqref{eq:tc_bio} evaluated at the stationary point $({\tilde{\chi}_0},{\tilde{\Phi}_0})$ and $\kappa^L=\kappa^R=0$
defines the \ac{TC}-\ac{BiO} energy
\begin{equation}
 \label{eq:tc_energy}
 \etcbio =
 \tilde{E}\big[\tilde{\chi}_0,\hat{\kappa}^L=0,\tilde{\Phi}_0,\hat{\kappa}^R=0\big].
\end{equation}


\subsubsection{Fock operator and iterative scheme}
\label{sec:fock-op}
Similar to the Hermitian case, satisfying these Brillouin conditions can be achieved through iterative diagonalization of a Fock-like operator. However, in the \ac{TC} framework, this Fock operator is non-Hermitian and constructed using a transition density between the left- and right-functions.
Starting from now on, matrices expressed in the \ac{AO} basis will be labeled by a tilde to distinguish them from matrices expressed in the biorthonormal basis.
The left- and right-functions are both expanded in the same \ac{AO} basis,
\begin{equation}
 \begin{aligned}
 & \chi_i(\br{}) = \sum_{\mu} {C}^L_{\mu i} \,\, \zeta_\mu(\br{}),\\
 & \phi_i(\br{}) = \sum_{\mu} {C}^R_{\mu i} \,\, \zeta_\mu(\br{}),\\
 \end{aligned}
\end{equation}
where $\zeta_\mu(\br{})$ are the atomic basis functions and ${C}^L_{\mu i}$ and ${C}^R_{\mu i}$ are the coefficients of the left- and right-functions expressed in the \ac{AO} basis.
Using matrix notations, the transition density matrix in the \ac{AO} basis $\tilde{\mathbf{P}} = \tilde{\mathbf{P}}^\uparrow + \tilde{\mathbf{P}}^\downarrow$ can be obtained from
\begin{equation}
  \label{eq:density_matrix}
  \tilde{\mathbf{P}}^\sigma = \mathbf{C}^{\text{occ},\sigma}_L \cdot {\mathbf{C}^{\text{occ},\sigma}_R}^\dagger , 
\end{equation}
where $\mathbf{C}^{\text{occ},\sigma}_L$ and $\mathbf{C}^{\text{occ},\sigma}_R$ are matrices that contain the coefficients of left- and right-orbitals, respectively, occupied by electrons with spin $\sigma$.
As the \ac{TC} Hamiltonian involves up to three-body terms, the Fock-matrix can be split into two types of contributions: one from the one- and two-body terms, denoted as $F_{ai}^{\sigma,2}$, and another from the three-body terms, denoted as $F_{ai}^{\sigma,3}$.
The total \ac{TC} Fock matrix element for electrons of spin $\sigma$ between the left-orbital $\chi_a(\br{})$ and right-orbital $\phi_i(\br{})$ is given by 
\begin{equation}
 F_{ai}^{\sigma,\text{TC}} = F_{ai}^{\sigma,2} + F_{ai}^{\sigma,3}.
\end{equation}
The matrix of one- and two-body terms is first computed in the \ac{AO} basis
\begin{equation}
 \tilde{F}_{\mu \nu}^{\sigma,2} = \tilde{h}_{\mu \nu} +
 \sum_{\zeta \lambda} \tilde{P}_{\zeta \lambda} \elemm{\mu \zeta}{\hat{V} - \hat{K}}{\nu \lambda}
  - \tilde{P}_{\zeta \lambda}^\sigma \elemm{\mu \zeta}{\hat{V} - \hat{K}}{\lambda \nu},
\end{equation}
where 
$\tilde{h}_{\mu \nu}$ are the usual one-electron integrals in the \ac{AO} basis and
$\elemm{\mu \zeta}{\hat{V} - \hat{K}}{\nu \lambda}$ are the total two-electron terms of the \ac{TC} operator in the \ac{AO} basis.
The matrix $\tilde{\mathbf{F}}^{\sigma,2}$ is then transformed into the biorthonormal basis as
\begin{equation}
 \mathbf{F}^{\sigma,2} = {\mathbf{C}_L}^\dagger \cdot \tilde{\mathbf{F}}^{\sigma,2}  \cdot \mathbf{C}_R.
\end{equation}
As the three-electron terms are computed using a mixed analytical-numerical scheme, the integrals can be directly evaluated in the biorthonormal \ac{MO} basis.
This saves a large amount of computational time as only $\Ne^2 \times M$ integrals are needed in the \ac{MO} basis, while $M^6$ are needed in the \ac{AO} basis.
Therefore, the contribution for the three-body term is directly written in the biorthonormal \ac{MO} basis.
The equations for the three-body contribution to the Fock matrix are
\begin{equation}
 \begin{aligned}
  F_{ai}^{\alpha,3} = F_{ai}^{\text{cs},3} + A_{ai}^{\beta \alpha}  + A_{ai}^{\alpha \alpha},\\
  F_{ai}^{\beta,3}  = F_{ai}^{\text{cs},3} + B_{ai}^{\beta \alpha}  + B_{ai}^{\alpha \alpha},\\
 \end{aligned}
\end{equation}
where
\begin{equation}
 \begin{aligned}
 F_{ai}^{\text{cs},3} = & \sum_{j,k=1}^{N_\beta} \bigg[2 \braket{akj}{ikj}
 - \frac{3}{2} \braket{akj}{jki} - \braket{akj}{ijk} \\
 &+ \frac{1}{2} \big(\braket{akj}{jik} + \braket{akj}{kji} - \braket{akj}{kij} \big)\bigg],
 \end{aligned}
\end{equation}
\begin{equation}
 \begin{aligned}
 A_{ai}^{\beta \alpha} = & \sum_{j=N_\beta +1}^{N_\alpha}\sum_{j=1}^{N_\beta} \bigg[
  \frac{3}{2} \big( \braket{akj}{ikj} - \braket{akj}{jki}\big) \\
 &+\frac{1}{2} \big( \braket{akj}{jik} + \braket{akj}{kji} - \braket{akj}{ijk} - \braket{akj}{kij} \big)\bigg],
 \end{aligned}
\end{equation}
\begin{equation}
 \begin{aligned}
 A_{ai}^{\alpha \alpha} =  \sum_{j=1}^{N_\alpha}\sum_{j=N_\beta + 1}^{N_\alpha} \bigg[&
    \braket{akj}{ikj} + \braket{akj}{jik} + \braket{akj}{kji} \\
& - \braket{akj}{jki} + \braket{akj}{ijk} + \braket{akj}{kij}  \bigg],
 \end{aligned}
\end{equation}
\begin{equation}
 \begin{aligned}
 B_{ai}^{\beta \alpha} =  \sum_{j=1}^{N_\beta}\sum_{j=N_\beta + 1}^{N_\alpha} \bigg[&
 \frac{3}{2}\braket{akj}{ikj} - \frac{1}{2} \braket{akj}{ijk} - \braket{akj}{jki} \bigg],
 \end{aligned}
\end{equation}
\begin{equation}
 \begin{aligned}
 B_{ai}^{\alpha \alpha} =  \sum_{j=N_\beta + 1}^{N_\alpha}\sum_{j= 1}^{N_\alpha} \frac{1}{2}\bigg[&
 \braket{akj}{ikj} - \braket{akj}{ijk} \bigg],
 \end{aligned}
\end{equation}
$N_\alpha$ and $N_\beta$ are the numbers of $\alpha$ and $\beta$ electrons, and $\braket{mnl}{ijk}$ are the matrix elements of the $-\hat{L}$ operator in the biorthonormal basis.


\subsubsection{Improved \ac{SCF} procedure}

To avoid the \ac{SCF} procedure from diverging due to orbital switching between closed, open, and virtual shells, we employ the \ac{LS} technique\cite{saunders1973level,guest1974methods} on the total Fock matrix expressed in the MO basis.
This involves introducing a shift parameter to the diagonal elements.


To speed up the optimization of orbitals, the \ac{DIIS} algorithm\cite{pulay1980convergence} is used.
In a \ac{DIIS} step, the Fock matrix (expressed in the \ac{AO} basis) is obtained as an interpolation of a set of Fock matrices from previous iterations,
\begin{equation}
	\tilde{\mathbf{F}}^{\text{DIIS}} = \sum_i \, c_i \, \tilde{\mathbf{F}}^{(i)}
	\text{,}
\end{equation}
with
\begin{equation}
	\sum_i \, c_i  = 1
	\text{.}
\end{equation}
The coefficients $c_i$ are obtained by minimizing, in the least-squares sense, the norm of an interpolated error matrix
\begin{equation}
	\mathbf{e}^{\text{DIIS}} = \sum_i \, c_i \, \mathbf{e}^{(i)}
	\text{.}
\end{equation}
Since the density matrix $\tilde{\mathbf{P}}$ is not symmetric,
the definition of the error matrix involves the transpose of the density matrix\cite{salvador2009convergence}
\begin{equation}
	\mathbf{e}^{(i)} = \tilde{\mathbf{F}}^{(i)} \cdot \tilde{\mathbf{P}}^\dagger \cdot \tilde{\mathbf{S}} 
			- \tilde{\mathbf{S}} \cdot \tilde{\mathbf{P}}^\dagger \cdot \tilde{\mathbf{F}}^{(i)}
	\text{,}
\end{equation}
where $\tilde{\mathbf{S}}$ is the \ac{AO}-overlap matrix.
When the error is zero, the right- and left-Brillouin's conditions are fulfilled.
%

\subsection{\ac{VMC} energies of various types of wavefunctions}
\label{sec:vmc}
To study the effect of orbital optimization, we computed the \ac{VMC} energies of three types of Slater-Jastrow wavefunctions that differed only in their orbitals. The first one was the
usual \ac{HF} wavefunction multiplied by the correlation factor
\begin{equation}
 \ket{\hfj} = \exp\qty({\sum_{i>j}\unew{i}{j}}) \ket{\text{HF}}.
\end{equation}
The second wavefunction, denoted as $\ket{\tcbioj}$, was obtained by taking the stationary point of the functional defined in Eq. \eqref{eq:tc_bio} for the right-determinant $\ket{\tilde{\Phi}_0}$ and multiplying it by the same correlation factor as the first wavefunction
\begin{equation}
 \label{def:tcbioj}
 \ket{\tcbioj} =  \exp\qty({\sum_{i>j}\unew{i}{j}}) \ket{\tilde{\Phi}_0}.
\end{equation}
Finally, we denote by $\ket{\varj}$ the wavefunction with the Slater determinant constructed from the optimal orbitals obtained from the \ac{TC}-\ac{Var} procedure\cite{Luo_2010,Luo_2011}.

\section{Results}
\label{sec:results}
\subsection{Computational details and general strategy for optimization}
We have implemented the $\tcbio$ SCF program, along with all the required integrals (detailed in the appendix), as a plugin in the Quantum Package software\cite{QP2-JCTC-19}.
For the calculations, we used the cc-pVXZ family of Dunning \ac{AO} basis sets\cite{Dun-JCP-89}.

The correlation factor $\unew{1}{2}$ introduced in this work (see Eq.~\eqref{def:unew}) depends on two types of parameters:
one single parameter $\mu$ controlling the typical range and depth of the universal two-electron correlation factor $\umu{1}{2}$ (see Eq.~\eqref{eq:def_j}),
and a set of parameters $\{ \alpha_m,m=1,\nuclnum\}$ controlling the range of the damping function around each nucleus at position $\brbig{m}$.
Our strategy in this work is to use a universal two-electron correlation factor that is designed for typical valence electrons, and then to optimize the nuclear parameters to ensure that the correlation factor is adapted to this valence correlation factor.
The value of $\mu=0.87$ was chosen because it has been shown previously\cite{Gin-JCP-21} that the correlation factor $\umu{1}{2}$ accurately reproduces the results of the FROGG\cite{TenNo-CPL-00-a} correlation factor, which was designed for valence electrons.
To optimize the nuclear parameters $\alpha_m$, we compute the \ac{VMC} energy of the $\ket{\tcbioj}$ right-wavefunction (Eq. \eqref{def:tcbioj}) for each value of $\alpha_m$.

\subsection{Comparison between the $\tcbio$ and \ac{TC}-\ac{Var} orbital optimization}
\label{sec:tcbio_vs_vartc}

\begin{figure}
	\centering
		\includegraphics[width=\linewidth]{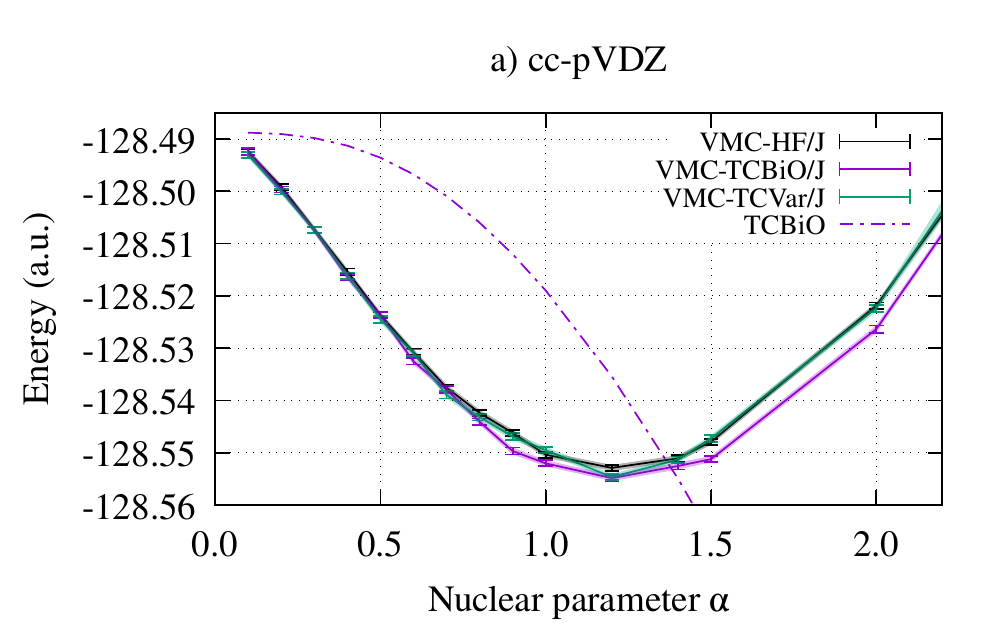}
		\includegraphics[width=\linewidth]{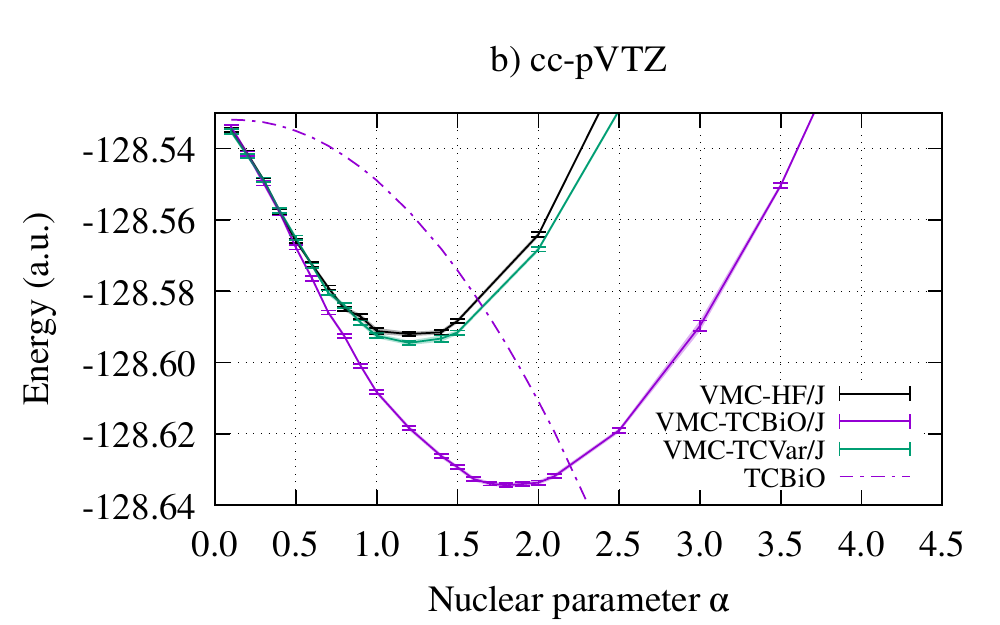}
		\includegraphics[width=\linewidth]{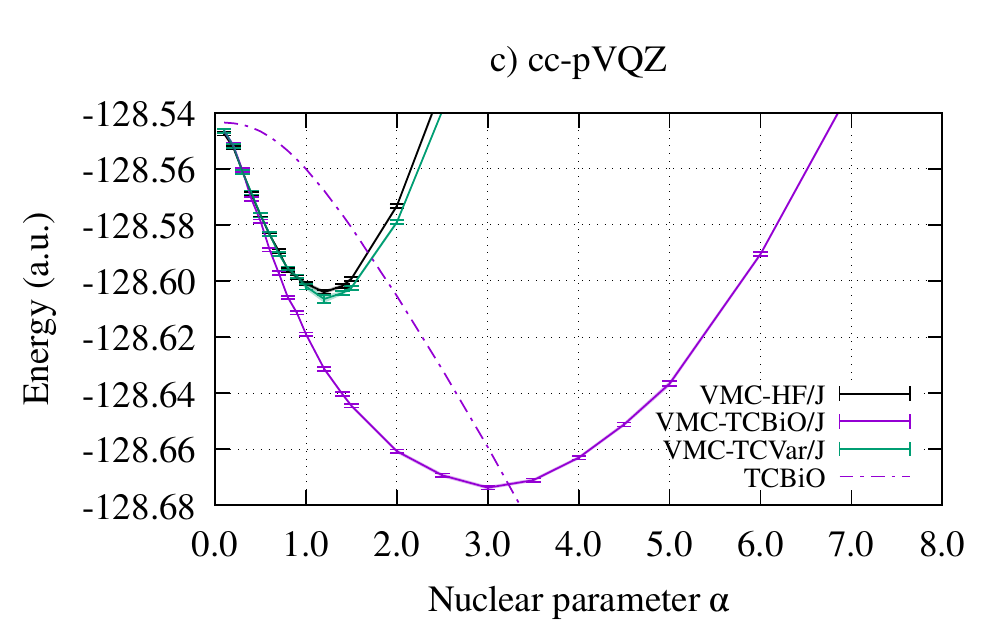}
	\caption{\label{fig:Neon-TCVar_vs_TCBiO}
Comparison of the \ac{VMC} energies obtained with the $\hfj$, $\tcbioj$, and $\varj$ right-wavefunctions (see Sec. \ref{sec:vmc}) for the Neon atom with the cc-pVDZ (a), cc-pVTZ (b), and cc-pVQZ (c) basis sets.
	}
\end{figure}

We start our analysis by examining the effect of the orbital optimization on the Ne atom with the cc-pVXZ (X=D,T,Q) basis set family.
In Fig.~\ref{fig:Neon-TCVar_vs_TCBiO}, we compare the \ac{VMC} energies (see Sec.~\ref{sec:vmc}) of the $\hfj$, $\tcbioj$, and $\varj$ right-wavefunctions as functions of the nuclear parameter $\alpha$.
Additionally, we report the value of the TC-energy $\etcbio$ (see Eq.~\eqref{eq:tc_energy}).

From the figure, we observe that the \ac{VMC} energies of the $\varj$ and $\hfj$ right-wavefunctions are very similar, indicating that the orbital optimization of the TC-Var approach does not lead to a significant improvement in \ac{VMC} energy. However, in the cc-pVTZ and cc-pVQZ basis sets, the \ac{VMC} energy of the $\tcbio$ right-wavefunction is clearly lower than that of the $\hfj$ right-wavefunction, indicating that the TC-biorthonormal framework can significantly improve the \ac{VMC} energy.

Regarding the variation of the \ac{VMC} energy with the $\alpha$ nuclear parameter, we observe that even without re-optimizing the orbitals, tuning the $\alpha$ parameter can result in a gain of about 60~mH. The re-optimization of orbitals within the $\tcbio$ framework allows for a further lowering of the energy by about 40~mH and~70 mH in the cc-pVTZ and cc-pVQZ basis sets, respectively. In total, the optimal $\tcbio$ right-wavefunction can lower the \ac{VMC} energy by about 60~mH, 100~mH, and 140~mH in the cc-pVDZ, cc-pVTZ, and cc-pVQZ basis sets, respectively.

Regarding the optimal $\alpha$ parameter within the $\tcbio$ framework, we observe that it tends to increase with the basis set. As the $\alpha \rightarrow 0$ limit corresponds to a vanishing correlation factor, and as the $\alpha \rightarrow \infty$ limit corresponds to the full universal two-electron correlation factor $\umu{1}{2}$, the increase in the optimal $\alpha$ parameter implies that allowing for more flexibility in the wavefunction to adapt to the correlation factor can enable the capture of more correlation effects from the two-electron correlation factor $\umu{1}{2}$.

Finally, we note that the value of $\alpha$ for which the $\etcbio$ energy matches the \ac{VMC} energy of the $\tcbioj$ right-wavefunction is very close to the optimal parameter $\alpha$.


\subsection{Optimization of nuclear parameters for second row atoms and transition elements}
\label{opt-atom}

\begin{figure*}
	\centering
	\begin{minipage}{0.42\textwidth}
		\centering
		\includegraphics[width=1.05\linewidth]{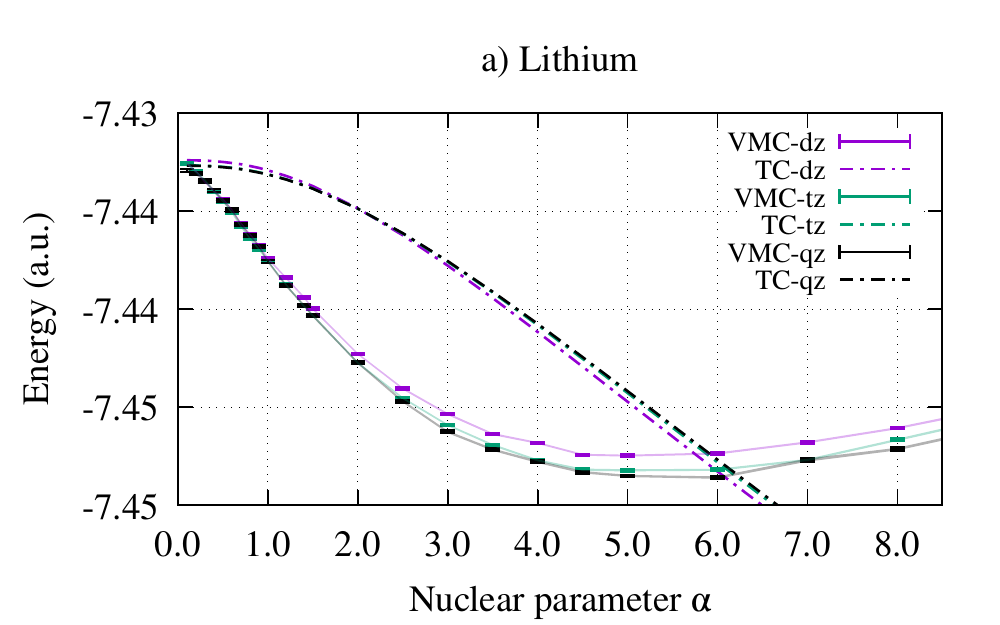}
	\end{minipage}
	\hfill
	\begin{minipage}{0.42\textwidth}
		\centering
		\includegraphics[width=1.05\linewidth]{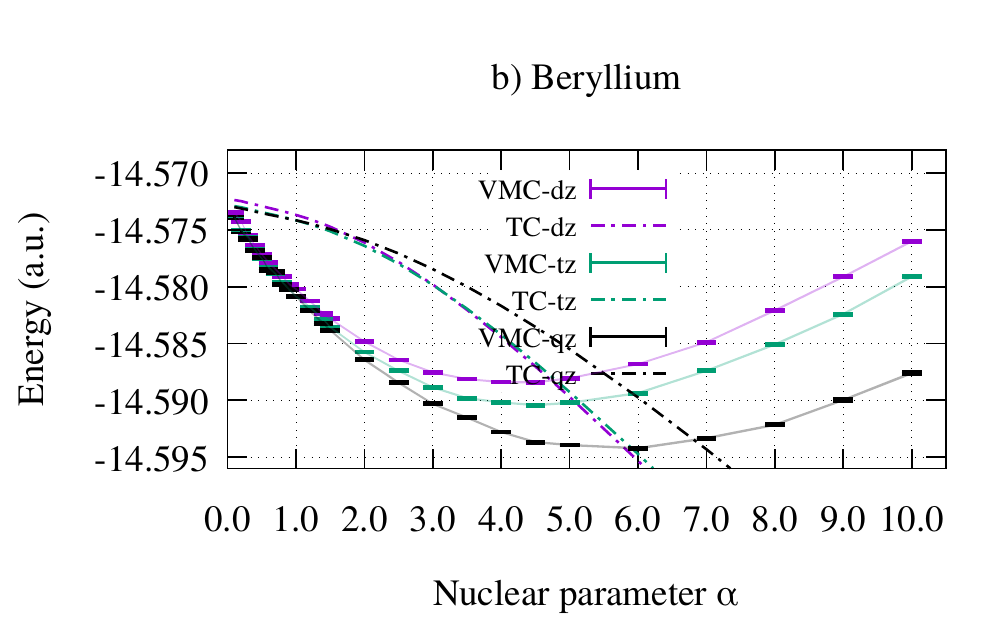}
	\end{minipage}
	\hfill
	\begin{minipage}{0.42\textwidth}
		\centering
		\includegraphics[width=1.05\linewidth]{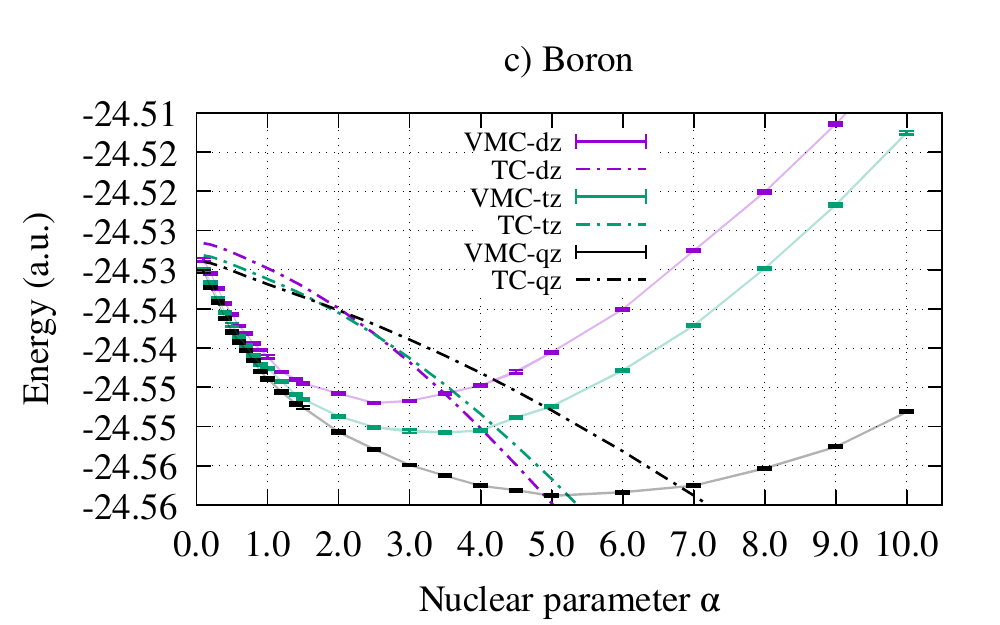}
	\end{minipage}
	\hfill
	\begin{minipage}{0.42\textwidth}
		\centering
		\includegraphics[width=1.05\linewidth]{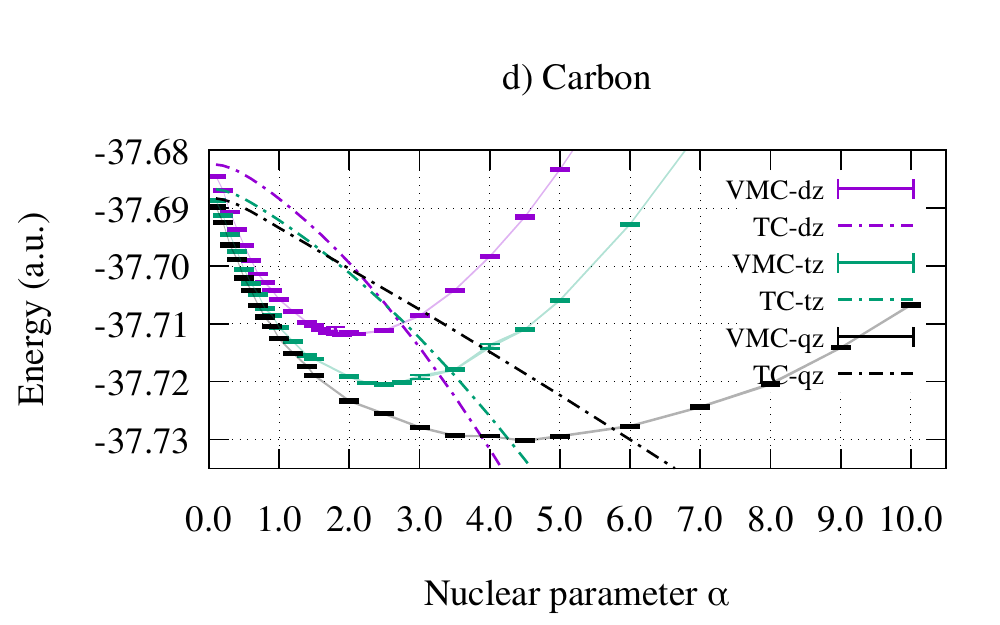}
	\end{minipage}
	\hfill
	\begin{minipage}{0.42\textwidth}
		\centering
		\includegraphics[width=1.05\linewidth]{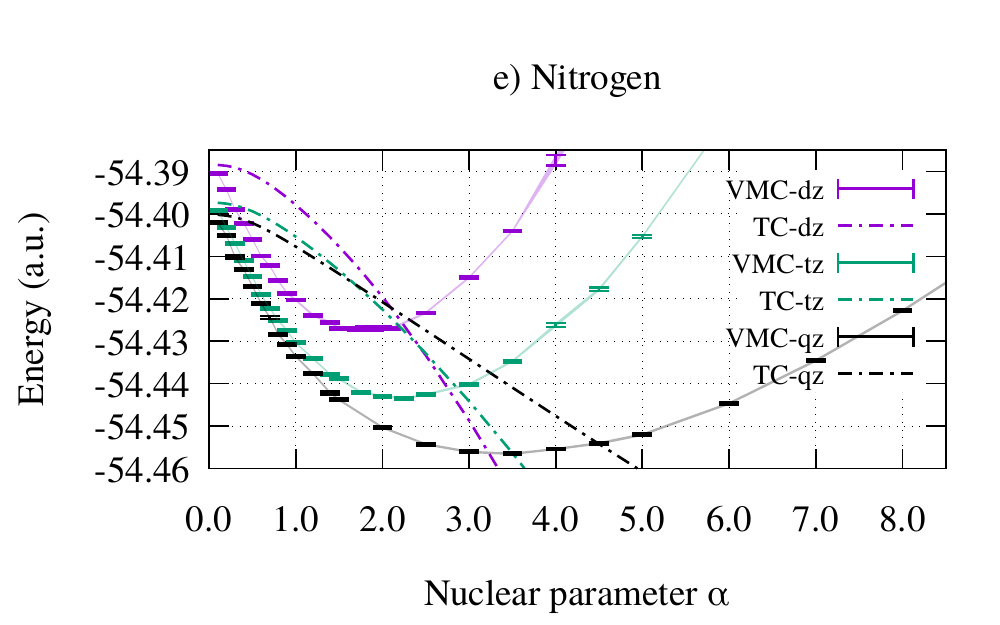}
	\end{minipage}
	\hfill
	\begin{minipage}{0.42\textwidth}
		\centering
		\includegraphics[width=1.05\linewidth]{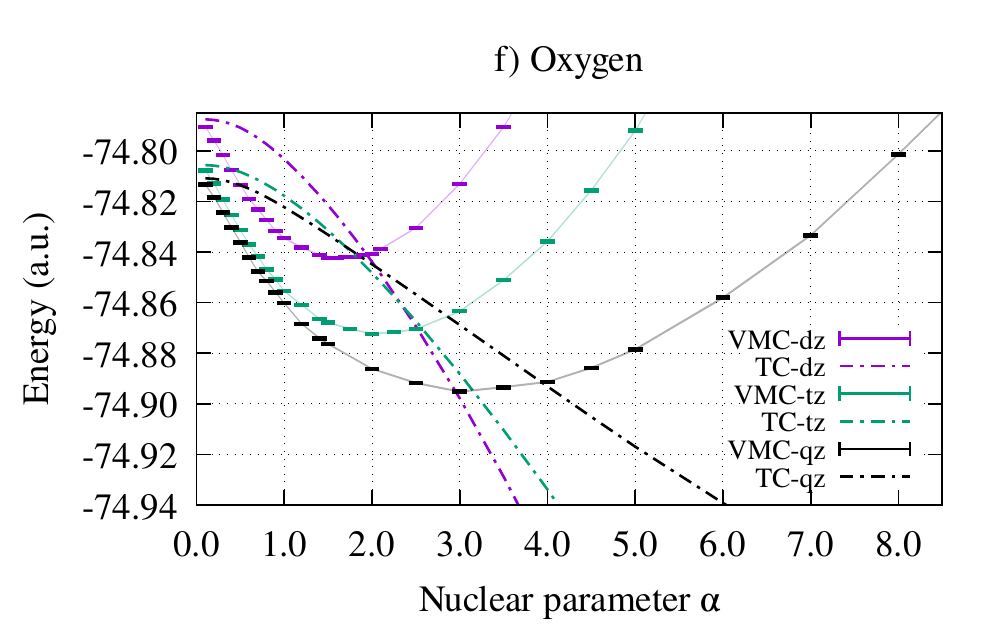}
	\end{minipage}
	\hfill
	\begin{minipage}{0.42\textwidth}
		\centering
		\includegraphics[width=1.05\linewidth]{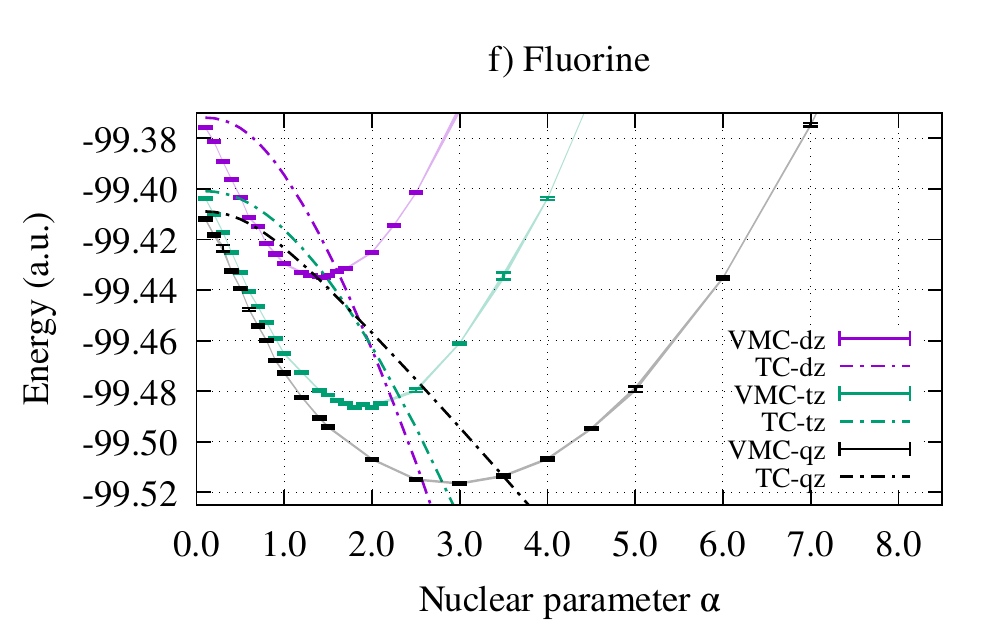}
	\end{minipage}
	\hfill
	\begin{minipage}{0.42\textwidth}
		\centering
		\includegraphics[width=1.05\linewidth]{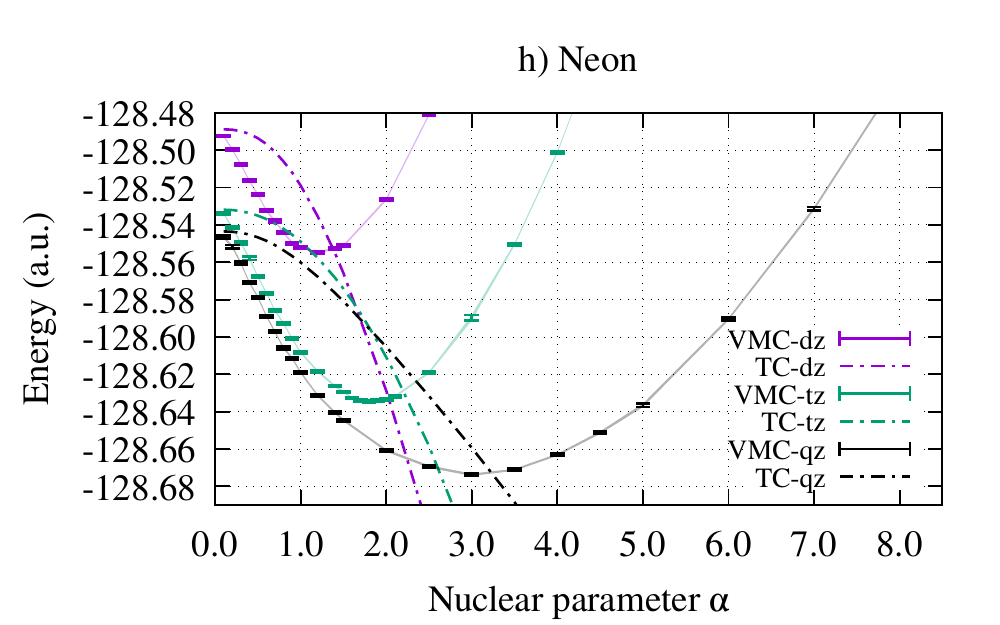}
	\end{minipage}
	\hfill
	\caption{\label{fig:opt_Li_Ne}
Comparison of the \ac{VMC} energies of the $\tcbioj$ right-wavefunctions and $\etcbio$ energies in the cc-pVXZ (X=D,T,Q) basis sets for the Li-Ne neutral species.
	}
\end{figure*}

\begin{figure}
	\centering
		\includegraphics[width=\linewidth]{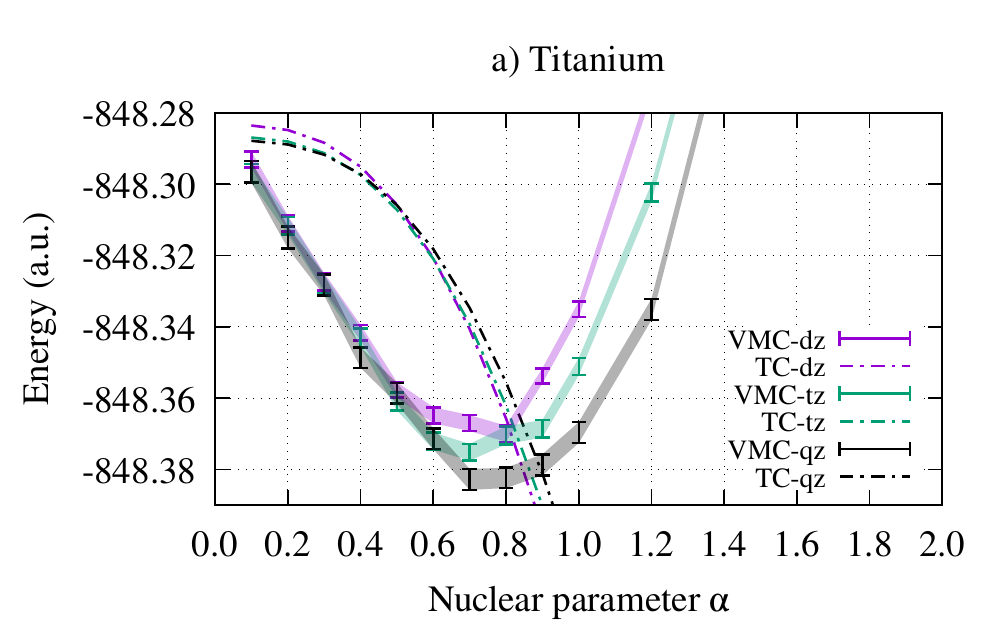}
		\includegraphics[width=\linewidth]{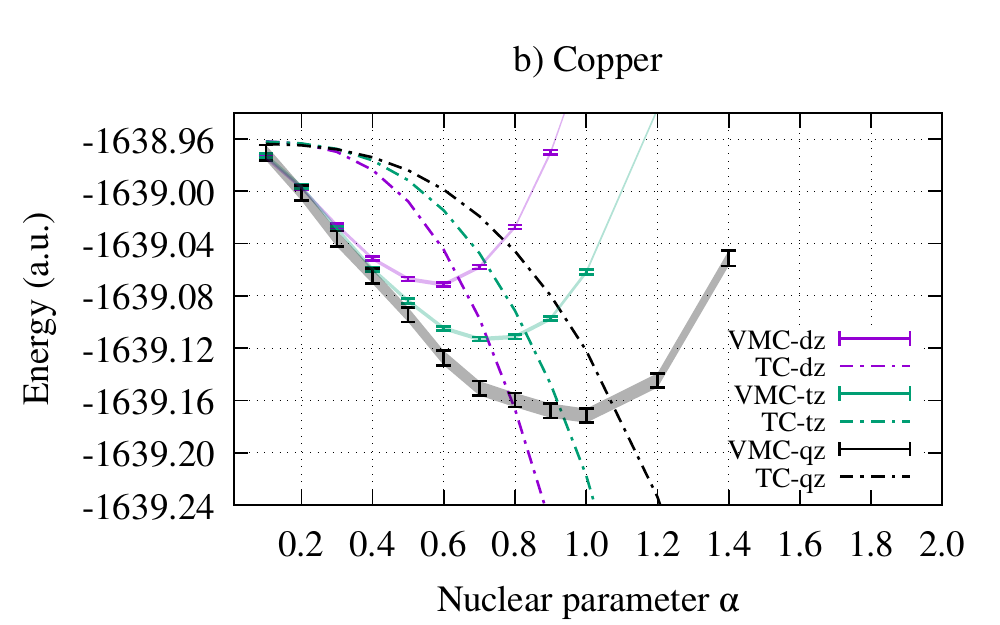}
	\caption{\label{fig:opt_Cu} Comparison of the \ac{VMC} energies of the $\tcbioj$ right-wavefunctions and $\etcbio$ energies in the cc-pVXZ (X=D,T) basis sets for the titanium and copper atoms.
	}
\end{figure}

In the cc-pVXZ basis sets (X=D,T,Q), we optimized the nuclear parameter $\alpha$ of the $\tcbioj$ right-wavefunction using the same approach as in Sec.~\ref{sec:tcbio_vs_vartc}.
Fig.~\ref{fig:opt_Li_Ne} and Fig.~\ref{fig:opt_Cu} show the \ac{VMC} energy of the $\tcbioj$ right-wavefunction (referred to as \emph{VMC}) and the $\etcbio$ energy (referred to as \emph{TC}) for the Li-Ne series, and the titanium and copper atoms. These figures reveal several trends:
i) as the basis set is improved, the optimal value of $\alpha$ increases, and more correlation energy is recovered at the \ac{VMC} level through orbital optimization;
ii) the value of $\alpha$ for which the $\etcbio$ energy matches the \ac{VMC} energy of the $\tcbioj$ right-wavefunction is always close to the optimal value of $\alpha$;
iii) in a given basis set, the optimal value of $\alpha$ decreases as the nuclear charge of the atom increases. This is expected as the core electrons are more strongly attracted to the nucleus.
It is noteworthy that even in the case of the copper atom, the orbital optimization still manages to lower the \ac{VMC} energy, allowing for a gain of about 110~mH and 28~mH with respect to the \ac{HF} and $\hfj$ right-wavefunctions in cc-pVDZ, respectively.
A similar improvement is observed in cc-pVTZ and cc-pVQZ.
The systematic nature of these observations for a broad set of atomic systems demonstrates the applicability of the newly proposed correlation factor.


\subsection{Transferability of optimal atomic $\alpha$ parameters to molecules}
\label{transfer-mol}

\begin{figure}
	\centering
		\includegraphics[width=\linewidth]{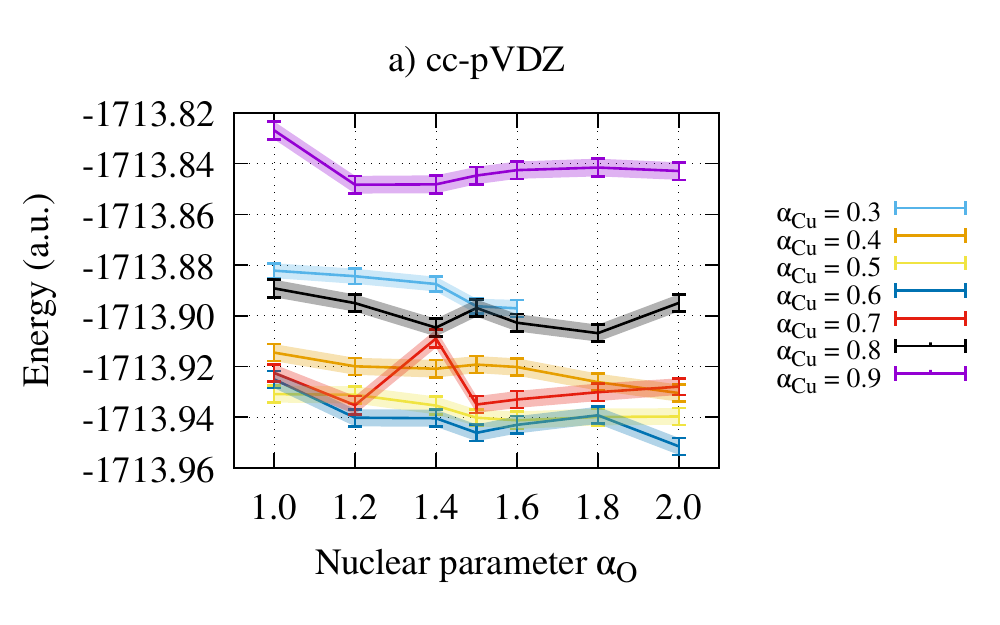}
		\includegraphics[width=\linewidth]{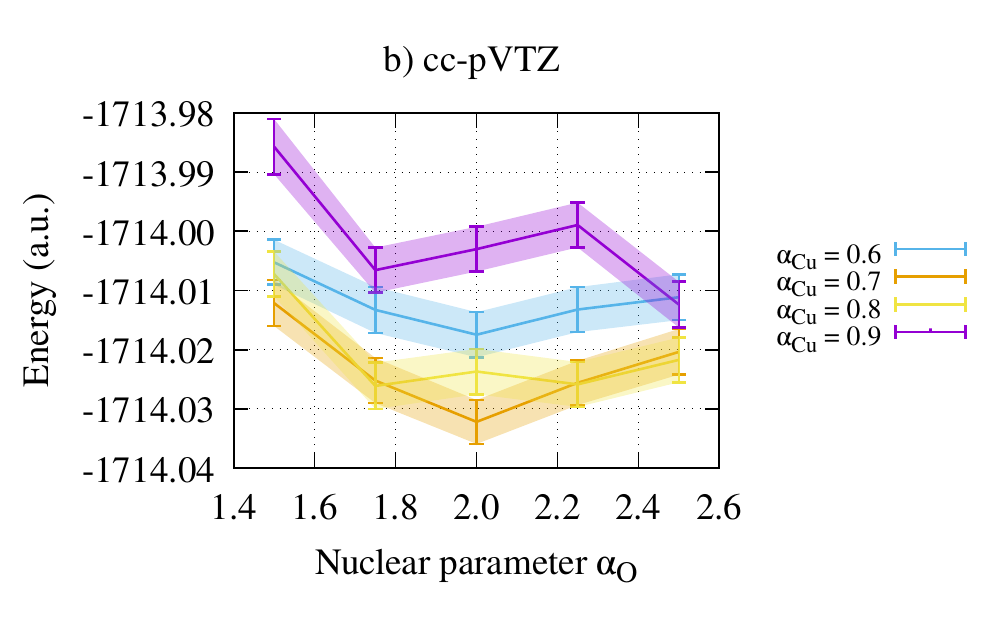}
		\includegraphics[width=\linewidth]{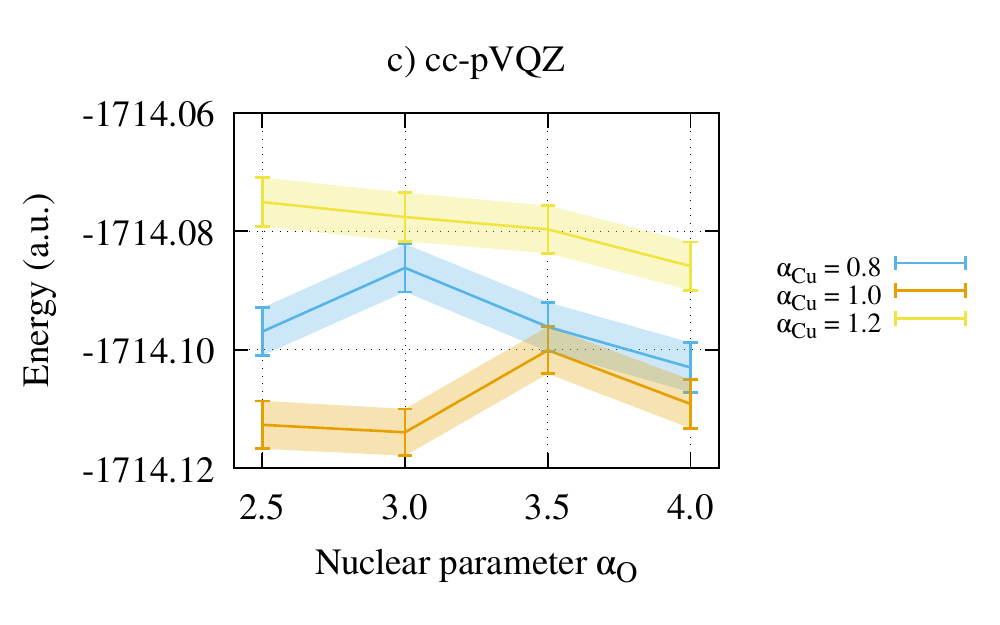}
	\caption{\label{fig:opt_CuO}\ac{VMC} energies of the $\tcbioj$ right-wavefunction (VMC) and $\etcbio$ energy
	in the cc-pVXZ (X=D,T,Q) for the \ce{CuO} molecule.
	}
\end{figure}

\begin{figure*}
	\centering
	\begin{minipage}{0.42\textwidth}
		\centering
		\includegraphics[width=1.05\linewidth]{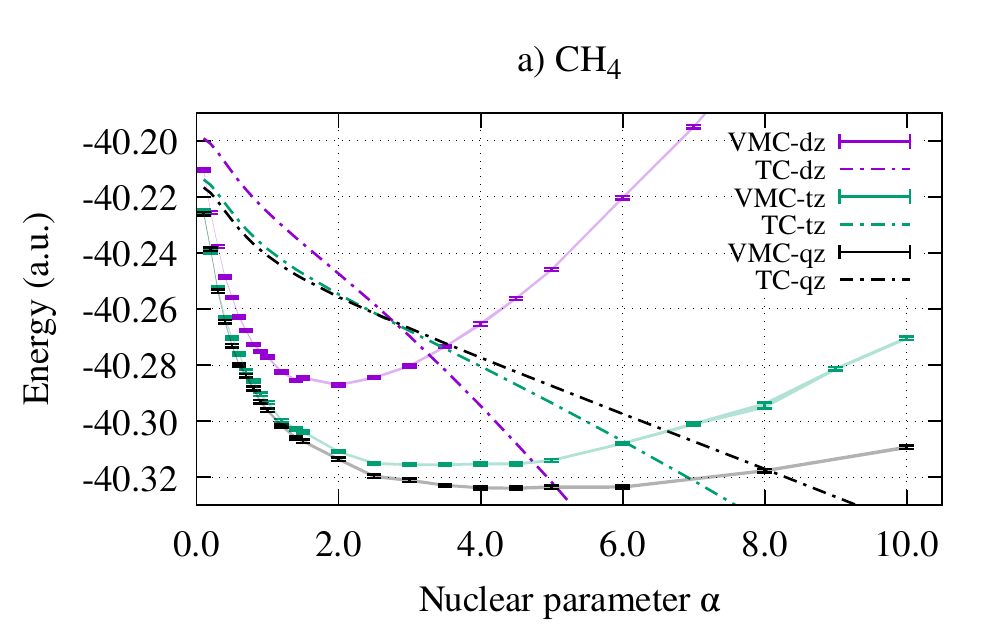}
	\end{minipage}
	\hfill
	\begin{minipage}{0.42\textwidth}
		\centering
		\includegraphics[width=1.05\linewidth]{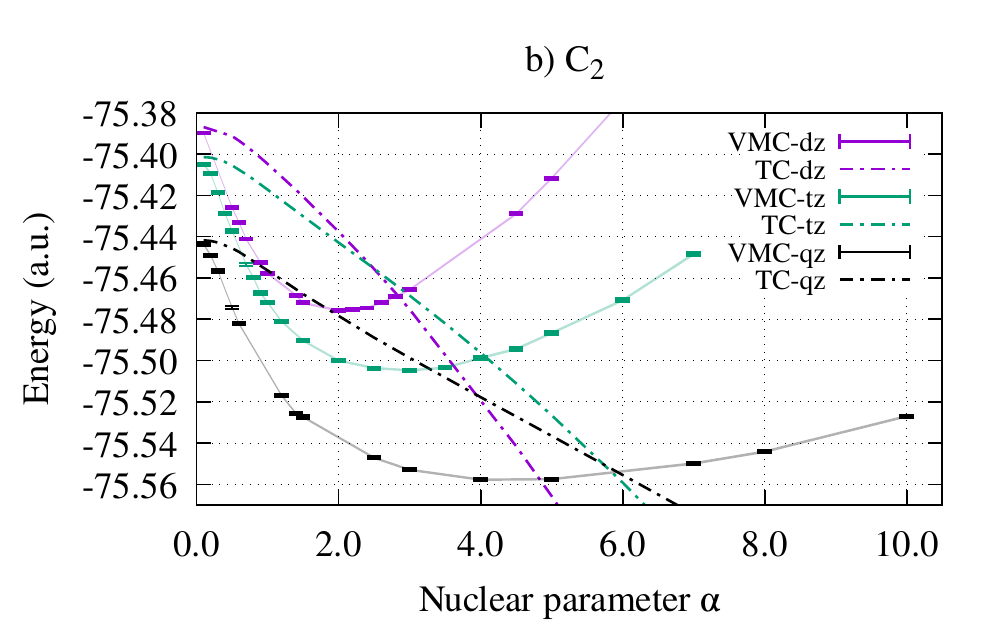}
	\end{minipage}
	\hfill
	\begin{minipage}{0.42\textwidth}
		\centering
		\includegraphics[width=1.05\linewidth]{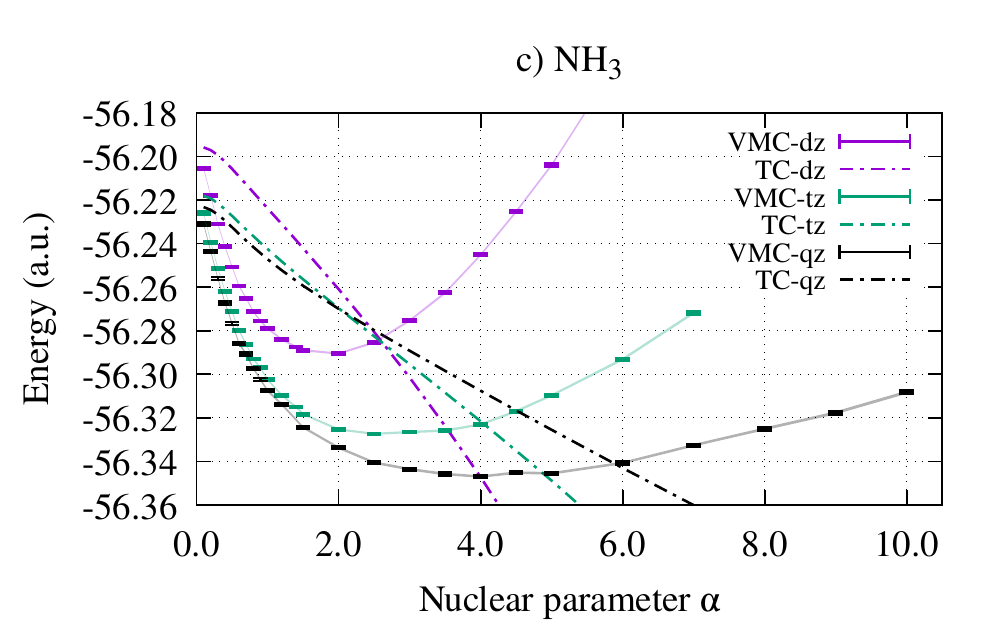}
	\end{minipage}
	\hfill
	\begin{minipage}{0.42\textwidth}
		\centering
		\includegraphics[width=1.05\linewidth]{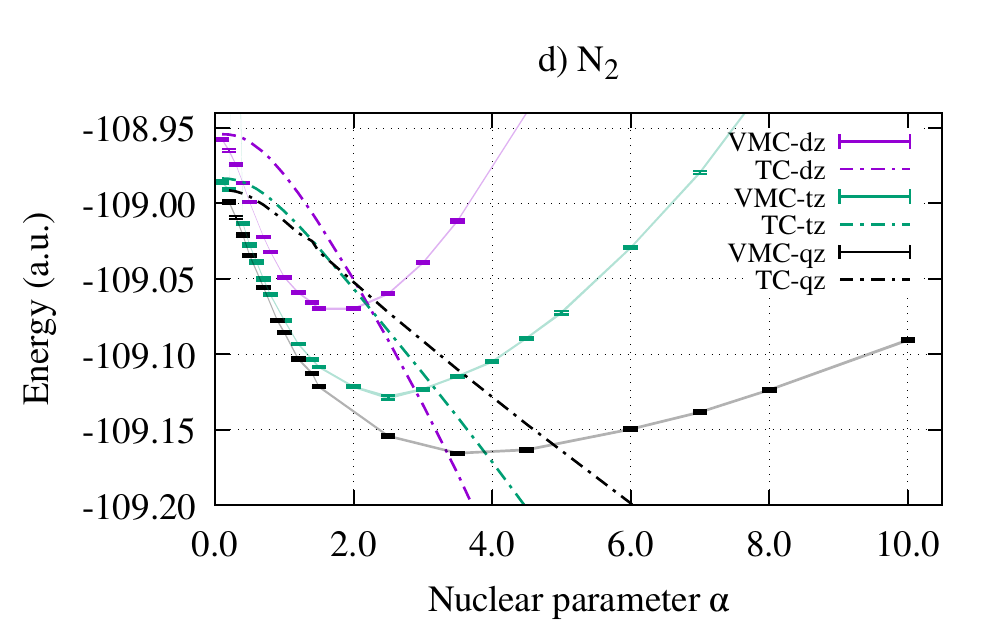}
	\end{minipage}
	\hfill
	\begin{minipage}{0.42\textwidth}
		\centering
		\includegraphics[width=1.05\linewidth]{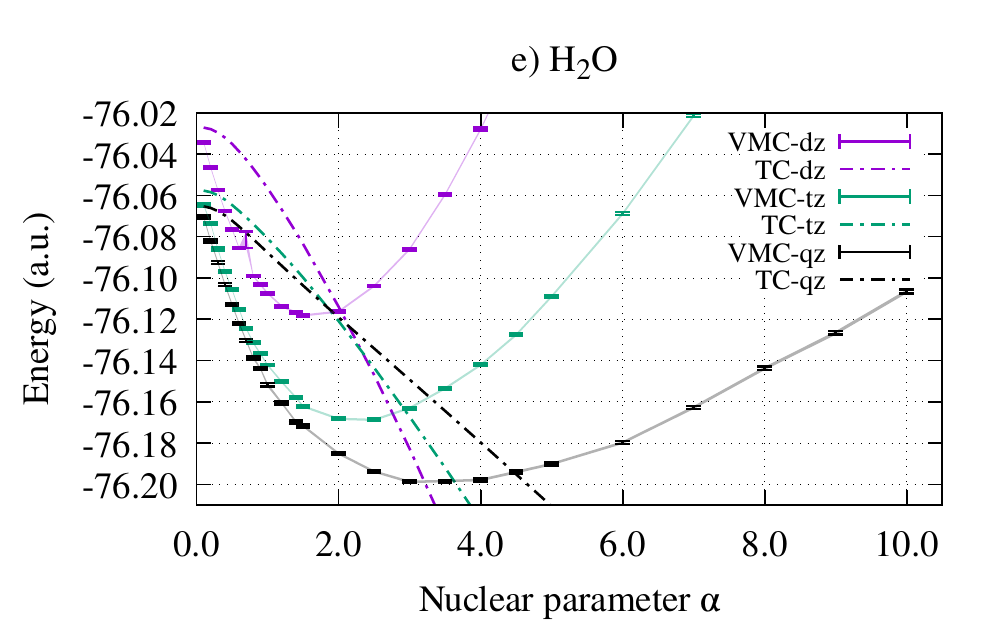}
	\end{minipage}
	\hfill
	\begin{minipage}{0.42\textwidth}
		\centering
		\includegraphics[width=1.05\linewidth]{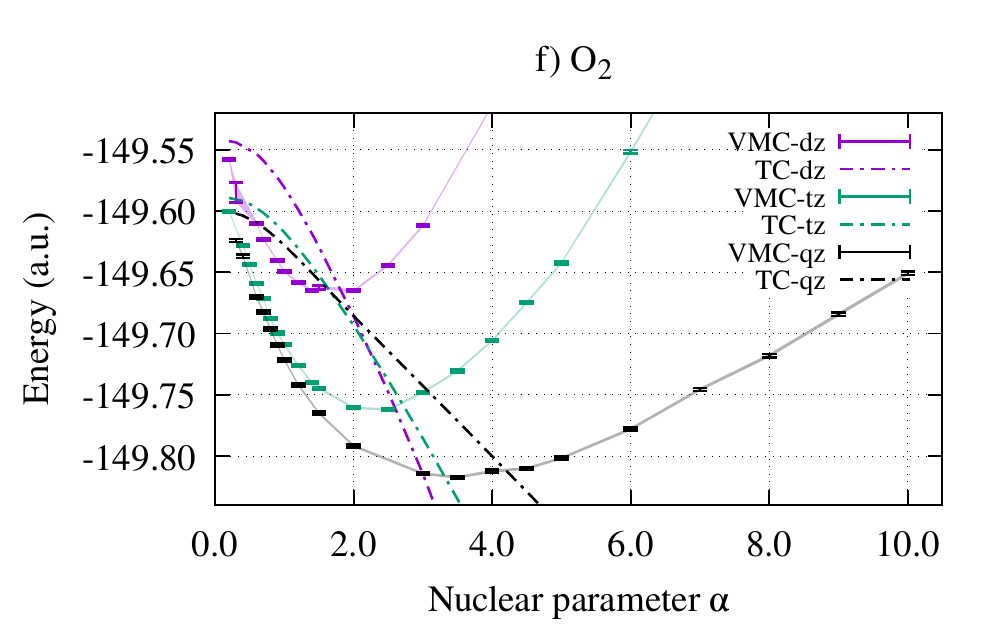}
	\end{minipage}
	\hfill
	\begin{minipage}{0.42\textwidth}
		\centering
		\includegraphics[width=1.05\linewidth]{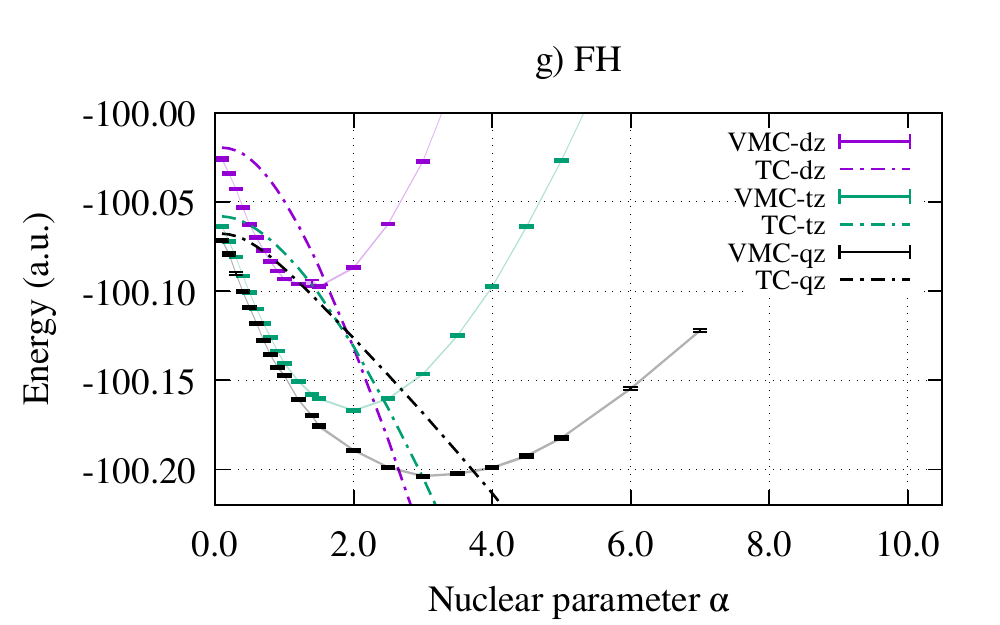}
	\end{minipage}
	\hfill
	\begin{minipage}{0.42\textwidth}
		\centering
		\includegraphics[width=1.05\linewidth]{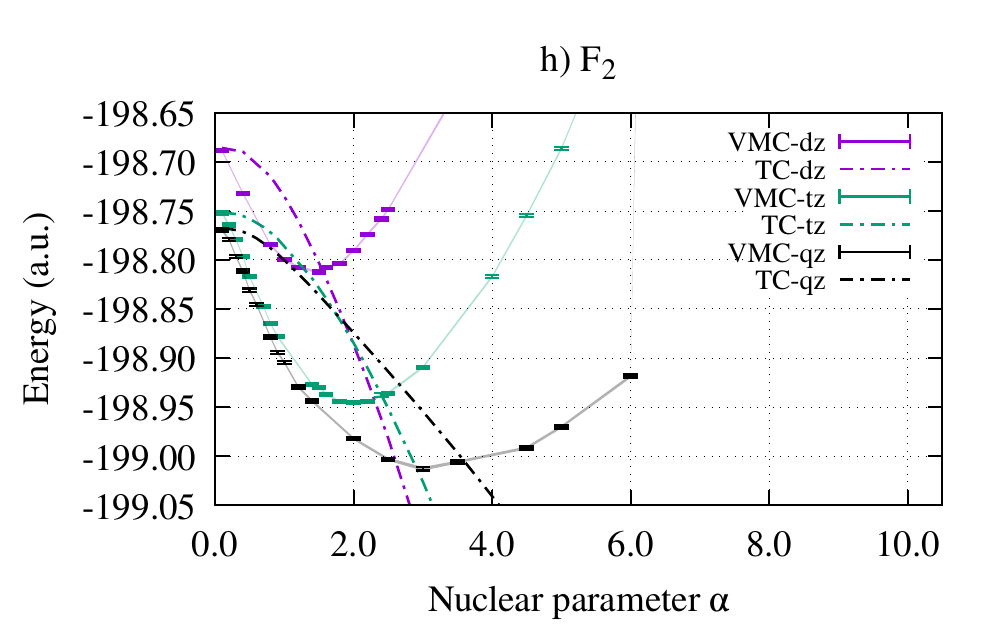}
	\end{minipage}
	\hfill
	\caption{\label{fig:molecules}\ac{VMC} energies of the $\tcbioj$ right-wavefunction (VMC) and $\etcbio$ energy
      in the cc-pVXZ basis sets (X=D,T,Q) for a set of molecules containing second-row elements.
	}
\end{figure*}

To investigate the transferability of the optimal $\alpha$ parameters obtained for purely atomic systems to molecules, we optimized the $\alpha$ parameter for a set of small molecules, namely \ce{C2}, \ce{CH4}, \ce{N2}, \ce{NH3}, \ce{O2}, \ce{H2O}, \ce{F2}, \ce{FH} and CuO, in the cc-pVXZ (X=D,T,Q) basis sets.
The geometries can be found in the supplementary material.
The $\alpha$ parameter was set to $\infty$ for all hydrogen atoms as they do not have core electrons.
Fig.~\ref{fig:opt_CuO} shows the simultaneous optimization of the $\alpha$ parameters for both copper and oxygen in the case of the \ce{CuO} molecule. The optimal values of the nuclear $\alpha$ parameters are found to be very close between isolated atoms and molecules, as observed in Fig.~\ref{fig:molecules}, indicating the transferability of atomic $\alpha$ values to molecules.
The optimal $\alpha$ ranges tend to be more shallow for molecules than for atomic systems, which further supports the idea of having an \emph{atomic} $\alpha$ transferred to molecular systems.
For example, after optimizing the orbitals of \ce{CH4} through a \ac{BiO}-\ac{TC} procedure, the optimal $\alpha$ ranges from approximately 1.5 to 2.5 for cc-pVDZ, 2.5 to 4.5 for cc-pVTZ, and 3.5 to 5.0 for cc-pVQZ.
Similarly, for C$_2$, the optimal $\alpha$ ranges from approximately 1.5 to 2.5 for cc-pVDZ, 2.5 to 3.5 for cc-pVTZ, and 3.0 to 5.0 for cc-pVQZ, which is in good agreement with the optimal $\alpha$ for the carbon atom in these basis sets.
These observations hold even for molecules containing very heterogeneous atomic elements such as \ce{CuO}.

\section{Conclusion}
\label{conclusion}
In this work, a new three-body Jastrow factor was introduced including electron-electron-nuclear coordinates.
Orbital optimization was performed in the presence of this Jastrow factor via the \ac{TC} Hamiltonian using a biorthonormal framework. 

One of the distinctive features of our approach is that it does not seek to capture dynamic correlation effects in the vicinity of the core electrons. 
Instead, we employ a universal two-body Jastrow factor\cite{Gin-JCP-21} with a range characteristic of valence electrons, and suppress the effect of the correlation factor in the core region by using a Gaussian-type envelope that limits the effect of the two-body correlation factor around each nucleus.
The nuclear envelope of the correlation factor, which depends on a set of parameters 
$\{\alpha_i,\,\,i=1, \nuclnum\}$, was optimized for both atomic and molecular systems, including elements from the Li-Ne series and $3d$ transition metals.
Interestingly, the optimization of nuclear parameters in molecules yielded very similar optimal $\alpha$ parameters to those obtained for isolated atomic systems, even in heterogeneous systems such as the \ce{CuO} molecule. This suggests that the optimal parameters obtained for atomic systems can be safely transferred to molecular systems, thus avoiding the need for Jastrow factor optimization in molecules.
Furthermore, the use of analytical integrals in combination with numerical integration significantly reduces the computational cost compared to conventional three-body Jastrow factors that rely solely on numerical integration.

The orbital optimization was performed by solving the TC-SCF equations at different levels, using either the non-Hermitian biorthonormal equations or its approximated Hermitian counterpart.
The results showed that the right-eigenvector of the biorthonormal framework significantly lowered the \ac{VMC} energy, while the Hermitian variant only had a weak improvement.
The optimization of the $\alpha$ parameter on atomic systems was carried out by varying its value and computing the \ac{VMC} energy of the right Slater-Jastrow wavefunction.
The results showed a systematic lowering of the energy by increasing the basis set, and an optimal parameter was found for all atoms and basis sets investigated.

The systematic nature of these observations indicates that the correlation factor introduced in this study can be applied to more complex molecular systems with ease, eliminating the need for Jastrow factor optimization in molecules. The simplicity and transferability of this approach make it highly practical for use in realistic chemical systems. Further research will explore the use of this framework with multi-determinant wavefunctions, inspired by our recent developments combining selected CI and TC approaches\cite{Ammar_etal_2022_extension}.

\begin{acknowledgments}
This work was performed using HPC resources from GENCI-TGCC
(gen1738,gen12363) and from CALMIP (Toulouse) under allocation
P22001, and was also supported by the European Centre of
Excellence in Exascale Computing TREX --- Targeting Real Chemical
Accuracy at the Exascale.
This project has received funding from the
European Union's Horizon 2020 --- Research and Innovation program ---
under grant agreement no.~952165.
A CC-BY 4.0 (\url{https://creativecommons.org/licenses/by/4.0/})
public copyright license has been applied by the authors to the present
document and will be applied to all subsequent versions up to the Author
Accepted Manuscript arising from this submission, in accordance with the
grant’s open access conditions.
The authors would like to thank Pierre-Fran\c cois Loos and Julien Toulouse for stimulating discussions.
\end{acknowledgments}

\appendix

\section{Integrals involved in the \ac{TC} matrix elements}

An analytical-numerical scheme was utilized in this work to compute the two- and three-electron integrals, $\kijkl$ and $\lmuijmkln$.
To perform these calculations, a $\mathbb{R}^3$ grid was employed, specifically a Becke numerical grid\cite{Bec-JCP-88b} that consisted of 30 radial points and a Lebedev angular grid consisting of 50 grid points. 
Numerical testing demonstrated that this relatively small number of grid points was sufficient to achieve sub-microhartree convergence of the total energies.

\subsection{Gradients of the one- and two-electron terms}

The integrals required for \ac{TC} matrix elements involve
the gradients of $\umu{1}{2}$~\eqref{eq:def_j} and $g(\mathbf{r})$~\eqref{def:envlop}.
The gradient of the two-electron correlation term~\eqref{eq:def_j} can be expressed as
\begin{equation}
	\nabla_1 \qty(  \uu{1}{2} )
	= \frac{  1 - \erf \qty( \mu \, \rab )}{2 \, \rab} \, \qty( \brb{1} - \brb{2} ) 
\text{,}
	\label{eq:grad1_u12}
\end{equation}
while the one-electron factor~\eqref{def:envlop} is transformed from a multiplication of envelopes into a sum of $1s$-Gaussians,
\begin{equation}
\begin{aligned}
	\envlop{}
	&= \prod_{m=1}^{\nuclnum} \bigg( 1 - \exp\big( - \alpha_m \big| \br{}-\brbig{m} \big|^2 \big)\bigg) \\
	&= \sum_{\mathbf{b} \in \Lset{2}}
		(-1)^{\sum_{\eta=1}^{\nuclnum} b_{\eta}} \,
		e^{- \sum_{\eta=1}^{\nuclnum}  b_{\eta} \alpha_{\eta} \, \big| \br{} - \brbig{\eta} \big|^2}
	\text{,}
	\label{eq:1e_sum_def}
\end{aligned}
\end{equation}
with $\mathcal{L}_2 = \{0,1\}$, and where $\mathbf{b} = (b_1, b_2, \dots, b_{\nuclnum}) \in \Lset{2}$ represents $2^{\nuclnum}$ possible combinations.
This form allows the gradient to be computed as
\begin{widetext}
\begin{equation}
\begin{aligned}
	\nabla \envlop{}
	&= \sum_{\mathbf{b} \in \Lset{2}}
		(-1)^{\sum_{\eta} b_{\eta}} \,
		\left( - 2 \sum_{\eta} b_{\eta} \alpha_{\eta} \, \left( \br{} - \brbig{\eta} \right) \right) \,
		e^{- \sum_{\eta} b_{\eta} \alpha_{\eta} \, \big| \br{} - \brbig{\eta} \big|^2} \text{.}
	\label{eq:grad_1e_sum_def}
\end{aligned}
\end{equation}
\end{widetext}
To simplify the evaluation of integrals, it is useful to write the summation in the exponential of~\eqref{eq:1e_sum_def} as a single $1s$-Gaussian using the generalized Gaussian product theorem\cite{besalu2011general}. 
\begin{equation}
\begin{aligned}
	(-1)^{\sum_{\eta} b_{\eta}} \, \Pi_{\eta=1}^{\nuclnum} \, e^{-b_{\eta} \, \alpha_{\eta} \, \big| \br{} - \brbig{\eta} \big|^2}
	&= K_{\mathbf{b}} \, e^{-\gamma_{\mathbf{b}} \, \big| \br{} - \brbig{\mathbf{b}} \big|^2} \text{,}
\end{aligned}
\end{equation}
with
\begin{equation}
\left\lbrace
\begin{aligned}
	\gamma_{\mathbf{b}} &= \sum_{\eta=1}^{\nuclnum} \, b_{\eta} \, \alpha_{\eta} \\
	\brbig{\mathbf{b}}  &= \frac{1}{\gamma_{\mathbf{b}}} \sum_{\eta=1}^{\nuclnum} \, b_{\eta} \, \alpha_{\eta} \, \brbig{\eta} \\
        K_{\mathbf{b}}      &= (-1)^{\sum_{\eta} b_{\eta}} \, \exp \left( -
		\frac{1}{\gamma_{\mathbf{b}}} \sum_{\eta < \delta} \, b_{\eta} \, \alpha_{\eta} \, b_{\delta} \, \alpha_{\delta} \,
		\big| \brbig{\eta} - \brbig{\delta} \big|^2 \right) \text{.}
\end{aligned}
\right.
\end{equation}
The one-electron factor \eqref{eq:1e_sum_def} and its gradient \eqref{eq:grad_1e_sum_def} can be simplified as 
\begin{align}
	\label{eq:1e_1s_Gauss}
	\envlop{}
	&= \sum_{\mathbf{b} \in \Lset{2}}
		K_{\mathbf{b}}  \,
		e^{-\gamma_{\mathbf{b}} \, \big| \br{} - \brbig{\mathbf{b}} \big|^2}
	\text{,} \\
	\label{eq:1e_1s_Gauss_deriv}
	\nabla \envlop{}
	&= -2 \, \sum_{\mathbf{b} \in \Lset{2}}
		\gamma_{\mathbf{b}} \, K_{\mathbf{b}}   \,
		\left( \br{} - \brbig{\mathbf{b}} \right) \,
		e^{-\gamma_{\mathbf{b}} \, \big| \br{} - \brbig{\mathbf{b}} \big|^2}
	\text{.}
\end{align}

The gradient of the total correlation factor $\unew{1}{2}$~\eqref{def:unew} becomes
\begin{equation}
\begin{aligned}
\label{eq:grad1_unew}
\nabla_1  \unew{1}{2} = \envlop{2} \big[ &
	\nabla_1 \, \uu{1}{2} \, \envlop{1}  \\
          &   + \uu{1}{2} \, \nabla_1 \, \envlop{1} \big] \\
\end{aligned}
\end{equation}
\begin{equation}
\label{eq:grad2_unew}
\begin{aligned}
\nabla_2  \unew{1}{2}= \envlop{1} \big[ &
              - \nabla_1 \, \uu{1}{2} \, \envlop{2} \\
            &  + \uu{1}{2} \,  \nabla_2 \, \envlop{2} \big] \text{,} 
\end{aligned}
\end{equation}
where the identity $\nabla_2 \, \uu{1}{2} = -\nabla_1 \, \uu{1}{2} $ is used.

\subsection{Two-electron integrals}

The integrals required are
\begin{equation}
\begin{aligned}
	\linearkijkl
	&=
	\int \text{d} \br{1} \, \int \text{d} \br{2} \, \chi_k(\br{1})  \chi_l(\br{2}) \\
	&\times
	\left[ \frac{1}{2} \, \Delta_1 \, \unew{1}{2} + \nabla_1 \, \unew{1}{2} \cdot \nabla_1 \right. \\
	&\left. \quad +
		\frac{1}{2} \, \Delta_2 \, \unew{1}{2} + \nabla_2 \, \unew{1}{2} \cdot \nabla_2
	\right] \,
	\phi_i(\br{1}) \, \phi_j(\br{2})
	\text{,}
\end{aligned}
\end{equation}
and
\begin{equation}
\begin{aligned}
	\quadrkijkl
	= \frac{1}{2}
	\int \text{d} \br{1} \, \int \text{d} \br{2} & \, \phi_i(\br{1}) \, \phi_j(\br{2}) \, \chi_k(\br{1}) \, \chi_l(\br{2}) \\
 	& \qty[ \qty[ \nabla_1 \, \unew{1}{2} ]^2 + \qty[ \nabla_2 \, \unew{1}{2} ]^2 ]
	\text{.}
\end{aligned}
\end{equation}
Using integration by parts, 
\begin{equation}
	\linearkijkl = a_{ij}^{kl} + a_{ji}^{lk} \text{,}
\end{equation}
with
\begin{equation}
\begin{aligned}
	a_{ij}^{kl}
	= \frac{1}{2} \,
	\int \text{d} \br{1} &\, \qty[ \chi_k(\br{1}) \, \nabla_1 \phi_i(\br{1}) - \phi_i(\br{1}) \, \nabla_1 \chi_k(\br{1}) ] \\
	& \times
	\int \text{d} \br{2} \, \phi_j(\br{2}) \, \chi_l(\br{2}) \, \nabla_1 \, \unew{1}{2}
	\text{.}
	\label{eq:aijkl_def}
\end{aligned}
\end{equation}

By substituting the gradients~\eqref{eq:grad1_unew} in~\eqref{eq:aijkl_def}, we obtain
\begin{equation}
\begin{aligned}
	a_{ij}^{kl}
	&= \frac{1}{2} \int \text{d} \br{} \,
	\qty[ \chi_k(\br{}) \, \nabla \phi_i(\br{}) - \phi_i(\br{}) \, \nabla \chi_k(\br{}) ] \, \mathbf{X}_j^l(\br{})
	\text{,}
\end{aligned}
\end{equation}
with 
\begin{equation}
\begin{aligned}
	\mathbf{X}_j^l(\br{1})
	&=
	\int \text{d} \br{2} \, \phi_j(\br{2}) \, \chi_l(\br{2}) \, \nabla_1 \, \unew{1}{2} \\
	&=
	\envlop{1} \, \mathbf{I}_{j}^{l}(\br{1}) + J_{j}^{l}(\br{1}) \, \nabla \, \envlop{1} 
	\text{,}
	\label{eq:X_def}
\end{aligned}
\end{equation}
where we have introduced the three-dimensional integrals
\begin{align}
	\label{eq:Ijl_def}
	\mathbf{I}_{j}^{l}(\br{1}) &= \int \text{d} \br{2} \,
	\chi_l(\br{2}) \, \phi_j(\br{2}) \,
	\nabla_1 \uu{1}{2} \, \envlop{2} \\
	\label{eq:Jjl_def}
	J_{j}^{l}(\br{1}) &= \int \text{d} \br{2} \,
	\chi_l(\br{2}) \, \phi_j(\br{2}) \, \uu{1}{2} \, \envlop{2} \text{.}
\end{align}
Substituting the gradient formula~\eqref{eq:grad1_u12} in the integral~\eqref{eq:Ijl_def},
and the definition~\eqref{eq:def_j} in the integral~\eqref{eq:Jjl_def},
we find,
\begin{align}
	\mathbf{I}_{j}^{l}(\br{})
	&= \frac{1}{2} \,
	\qty( A_j^l(\br{}) \, \br{} + \mathbf{B}_j^l(\br{}) )
	\text{,} \\
	J_{j}^{l}(\br{1})
	&=  \frac{{r_1}^2}{2} \,
	A_j^l(\br{1}) + \mathbf{B}_j^l(\br{1}) \cdot \br{1} + \frac{C_j^l(\br{1})}{2} - \frac{D_j^l(\br{1})}{2\,\sqrt{\pi}\,\mu}
	\text{,}
\end{align}
where
\begin{align}
	A_j^l(\br{1}) &=
 	\int \text{d} \br{2} \,  \chi_l(\br{2}) \, \phi_j(\br{2}) \, \envlop{2} \,
	\frac{\left[  1 - \erf \left( \mu \, r_{12} \right) \right]}{r_{12}}
	\text{,} \\
	\mathbf{B}_j^l(\br{1})  &=
 	\int \text{d} \br{2} \,  \chi_l(\br{2}) \, \phi_j(\br{2}) \, \envlop{2}
	\frac{\left[  1 - \erf \left( \mu \, r_{12} \right) \right]}{r_{12}} \, \mathbf{r}_2
	\text{,} \\
	C_j^l(\br{1})  &=
 	\int \text{d} \br{2} \,  \chi_l(\br{2}) \, \phi_j(\br{2}) \, \envlop{2}
	\frac{\left[  1 - \erf \left( \mu \, r_{12} \right) \right]}{r_{12}} \, {r_2}^2
	\text{,} \\
	D_j^l(\br{1})  &=
 	\int \text{d} \br{2} \,  \chi_l(\br{2}) \, \phi_j(\br{2}) \, \envlop{2} \, e^{-(r_{12}\mu)^2}
	\text{.}
\end{align}
The integrals $A_j^l(\br{1})$, $\mathbf{B}_j^l(\br{1})$, $C_j^l(\br{1})$ and $D_j^l(\br{1})$ are analytical\cite{Gin-JCP-21}.

On the other hand, using the gradients
\begin{equation}
\begin{aligned}
	\qty[ \nabla_1 \, \unew12 ]^2
	=&
	\qty[ \nabla_1 \uu12 ]^2 \, \envlop{1}^2 \, \envlop{2}^2 \\
	& +
	\uu12^2 \, \qty[ \nabla_1 \, \envlop{1} ]^2 \, \envlop{2}^2 \\
	& +
	2 \, \uu12 \, \qty[\nabla_1 \, \uu12] \, \envlop{1} \\
	& \quad \nabla_1 \, \envlop{1} \, \envlop{2}^2
	\text{,}
\end{aligned}
\end{equation}
\begin{equation}
\begin{aligned}
	\qty[ \nabla_2 \, \unew12 ]^2
	=& \qty[ \nabla_2 \uu12 ]^2 \, \envlop{1}^2 \, \envlop{2}^2 \\
	& +
	\uu12^2 \, [ \nabla_2 \, \envlop{2} ]^2 \, \envlop{1}^2 \\
	& +
	2 \, \uu12 \, [ \nabla_2 \, \uu12] \, \envlop{2} \\
	& \quad \nabla_2 \, \envlop{2} \, \envlop{1}^2
	\text{,}
\end{aligned}
\end{equation}
the integral $\quadrkijkl$ becomes
\begin{equation}
	\quadrkijkl = \frac{1}{2}  \qty( \alpha_{ij}^{kl} + \alpha_{ji}^{lk} )
	\text{,}
\end{equation}
\begin{equation}
	\alpha_{ij}^{kl} = \beta_{ij}^{kl} + \gamma_{ij}^{kl} + 2 \, \delta_{ij}^{kl}
	\text{,}
\end{equation}
\begin{equation}
\begin{aligned}
	\beta_{ij}^{kl}
	=
	\int \text{d} \br{1} & \, \phi_i(\br{1}) \chi_k(\br{1}) \, \envlop{1}^2  \\
	& \int \text{d} \br{2} \, \phi_j(\br{2}) \chi_l(\br{2}) 
          \qty[ \nabla_1 \uu{1}{2} ]^2 \, \envlop{2}^2
	\text{,}
	\label{eq:def_beta_ijkl}
\end{aligned}
\end{equation}
\begin{equation}
\begin{aligned}
	\gamma_{ij}^{kl}
	=
	\int \text{d} \br{1} &\, \phi_i(\br{1}) \chi_k(\br{1}) \, \qty[ \nabla_1 \, \envlop{1} ]^2  \\
	& \int \text{d} \br{2} \, \phi_j(\br{2}) \chi_l(\br{2})
	\qty[ \uu{1}{2} ]^2 \, \envlop{2}^2
	\text{,}
	\label{eq:def_gamma_ijkl}
\end{aligned}
\end{equation}
\begin{equation}
\begin{aligned}
	\delta_{ij}^{kl}
	=
	\int \text{d} \br{1} \, & \phi_i(\br{1}) \chi_k(\br{1}) \, \qty[ \nabla_1 \, \envlop{1} ] \, \envlop{1} \\
	& \int \text{d} \br{2} \, \phi_j(\br{2}) \chi_l(\br{2}) \,
	\uu{1}{2} \, \qty[ \nabla_1 \uu{1}{2} ] \, \envlop{2}^2
	\text{.}
	\label{eq:def_delta_ijkl}
\end{aligned}
\end{equation}
The integrals over the gradients involve
\begin{equation}
\begin{aligned}
	f_1^{\mu}(r_{12})
	&= \qty[ \nabla_1 \umu12 ]^2 \\
	&= \qty[ \nabla_2 \umu12 ]^2 \\
	&= \frac{\left[ 1 - \erf\left(\mu\, r_{12} \right) \right]^2 }{4}
	\text{,}
\end{aligned}
\end{equation}
\begin{equation}
\begin{aligned}
	f_2^{\mu}(r_{12})
	&= \umu{1}{2}^2  \\
	&= \frac{\qty[ 1 - \erf \qty(\mu\, r_{12}) ]^2}{4}  \, {r_{12}}^2
	+ \frac{e^{-2(\mu\, r_{12})^2}}{4 \, \pi \, \mu^2} \\
	& \quad -
	\frac{\qty[ 1 - \erf \qty(\mu\, r_{12} ) ]}{2 \, \sqrt{\pi} \, \mu}  \, r_{12} \, e^{-(\mu\, r_{12})^2}
	\text{,}
\end{aligned}
\end{equation}
\begin{equation}
\begin{aligned}
	\mathbf{f}_3^{\mu}(\br{1}, \br{2})
	&= \phantom{-}\uu{1}{2} \, \nabla_1 \uu{1}{2} \\
	&= -\uu{1}{2} \, \nabla_2 \uu{1}{2} \\
	&= \frac{\qty[ 1 - \erf \qty(\mu\, r_{12} ) ]^2}{4}  \, \qty( \br{1} - \br{2} ) \\
	& \phantom{=} -
	\frac{\qty[ 1 - \erf \qty(\mu\, r_{12} ) ]}{r_{12}}  \,
	\frac{e^{-(\mu\, r_{12})^2}}{4 \, \sqrt{\pi} \, \mu} \,
	\qty( \br{1} - \br{2} )
	\text{.}
\end{aligned}
\end{equation}

To facilitate the evaluation of the integrals we fit $f_1,f_2,\mathbf{f}_3$ with a combination
of Gaussians.
The first function is fitted directly as
\begin{equation}
\begin{aligned}
	f_1^{\mu}(r_{12})
	& =
	\frac{1}{4} g_1(\mu \, r_{12})
	\text{,}
\end{aligned}
\end{equation}
\begin{equation}
\begin{aligned}
	g_1(x)
	&=  \qty[ 1 - \erf(x) ]^2
	\approx \sum_{i=1}^{N_1} c_i^{g_1} \, e^{-\alpha_i^{g_1} \, x^2}
	\text{.}
\end{aligned}
\end{equation}
And by noticing that
\begin{equation}
	\umu{1}{2} = \frac{1}{2 \, \mu} \, F(\mu \, r_{12})
	\text{,}
\end{equation}
\begin{equation}
	F(x) = \qty[1 - \erf(x)] \, x - \frac{1}{\sqrt{\pi}} e^{-x^2}
	\text{,}
\end{equation}
we can write the second function $f_2$ as
\begin{equation}
\begin{aligned}
	f_2^{\mu}(r_{12}) = \left( \frac{1}{2 \, \mu} \right)^2 \, g_2(\mu \, r_{12})
	\text{,}
\end{aligned}
\end{equation}
\begin{equation}
	g_2(x)
	= \qty[ F(x) ]^2
	\approx \sum_{i=1}^{N_2} c_i^{g_2} \, e^{-\alpha_i^{ g_2 } \, x^2}
	\text{.}
\end{equation}
The last function is written as
\begin{equation}
\begin{aligned}
	\mathbf{f}_3^{\mu}(\br1, \br2)
	&= \frac{1}{4 \, \mu} \, G(\mu \, r_{12}) \, \frac{\left( \br{1} - \br{2} \right)}{r_{12}}
\end{aligned}
\end{equation}
\begin{equation}
	G(x) = F(x) \, \left[ 1 - \erf(x) \right]
	\approx \sum_{i=1}^{N_3} c_i^{ g_3 } \, e^{-\alpha_i^{ g_3 } \, x^2}
	\text{.}
\end{equation}
The terms $ \envlop{2}^2$ in~\eqref{eq:def_beta_ijkl}, ~\eqref{eq:def_gamma_ijkl},
and~\eqref{eq:def_delta_ijkl} can be evaluated by considering the square of the sum~\eqref{eq:1e_1s_Gauss}.
However this involves $4^{\nuclnum}$ integrals, but it can be reduced to $3^{\nuclnum}$ using
\begin{equation}
\begin{aligned}
	\envlop{}
	&= \prod_{m=1}^{\nuclnum} \qty[ 1 - \exp\qty( - \alpha_m \big| \br{}-\brbig{m} \big|^2 )]^2 \\
	&= \sum_{\mathbf{b} \in \Lset{3}}
		(-1)^{\sum_{\eta=1}^{\nuclnum} b_{\eta}} \, {2 \choose b_{\eta}} \,
		e^{- \sum_{\eta=1}^{\nuclnum}  b_{\eta} \alpha_{\eta} \, \big| \br{} - \brbig{\eta} \big|^2}
	\text{,}
	\label{eq:1e_sum_def_squared}
\end{aligned}
\end{equation}
with $\mathbf{b} = (b_1, b_2, \dots, b_{\nuclnum}) \in \Lset{3}$ and 
$\mathcal{L}_3 = \{0,1,2\}$.
As in the case of $\envlop{}$,
this expression can be written as a sum over $1s-$Gaussian. This leads to,
\begin{equation}
\begin{aligned}
	(-1)^{\sum_{\eta} b_{\eta}} \, {2 \choose b_{\eta}} \,
	\Pi_{\eta=1}^{\nuclnum} \, e^{-b_{\eta} \, \alpha_{\eta} \, \big| \br{} - \brbig{\eta} \big|^2}
	&= K_{\mathbf{b}} \, e^{-\gamma_{\mathbf{b}} \, \big| \br{} - \brbig{\mathbf{b}} \big|^2} \text{,}
\end{aligned}
\end{equation}
with
\begin{equation}
\left\lbrace
\begin{aligned}
	\gamma_{\mathbf{b}} &= \sum_{\eta=1}^{\nuclnum} \, b_{\eta} \, \alpha_{\eta} \\
	\brbig{\mathbf{b}}  &= \frac{1}{\gamma_{\mathbf{b}}} \sum_{\eta=1}^{\nuclnum} \, b_{\eta} \, \alpha_{\eta} \, \brbig{\eta} \\
        K_{\mathbf{b}}      &= (-1)^{\sum_{\eta} b_{\eta}} \, {2 \choose b_{\eta}} \\
		& \phantom{=} \times \exp \left( -
		\frac{1}{\gamma_{\mathbf{b}}} \sum_{\eta < \delta} \, b_{\eta} \, \alpha_{\eta} \, b_{\delta} \, \alpha_{\delta} \,
		\big| \brbig{\eta} - \brbig{\delta} \big|^2 \right) \text{.}
\end{aligned}
\right.
\end{equation}
This leads to,
\begin{equation}
	\envlop{}^2 =
	\sum_{\mathbf{b} \in \Lset{3}} \, K_{\mathbf{b}}  \,
		e^{-\gamma_{\mathbf{b}} \, \big| \br{} - \brbig{\mathbf{b}} \big|^2}
	\text{.}
	\label{eq:1e_1s_Gauss_squared}
\end{equation}

The final expressions of $\beta_{ij}^{kl}$~\eqref{eq:def_beta_ijkl},
$\gamma_{ij}^{kl}$~\eqref{eq:def_gamma_ijkl}, and $\delta_{ij}^{kl}$~\eqref{eq:def_delta_ijkl}
become,
\begin{align}
	\beta_{ij}^{kl}  &= \int \text{d} \br{} \, \phi_i(\br{}) \, \chi_k(\br{}) \, \envlop{}^2  \, E_j^{l}(\br{})                          \text{,} \\
	\gamma_{ij}^{kl} &= \int \text{d} \br{} \, \phi_i(\br{}) \, \chi_k(\br{}) \, \qty[ \nabla \, \envlop{} ]^2  \, F_j^{l}(\br{}) \text{,} \\
	\delta_{ij}^{kl} &= \int \text{d} \br{} \, \phi_i(\br{}) \, \chi_k(\br{}) \, \envlop{} \, \qty[ \nabla \, \envlop{} ] \cdot \mathbf{G}_j^{l}(\br{}) \text{,}
\end{align}
where the following integrals are analytical (see Appendix A in Ref.~\onlinecite{toulouse2004long})
\begin{equation}
\begin{aligned}
	E_j^{l}(\br{1})
	&= \frac{1}{4}
	\sum_{i=1}^{N_1} \, c_i^{ g_1 } \,
	\sum_{\mathbf{b} \in \Lset{3}} \, K_{\mathbf{b}}  \\
	& \phantom{=} \times
	\int \text{d} \br{2} \, \phi_j(\br{2}) \chi_l(\br{2}) \,
	e^{-\gamma_{\mathbf{b}} \, \big| \br{} - \brbig{\mathbf{b}} \big|^2 - \alpha_i^{ g_1 } \, \mu^2 \, r_{12}^{2}}
	\text{,}
\end{aligned}
\end{equation}
\begin{equation}
\begin{aligned}
	F_j^{l}(\br{1})
	&=
	\left( \frac{1}{2 \, \mu} \right)^2 \,
	\sum_{i=1}^{N_2} \, c_i^{ g_2 } \,
	\sum_{\mathbf{b} \in \Lset{3}} \, K_{\mathbf{b}}  \\
	& \times
	\int \text{d} \br{2} \, \phi_j(\br{2}) \, \chi_l(\br{2}) \,
	e^{-\gamma_{\mathbf{b}} \, \big| \br{} - \brbig{\mathbf{b}} \big|^2 - \alpha_i^{ g_2 } \, \mu^2 \, r_{12}^2}
	\text{,}
\end{aligned}
\end{equation}
\begin{equation}
\begin{aligned}
	&\mathbf{G}_j^{l}(\br{1})
	=
	\frac{1}{4 \, \mu} \,
	\sum_{i=1}^{N_3} \, c_i^{ g_3 }
	\sum_{\mathbf{b} \in \Lset{3}} \, K_{\mathbf{b}}  \\
	& \times
	\int \text{d} \br{2} \, \phi_j(\br{2}) \, \chi_l(\br{2}) \,
	\frac{\left( \br{1} - \br{2} \right)}{r_{12}} \,
	e^{-\gamma_{\mathbf{b}} \, \big| \br{} - \brbig{\mathbf{b}} \big|^2 - \alpha_i^{ g_3 } \, \mu^2 \, r_{12}^2}
	\text{.}
\end{aligned}
\end{equation}

\subsection{Three-electron integrals}

We turn now to the calculation of the $3-$electron integrals
\begin{equation}
\begin{aligned}
	L_{ijk}^{lmn}
	= &\int \text{d} \br{1} \, \int \text{d} \br{2} \, \int \text{d} \br{3} \\
	& \phi_i(\br{1}) \, \phi_j(\br{2}) \, \phi_k(\br{3}) \, \chi_l(\br{1}) \, \chi_m(\br{2}) \, \chi_n(\br{3}) \\
	& \times \Big[  \nabla_1 \unew{1}{2}  \cdot  \nabla_1 \unew{1}{3}  \\
	& \quad + \nabla_2 \unew{2}{1}  \cdot  \nabla_2 \unew{2}{3}  \\
	& \quad + \nabla_3 \unew{3}{1}  \cdot  \nabla_3 \unew{3}{2}  \Big]
	\text{.}
\end{aligned}
\end{equation}
We can show that
\begin{equation}
	L_{ijk}^{lmn} = M_{ijk}^{lmn} + M_{jik}^{mln} + M_{kji}^{nml} 
	\text{,}
\end{equation}
\begin{equation}
\begin{aligned}
	M_{ijk}^{lmn}
	=&       \int \text{d} \br{1} \, \phi_i(\br{1}) \, \chi_l(\br{1}) \\
	 & \quad \int \text{d} \br{2} \, \phi_j(\br{2}) \, \chi_m(\br{2}) \, \nabla_1 \unew{1}{2} \\
	& \quad \cdot  \int \text{d} \br{3} \, \phi_k(\br{3}) \, \chi_n(\br{3}) \, \nabla_1 \unew{1}{3} 
	\text{.}
\end{aligned}
\end{equation}
By using the analytical expression~\eqref{eq:X_def}, this integral can be rewritten as the following three-dimensional integral,
\begin{equation}
	M_{ijk}^{lmn} = \int \text{d} \br{} \, \phi_i(\br{}) \, \chi_l(\br{}) \, \mathbf{X}_j^m(\br{})  \, \mathbf{X}_k^n(\br{})
\end{equation}
which is evaluated numerically.

\bibliography{paper_tc_scf}

\begin{thebibliography}{102}%
\makeatletter
\providecommand \@ifxundefined [1]{%
 \@ifx{#1\undefined}
}%
\providecommand \@ifnum [1]{%
 \ifnum #1\expandafter \@firstoftwo
 \else \expandafter \@secondoftwo
 \fi
}%
\providecommand \@ifx [1]{%
 \ifx #1\expandafter \@firstoftwo
 \else \expandafter \@secondoftwo
 \fi
}%
\providecommand \natexlab [1]{#1}%
\providecommand \enquote  [1]{``#1''}%
\providecommand \bibnamefont  [1]{#1}%
\providecommand \bibfnamefont [1]{#1}%
\providecommand \citenamefont [1]{#1}%
\providecommand \href@noop [0]{\@secondoftwo}%
\providecommand \href [0]{\begingroup \@sanitize@url \@href}%
\providecommand \@href[1]{\@@startlink{#1}\@@href}%
\providecommand \@@href[1]{\endgroup#1\@@endlink}%
\providecommand \@sanitize@url [0]{\catcode `\\12\catcode `\$12\catcode
  `\&12\catcode `\#12\catcode `\^12\catcode `\_12\catcode `\%12\relax}%
\providecommand \@@startlink[1]{}%
\providecommand \@@endlink[0]{}%
\providecommand \url  [0]{\begingroup\@sanitize@url \@url }%
\providecommand \@url [1]{\endgroup\@href {#1}{\urlprefix }}%
\providecommand \urlprefix  [0]{URL }%
\providecommand \Eprint [0]{\href }%
\providecommand \doibase [0]{http://dx.doi.org/}%
\providecommand \selectlanguage [0]{\@gobble}%
\providecommand \bibinfo  [0]{\@secondoftwo}%
\providecommand \bibfield  [0]{\@secondoftwo}%
\providecommand \translation [1]{[#1]}%
\providecommand \BibitemOpen [0]{}%
\providecommand \bibitemStop [0]{}%
\providecommand \bibitemNoStop [0]{.\EOS\space}%
\providecommand \EOS [0]{\spacefactor3000\relax}%
\providecommand \BibitemShut  [1]{\csname bibitem#1\endcsname}%
\let\auto@bib@innerbib\@empty
\bibitem [{\citenamefont {Boys}\ and\ \citenamefont
  {Handy}(1969{\natexlab{a}})}]{Boys_Handy_1969_determination}%
  \BibitemOpen
  \bibfield  {author} {\bibinfo {author} {\bibfnamefont {S.~F.}\ \bibnamefont
  {Boys}}\ and\ \bibinfo {author} {\bibfnamefont {N.~C.}\ \bibnamefont
  {Handy}},\ }\href {\doibase 10.1098/rspa.1969.0061} {\bibfield  {journal}
  {\bibinfo  {journal} {Proceedings of the Royal Society of London. A.
  Mathematical and Physical Sciences}\ }\textbf {\bibinfo {volume} {310}},\
  \bibinfo {pages} {43} (\bibinfo {year} {1969}{\natexlab{a}})}\BibitemShut
  {NoStop}%
\bibitem [{\citenamefont {Hino}, \citenamefont {Tanimura},\ and\ \citenamefont
  {Ten-no}(2001)}]{Hino_etal_2001}%
  \BibitemOpen
  \bibfield  {author} {\bibinfo {author} {\bibfnamefont {O.}~\bibnamefont
  {Hino}}, \bibinfo {author} {\bibfnamefont {Y.}~\bibnamefont {Tanimura}}, \
  and\ \bibinfo {author} {\bibfnamefont {S.}~\bibnamefont {Ten-no}},\ }\href
  {\doibase 10.1063/1.1408299} {\bibfield  {journal} {\bibinfo  {journal} {The
  Journal of Chemical Physics}\ }\textbf {\bibinfo {volume} {115}},\ \bibinfo
  {pages} {7865} (\bibinfo {year} {2001})}\BibitemShut {NoStop}%
\bibitem [{\citenamefont {Ochi}\ and\ \citenamefont
  {Tsuneyuki}(2015)}]{Ochi_Tsuneyuki_2015}%
  \BibitemOpen
  \bibfield  {author} {\bibinfo {author} {\bibfnamefont {M.}~\bibnamefont
  {Ochi}}\ and\ \bibinfo {author} {\bibfnamefont {S.}~\bibnamefont
  {Tsuneyuki}},\ }\href {\doibase https://doi.org/10.1016/j.cplett.2015.01.009}
  {\bibfield  {journal} {\bibinfo  {journal} {Chemical Physics Letters}\
  }\textbf {\bibinfo {volume} {621}},\ \bibinfo {pages} {177} (\bibinfo {year}
  {2015})}\BibitemShut {NoStop}%
\bibitem [{\citenamefont {Prasad}\ \emph {et~al.}(2007)\citenamefont {Prasad},
  \citenamefont {Umezawa}, \citenamefont {Domin}, \citenamefont
  {Salomon-Ferrer},\ and\ \citenamefont {Lester}}]{Prasad_etal_2007}%
  \BibitemOpen
  \bibfield  {author} {\bibinfo {author} {\bibfnamefont {R.}~\bibnamefont
  {Prasad}}, \bibinfo {author} {\bibfnamefont {N.}~\bibnamefont {Umezawa}},
  \bibinfo {author} {\bibfnamefont {D.}~\bibnamefont {Domin}}, \bibinfo
  {author} {\bibfnamefont {R.}~\bibnamefont {Salomon-Ferrer}}, \ and\ \bibinfo
  {author} {\bibfnamefont {W.~A.}\ \bibnamefont {Lester}},\ }\href {\doibase
  10.1063/1.2715581} {\bibfield  {journal} {\bibinfo  {journal} {The Journal of
  Chemical Physics}\ }\textbf {\bibinfo {volume} {126}},\ \bibinfo {pages}
  {164109} (\bibinfo {year} {2007})}\BibitemShut {NoStop}%
\bibitem [{\citenamefont {Zweistra}, \citenamefont {Samson},\ and\
  \citenamefont {Klopper}(2003)}]{Zweistra_etal_2003}%
  \BibitemOpen
  \bibfield  {author} {\bibinfo {author} {\bibfnamefont {H.~J.}\ \bibnamefont
  {Zweistra}}, \bibinfo {author} {\bibfnamefont {C.}~\bibnamefont {Samson}}, \
  and\ \bibinfo {author} {\bibfnamefont {W.}~\bibnamefont {Klopper}},\ }\href
  {\doibase 10.1135/cccc20030374} {\bibfield  {journal} {\bibinfo  {journal}
  {Collection of Czechoslovak chemical communications}\ }\textbf {\bibinfo
  {volume} {68}},\ \bibinfo {pages} {374} (\bibinfo {year} {2003})}\BibitemShut
  {NoStop}%
\bibitem [{\citenamefont {Tsuneyuki}(2008)}]{Tsuneyuki_2008}%
  \BibitemOpen
  \bibfield  {author} {\bibinfo {author} {\bibfnamefont {S.}~\bibnamefont
  {Tsuneyuki}},\ }\href {\doibase 10.1143/PTPS.176.134} {\bibfield  {journal}
  {\bibinfo  {journal} {Progress of Theoretical Physics Supplement}\ }\textbf
  {\bibinfo {volume} {176}},\ \bibinfo {pages} {134} (\bibinfo {year}
  {2008})}\BibitemShut {NoStop}%
\bibitem [{\citenamefont {Luo}(2011)}]{Luo_2011}%
  \BibitemOpen
  \bibfield  {author} {\bibinfo {author} {\bibfnamefont {H.}~\bibnamefont
  {Luo}},\ }\href {\doibase 10.1063/1.3607990} {\bibfield  {journal} {\bibinfo
  {journal} {The Journal of Chemical Physics}\ }\textbf {\bibinfo {volume}
  {135}},\ \bibinfo {pages} {024109} (\bibinfo {year} {2011})}\BibitemShut
  {NoStop}%
\bibitem [{\citenamefont {Ochi}\ and\ \citenamefont
  {Tsuneyuki}(2014)}]{Ochi_Tsuneyuki_2014}%
  \BibitemOpen
  \bibfield  {author} {\bibinfo {author} {\bibfnamefont {M.}~\bibnamefont
  {Ochi}}\ and\ \bibinfo {author} {\bibfnamefont {S.}~\bibnamefont
  {Tsuneyuki}},\ }\href {\doibase 10.1021/ct500485b} {\bibfield  {journal}
  {\bibinfo  {journal} {Journal of Chemical Theory and Computation}\ }\textbf
  {\bibinfo {volume} {10}},\ \bibinfo {pages} {4098} (\bibinfo {year}
  {2014})}\BibitemShut {NoStop}%
\bibitem [{\citenamefont {Giner}(2021)}]{Gin-JCP-21}%
  \BibitemOpen
  \bibfield  {author} {\bibinfo {author} {\bibfnamefont {E.}~\bibnamefont
  {Giner}},\ }\href {\doibase 10.1063/5.0044683} {\bibfield  {journal}
  {\bibinfo  {journal} {The Journal of Chemical Physics}\ }\textbf {\bibinfo
  {volume} {154}},\ \bibinfo {pages} {084119} (\bibinfo {year}
  {2021})}\BibitemShut {NoStop}%
\bibitem [{\citenamefont {Ammar}, \citenamefont {Scemama},\ and\ \citenamefont
  {Giner}(2022)}]{Ammar_etal_2022_extension}%
  \BibitemOpen
  \bibfield  {author} {\bibinfo {author} {\bibfnamefont {A.}~\bibnamefont
  {Ammar}}, \bibinfo {author} {\bibfnamefont {A.}~\bibnamefont {Scemama}}, \
  and\ \bibinfo {author} {\bibfnamefont {E.}~\bibnamefont {Giner}},\ }\href
  {\doibase 10.1063/5.0115524} {\bibfield  {journal} {\bibinfo  {journal} {The
  Journal of Chemical Physics}\ }\textbf {\bibinfo {volume} {157}},\ \bibinfo
  {pages} {134107} (\bibinfo {year} {2022})}\BibitemShut {NoStop}%
\bibitem [{\citenamefont {Ammar}, \citenamefont {Giner},\ and\ \citenamefont
  {Scemama}(2022)}]{Ammar_etal_2022_optimization}%
  \BibitemOpen
  \bibfield  {author} {\bibinfo {author} {\bibfnamefont {A.}~\bibnamefont
  {Ammar}}, \bibinfo {author} {\bibfnamefont {E.}~\bibnamefont {Giner}}, \ and\
  \bibinfo {author} {\bibfnamefont {A.}~\bibnamefont {Scemama}},\ }\href
  {\doibase 10.1021/acs.jctc.2c00556} {\bibfield  {journal} {\bibinfo
  {journal} {Journal of Chemical Theory and Computation}\ }\textbf {\bibinfo
  {volume} {18}},\ \bibinfo {pages} {5325} (\bibinfo {year}
  {2022})}\BibitemShut {NoStop}%
\bibitem [{\citenamefont {Luo}\ and\ \citenamefont
  {Alavi}(2018)}]{Luo_Alavi_2018}%
  \BibitemOpen
  \bibfield  {author} {\bibinfo {author} {\bibfnamefont {H.}~\bibnamefont
  {Luo}}\ and\ \bibinfo {author} {\bibfnamefont {A.}~\bibnamefont {Alavi}},\
  }\href {\doibase 10.1021/acs.jctc.7b01257} {\bibfield  {journal} {\bibinfo
  {journal} {Journal of Chemical Theory and Computation}\ }\textbf {\bibinfo
  {volume} {14}},\ \bibinfo {pages} {1403} (\bibinfo {year}
  {2018})}\BibitemShut {NoStop}%
\bibitem [{\citenamefont {Dobrautz}, \citenamefont {Luo},\ and\ \citenamefont
  {Alavi}(2019)}]{Dobrautz_etal_2019}%
  \BibitemOpen
  \bibfield  {author} {\bibinfo {author} {\bibfnamefont {W.}~\bibnamefont
  {Dobrautz}}, \bibinfo {author} {\bibfnamefont {H.}~\bibnamefont {Luo}}, \
  and\ \bibinfo {author} {\bibfnamefont {A.}~\bibnamefont {Alavi}},\ }\href
  {\doibase 10.1103/PhysRevB.99.075119} {\bibfield  {journal} {\bibinfo
  {journal} {Phys. Rev. B}\ }\textbf {\bibinfo {volume} {99}},\ \bibinfo
  {pages} {075119} (\bibinfo {year} {2019})}\BibitemShut {NoStop}%
\bibitem [{\citenamefont {Cohen}\ \emph {et~al.}(2019)\citenamefont {Cohen},
  \citenamefont {Luo}, \citenamefont {Guther}, \citenamefont {Dobrautz},
  \citenamefont {Tew},\ and\ \citenamefont {Alavi}}]{Cohen_etal_2019}%
  \BibitemOpen
  \bibfield  {author} {\bibinfo {author} {\bibfnamefont {A.~J.}\ \bibnamefont
  {Cohen}}, \bibinfo {author} {\bibfnamefont {H.}~\bibnamefont {Luo}}, \bibinfo
  {author} {\bibfnamefont {K.}~\bibnamefont {Guther}}, \bibinfo {author}
  {\bibfnamefont {W.}~\bibnamefont {Dobrautz}}, \bibinfo {author}
  {\bibfnamefont {D.~P.}\ \bibnamefont {Tew}}, \ and\ \bibinfo {author}
  {\bibfnamefont {A.}~\bibnamefont {Alavi}},\ }\href {\doibase
  10.1063/1.5116024} {\bibfield  {journal} {\bibinfo  {journal} {The Journal of
  Chemical Physics}\ }\textbf {\bibinfo {volume} {151}},\ \bibinfo {pages}
  {061101} (\bibinfo {year} {2019})}\BibitemShut {NoStop}%
\bibitem [{\citenamefont {Dobrautz}\ \emph {et~al.}(2022)\citenamefont
  {Dobrautz}, \citenamefont {Cohen}, \citenamefont {Alavi},\ and\ \citenamefont
  {Giner}}]{Dobrautz_etal_2022}%
  \BibitemOpen
  \bibfield  {author} {\bibinfo {author} {\bibfnamefont {W.}~\bibnamefont
  {Dobrautz}}, \bibinfo {author} {\bibfnamefont {A.~J.}\ \bibnamefont {Cohen}},
  \bibinfo {author} {\bibfnamefont {A.}~\bibnamefont {Alavi}}, \ and\ \bibinfo
  {author} {\bibfnamefont {E.}~\bibnamefont {Giner}},\ }\href {\doibase
  10.1063/5.0088981} {\bibfield  {journal} {\bibinfo  {journal} {The Journal of
  Chemical Physics}\ }\textbf {\bibinfo {volume} {156}},\ \bibinfo {pages}
  {234108} (\bibinfo {year} {2022})}\BibitemShut {NoStop}%
\bibitem [{\citenamefont {Jeszenszki}\ \emph {et~al.}(2020)\citenamefont
  {Jeszenszki}, \citenamefont {Ebling}, \citenamefont {Luo}, \citenamefont
  {Alavi},\ and\ \citenamefont {Brand}}]{Jeszenszki_etal_2020}%
  \BibitemOpen
  \bibfield  {author} {\bibinfo {author} {\bibfnamefont {P.}~\bibnamefont
  {Jeszenszki}}, \bibinfo {author} {\bibfnamefont {U.}~\bibnamefont {Ebling}},
  \bibinfo {author} {\bibfnamefont {H.}~\bibnamefont {Luo}}, \bibinfo {author}
  {\bibfnamefont {A.}~\bibnamefont {Alavi}}, \ and\ \bibinfo {author}
  {\bibfnamefont {J.}~\bibnamefont {Brand}},\ }\href {\doibase
  10.1103/PhysRevResearch.2.043270} {\bibfield  {journal} {\bibinfo  {journal}
  {Phys. Rev. Res.}\ }\textbf {\bibinfo {volume} {2}},\ \bibinfo {pages}
  {043270} (\bibinfo {year} {2020})}\BibitemShut {NoStop}%
\bibitem [{\citenamefont {Guther}\ \emph {et~al.}(2021)\citenamefont {Guther},
  \citenamefont {Cohen}, \citenamefont {Luo},\ and\ \citenamefont
  {Alavi}}]{Guther_etal_2021}%
  \BibitemOpen
  \bibfield  {author} {\bibinfo {author} {\bibfnamefont {K.}~\bibnamefont
  {Guther}}, \bibinfo {author} {\bibfnamefont {A.~J.}\ \bibnamefont {Cohen}},
  \bibinfo {author} {\bibfnamefont {H.}~\bibnamefont {Luo}}, \ and\ \bibinfo
  {author} {\bibfnamefont {A.}~\bibnamefont {Alavi}},\ }\href {\doibase
  10.1063/5.0055575} {\bibfield  {journal} {\bibinfo  {journal} {The Journal of
  Chemical Physics}\ }\textbf {\bibinfo {volume} {155}},\ \bibinfo {pages}
  {011102} (\bibinfo {year} {2021})}\BibitemShut {NoStop}%
\bibitem [{\citenamefont {Haupt}\ \emph {et~al.}(2023)\citenamefont {Haupt},
  \citenamefont {Hosseini}, \citenamefont {Rios}, \citenamefont {Dobrautz},
  \citenamefont {Cohen},\ and\ \citenamefont {Alavi}}]{Haupt_etal_2023}%
  \BibitemOpen
  \bibfield  {author} {\bibinfo {author} {\bibfnamefont {J.~P.}\ \bibnamefont
  {Haupt}}, \bibinfo {author} {\bibfnamefont {S.~M.}\ \bibnamefont {Hosseini}},
  \bibinfo {author} {\bibfnamefont {P.~L.}\ \bibnamefont {Rios}}, \bibinfo
  {author} {\bibfnamefont {W.}~\bibnamefont {Dobrautz}}, \bibinfo {author}
  {\bibfnamefont {A.}~\bibnamefont {Cohen}}, \ and\ \bibinfo {author}
  {\bibfnamefont {A.}~\bibnamefont {Alavi}},\ }\href
  {https://arxiv.org/abs/2302.13683} {\bibfield  {journal} {\bibinfo  {journal}
  {arXiv}\ } (\bibinfo {year} {2023})}\BibitemShut {NoStop}%
\bibitem [{\citenamefont {Hino}, \citenamefont {Tanimura},\ and\ \citenamefont
  {Ten-no}(2002)}]{Hino_etal_2002}%
  \BibitemOpen
  \bibfield  {author} {\bibinfo {author} {\bibfnamefont {O.}~\bibnamefont
  {Hino}}, \bibinfo {author} {\bibfnamefont {Y.}~\bibnamefont {Tanimura}}, \
  and\ \bibinfo {author} {\bibfnamefont {S.}~\bibnamefont {Ten-no}},\ }\href
  {\doibase 10.1016/S0009-2614(02)00042-8} {\bibfield  {journal} {\bibinfo
  {journal} {Chemical Physics Letters}\ }\textbf {\bibinfo {volume} {353}},\
  \bibinfo {pages} {317} (\bibinfo {year} {2002})}\BibitemShut {NoStop}%
\bibitem [{\citenamefont {Ten-no}\ and\ \citenamefont
  {Hino}(2002)}]{TenNo_Hino_2002}%
  \BibitemOpen
  \bibfield  {author} {\bibinfo {author} {\bibfnamefont {S.}~\bibnamefont
  {Ten-no}}\ and\ \bibinfo {author} {\bibfnamefont {O.}~\bibnamefont {Hino}},\
  }\href {\doibase 10.3390/i3050459} {\bibfield  {journal} {\bibinfo  {journal}
  {International Journal of Molecular Sciences}\ }\textbf {\bibinfo {volume}
  {3}},\ \bibinfo {pages} {459} (\bibinfo {year} {2002})}\BibitemShut {NoStop}%
\bibitem [{\citenamefont {Liao}\ \emph {et~al.}(2021)\citenamefont {Liao},
  \citenamefont {Schraivogel}, \citenamefont {Luo}, \citenamefont {Kats},\ and\
  \citenamefont {Alavi}}]{Liao_etal_2021}%
  \BibitemOpen
  \bibfield  {author} {\bibinfo {author} {\bibfnamefont {K.}~\bibnamefont
  {Liao}}, \bibinfo {author} {\bibfnamefont {T.}~\bibnamefont {Schraivogel}},
  \bibinfo {author} {\bibfnamefont {H.}~\bibnamefont {Luo}}, \bibinfo {author}
  {\bibfnamefont {D.}~\bibnamefont {Kats}}, \ and\ \bibinfo {author}
  {\bibfnamefont {A.}~\bibnamefont {Alavi}},\ }\href {\doibase
  10.1103/PhysRevResearch.3.033072} {\bibfield  {journal} {\bibinfo  {journal}
  {Phys. Rev. Res.}\ }\textbf {\bibinfo {volume} {3}},\ \bibinfo {pages}
  {033072} (\bibinfo {year} {2021})}\BibitemShut {NoStop}%
\bibitem [{\citenamefont {Schraivogel}\ \emph {et~al.}(2021)\citenamefont
  {Schraivogel}, \citenamefont {Cohen}, \citenamefont {Alavi},\ and\
  \citenamefont {Kats}}]{Schraivogel_etal_2021}%
  \BibitemOpen
  \bibfield  {author} {\bibinfo {author} {\bibfnamefont {T.}~\bibnamefont
  {Schraivogel}}, \bibinfo {author} {\bibfnamefont {A.~J.}\ \bibnamefont
  {Cohen}}, \bibinfo {author} {\bibfnamefont {A.}~\bibnamefont {Alavi}}, \ and\
  \bibinfo {author} {\bibfnamefont {D.}~\bibnamefont {Kats}},\ }\href {\doibase
  10.1063/5.0072495} {\bibfield  {journal} {\bibinfo  {journal} {The Journal of
  Chemical Physics}\ }\textbf {\bibinfo {volume} {155}},\ \bibinfo {pages}
  {191101} (\bibinfo {year} {2021})}\BibitemShut {NoStop}%
\bibitem [{\citenamefont {Baiardi}\ and\ \citenamefont
  {Reiher}(2020)}]{Baiardi_Reither_2020}%
  \BibitemOpen
  \bibfield  {author} {\bibinfo {author} {\bibfnamefont {A.}~\bibnamefont
  {Baiardi}}\ and\ \bibinfo {author} {\bibfnamefont {M.}~\bibnamefont
  {Reiher}},\ }\href {\doibase 10.1063/5.0028608} {\bibfield  {journal}
  {\bibinfo  {journal} {The Journal of Chemical Physics}\ }\textbf {\bibinfo
  {volume} {153}},\ \bibinfo {pages} {164115} (\bibinfo {year}
  {2020})}\BibitemShut {NoStop}%
\bibitem [{\citenamefont {Baiardi}, \citenamefont {Lesiuk},\ and\ \citenamefont
  {Reiher}(2022)}]{Baiardi_etal_2022}%
  \BibitemOpen
  \bibfield  {author} {\bibinfo {author} {\bibfnamefont {A.}~\bibnamefont
  {Baiardi}}, \bibinfo {author} {\bibfnamefont {M.}~\bibnamefont {Lesiuk}}, \
  and\ \bibinfo {author} {\bibfnamefont {M.}~\bibnamefont {Reiher}},\ }\href
  {\doibase 10.1021/acs.jctc.2c00167} {\bibfield  {journal} {\bibinfo
  {journal} {Journal of Chemical Theory and Computation}\ }\textbf {\bibinfo
  {volume} {18}},\ \bibinfo {pages} {4203} (\bibinfo {year}
  {2022})}\BibitemShut {NoStop}%
\bibitem [{\citenamefont {Imamura}\ and\ \citenamefont
  {Scuseria}(2003)}]{Imamura_Scuseria_2003}%
  \BibitemOpen
  \bibfield  {author} {\bibinfo {author} {\bibfnamefont {Y.}~\bibnamefont
  {Imamura}}\ and\ \bibinfo {author} {\bibfnamefont {G.~E.}\ \bibnamefont
  {Scuseria}},\ }\href {\doibase 10.1063/1.1535442} {\bibfield  {journal}
  {\bibinfo  {journal} {The Journal of Chemical Physics}\ }\textbf {\bibinfo
  {volume} {118}},\ \bibinfo {pages} {2464} (\bibinfo {year}
  {2003})}\BibitemShut {NoStop}%
\bibitem [{\citenamefont {Umezawa}(2017)}]{Umezawa_2017}%
  \BibitemOpen
  \bibfield  {author} {\bibinfo {author} {\bibfnamefont {N.}~\bibnamefont
  {Umezawa}},\ }\href {\doibase 10.1063/1.5001260} {\bibfield  {journal}
  {\bibinfo  {journal} {The Journal of Chemical Physics}\ }\textbf {\bibinfo
  {volume} {147}},\ \bibinfo {pages} {104104} (\bibinfo {year}
  {2017})}\BibitemShut {NoStop}%
\bibitem [{\citenamefont {Suzuki}\ and\ \citenamefont
  {Matsumura}(2005)}]{Suzuki_Matsumura_2005}%
  \BibitemOpen
  \bibfield  {author} {\bibinfo {author} {\bibfnamefont {Y.}~\bibnamefont
  {Suzuki}}\ and\ \bibinfo {author} {\bibfnamefont {H.}~\bibnamefont
  {Matsumura}},\ }\href {\doibase 10.1143/PTP.113.87} {\bibfield  {journal}
  {\bibinfo  {journal} {Progress of Theoretical Physics}\ }\textbf {\bibinfo
  {volume} {113}},\ \bibinfo {pages} {87} (\bibinfo {year} {2005})}\BibitemShut
  {NoStop}%
\bibitem [{\citenamefont {Jeszenszki}\ \emph {et~al.}(2018)\citenamefont
  {Jeszenszki}, \citenamefont {Luo}, \citenamefont {Alavi},\ and\ \citenamefont
  {Brand}}]{Jeszenszki_etal_2018}%
  \BibitemOpen
  \bibfield  {author} {\bibinfo {author} {\bibfnamefont {P.}~\bibnamefont
  {Jeszenszki}}, \bibinfo {author} {\bibfnamefont {H.}~\bibnamefont {Luo}},
  \bibinfo {author} {\bibfnamefont {A.}~\bibnamefont {Alavi}}, \ and\ \bibinfo
  {author} {\bibfnamefont {J.}~\bibnamefont {Brand}},\ }\href {\doibase
  10.1103/PhysRevA.98.053627} {\bibfield  {journal} {\bibinfo  {journal} {Phys.
  Rev. A}\ }\textbf {\bibinfo {volume} {98}},\ \bibinfo {pages} {053627}
  (\bibinfo {year} {2018})}\BibitemShut {NoStop}%
\bibitem [{\citenamefont {{McArdle}}\ and\ \citenamefont
  {Tew}(2020)}]{McArdle_Tew_2020}%
  \BibitemOpen
  \bibfield  {author} {\bibinfo {author} {\bibfnamefont {S.}~\bibnamefont
  {{McArdle}}}\ and\ \bibinfo {author} {\bibfnamefont {D.~P.}\ \bibnamefont
  {Tew}},\ }\href {https://arxiv.org/abs/2006.11181} {\bibfield  {journal}
  {\bibinfo  {journal} {arXiv preprint arXiv:2006.11181}\ } (\bibinfo {year}
  {2020})}\BibitemShut {NoStop}%
\bibitem [{\citenamefont {Sokolov}\ \emph {et~al.}(2022)\citenamefont
  {Sokolov}, \citenamefont {Dobrautz}, \citenamefont {Luo}, \citenamefont
  {Alavi},\ and\ \citenamefont {Tavernelli}}]{Sokolov_etal_2022}%
  \BibitemOpen
  \bibfield  {author} {\bibinfo {author} {\bibfnamefont {I.~O.}\ \bibnamefont
  {Sokolov}}, \bibinfo {author} {\bibfnamefont {W.}~\bibnamefont {Dobrautz}},
  \bibinfo {author} {\bibfnamefont {H.}~\bibnamefont {Luo}}, \bibinfo {author}
  {\bibfnamefont {A.}~\bibnamefont {Alavi}}, \ and\ \bibinfo {author}
  {\bibfnamefont {I.}~\bibnamefont {Tavernelli}},\ }\href
  {https://arxiv.org/abs/2201.03049} {\bibfield  {journal} {\bibinfo  {journal}
  {arXiv preprint arXiv:2201.03049}\ } (\bibinfo {year} {2022})}\BibitemShut
  {NoStop}%
\bibitem [{\citenamefont {Hirschfelder}(1963)}]{Hirschfelder-JCP-63}%
  \BibitemOpen
  \bibfield  {author} {\bibinfo {author} {\bibfnamefont {J.~O.}\ \bibnamefont
  {Hirschfelder}},\ }\href {\doibase 10.1063/1.1734157} {\bibfield  {journal}
  {\bibinfo  {journal} {The Journal of Chemical Physics}\ }\textbf {\bibinfo
  {volume} {39}},\ \bibinfo {pages} {3145} (\bibinfo {year}
  {1963})}\BibitemShut {NoStop}%
\bibitem [{\citenamefont {Boys}\ and\ \citenamefont
  {Handy}(1969{\natexlab{b}})}]{Boys_Handy_1969_condition}%
  \BibitemOpen
  \bibfield  {author} {\bibinfo {author} {\bibfnamefont {S.~F.}\ \bibnamefont
  {Boys}}\ and\ \bibinfo {author} {\bibfnamefont {N.~C.}\ \bibnamefont
  {Handy}},\ }\href {\doibase 10.1098/rspa.1969.0038} {\bibfield  {journal}
  {\bibinfo  {journal} {Proceedings of the Royal Society of London. A.
  Mathematical and Physical Sciences}\ }\textbf {\bibinfo {volume} {309}},\
  \bibinfo {pages} {209} (\bibinfo {year} {1969}{\natexlab{b}})}\BibitemShut
  {NoStop}%
\bibitem [{\citenamefont {Bernardi}\ and\ \citenamefont
  {Boys}(1973)}]{Bernardi_Boys_1973}%
  \BibitemOpen
  \bibfield  {author} {\bibinfo {author} {\bibfnamefont {F.}~\bibnamefont
  {Bernardi}}\ and\ \bibinfo {author} {\bibfnamefont {S.~F.}\ \bibnamefont
  {Boys}},\ }\href {\doibase 10.1080/00268977300100051} {\bibfield  {journal}
  {\bibinfo  {journal} {Molecular Physics}\ }\textbf {\bibinfo {volume} {25}},\
  \bibinfo {pages} {35} (\bibinfo {year} {1973})}\BibitemShut {NoStop}%
\bibitem [{\citenamefont {Boys}\ and\ \citenamefont
  {Handy}(1969{\natexlab{c}})}]{Boys_Handy_1969_calculation}%
  \BibitemOpen
  \bibfield  {author} {\bibinfo {author} {\bibfnamefont {S.~F.}\ \bibnamefont
  {Boys}}\ and\ \bibinfo {author} {\bibfnamefont {N.~C.}\ \bibnamefont
  {Handy}},\ }\href {\doibase 10.1098/rspa.1969.0062} {\bibfield  {journal}
  {\bibinfo  {journal} {Proceedings of the Royal Society of London. A.
  Mathematical and Physical Sciences}\ }\textbf {\bibinfo {volume} {310}},\
  \bibinfo {pages} {63} (\bibinfo {year} {1969}{\natexlab{c}})}\BibitemShut
  {NoStop}%
\bibitem [{\citenamefont {Boys}\ and\ \citenamefont
  {Handy}(1969{\natexlab{d}})}]{Boys_Handy_1969_first}%
  \BibitemOpen
  \bibfield  {author} {\bibinfo {author} {\bibfnamefont {S.~F.}\ \bibnamefont
  {Boys}}\ and\ \bibinfo {author} {\bibfnamefont {N.~C.}\ \bibnamefont
  {Handy}},\ }\href {\doibase 10.1098/rspa.1969.0120} {\bibfield  {journal}
  {\bibinfo  {journal} {Proceedings of the Royal Society of London Series A}\
  }\textbf {\bibinfo {volume} {311}},\ \bibinfo {pages} {309} (\bibinfo {year}
  {1969}{\natexlab{d}})}\BibitemShut {NoStop}%
\bibitem [{\citenamefont {Handy}(1969)}]{Handy_1969}%
  \BibitemOpen
  \bibfield  {author} {\bibinfo {author} {\bibfnamefont {N.~C.}\ \bibnamefont
  {Handy}},\ }\href {\doibase 10.1063/1.1672496} {\bibfield  {journal}
  {\bibinfo  {journal} {The Journal of Chemical Physics}\ }\textbf {\bibinfo
  {volume} {51}},\ \bibinfo {pages} {3205} (\bibinfo {year}
  {1969})}\BibitemShut {NoStop}%
\bibitem [{\citenamefont {Hall}\ and\ \citenamefont
  {Solomon}(1969)}]{Hall_Solomon_1969}%
  \BibitemOpen
  \bibfield  {author} {\bibinfo {author} {\bibfnamefont {G.}~\bibnamefont
  {Hall}}\ and\ \bibinfo {author} {\bibfnamefont {C.}~\bibnamefont {Solomon}},\
  }\href {\doibase https://doi.org/10.1016/0009-2614(69)80260-5} {\bibfield
  {journal} {\bibinfo  {journal} {Chemical Physics Letters}\ }\textbf {\bibinfo
  {volume} {4}},\ \bibinfo {pages} {352} (\bibinfo {year} {1969})}\BibitemShut
  {NoStop}%
\bibitem [{\citenamefont {Lim}(1974)}]{Lim_1974}%
  \BibitemOpen
  \bibfield  {author} {\bibinfo {author} {\bibfnamefont {T.~K.}\ \bibnamefont
  {Lim}},\ }\href {\doibase doi.org/10.1002/qua.560080405} {\bibfield
  {journal} {\bibinfo  {journal} {International Journal of Quantum Chemistry}\
  }\textbf {\bibinfo {volume} {8}},\ \bibinfo {pages} {539} (\bibinfo {year}
  {1974})}\BibitemShut {NoStop}%
\bibitem [{\citenamefont {Boys}(1969)}]{Boys_1969}%
  \BibitemOpen
  \bibfield  {author} {\bibinfo {author} {\bibfnamefont {S.~F.}\ \bibnamefont
  {Boys}},\ }\href {\doibase 10.1098/rspa.1969.0037} {\bibfield  {journal}
  {\bibinfo  {journal} {Proceedings of the Royal Society of London. A.
  Mathematical and Physical Sciences}\ }\textbf {\bibinfo {volume} {309}},\
  \bibinfo {pages} {195} (\bibinfo {year} {1969})}\BibitemShut {NoStop}%
\bibitem [{\citenamefont {Boys}\ and\ \citenamefont
  {Bernardi}(1970)}]{Boys_Bernardi_1970}%
  \BibitemOpen
  \bibfield  {author} {\bibinfo {author} {\bibfnamefont {S.~F.}\ \bibnamefont
  {Boys}}\ and\ \bibinfo {author} {\bibfnamefont {F.}~\bibnamefont
  {Bernardi}},\ }\href {\doibase 10.1080/00268977000101561} {\bibfield
  {journal} {\bibinfo  {journal} {Molecular Physics}\ }\textbf {\bibinfo
  {volume} {19}},\ \bibinfo {pages} {553} (\bibinfo {year} {1970})}\BibitemShut
  {NoStop}%
\bibitem [{\citenamefont {Handy}\ and\ \citenamefont
  {Epstein}(1970)}]{Handy_Epstein_1970}%
  \BibitemOpen
  \bibfield  {author} {\bibinfo {author} {\bibfnamefont {N.~C.}\ \bibnamefont
  {Handy}}\ and\ \bibinfo {author} {\bibfnamefont {S.~T.}\ \bibnamefont
  {Epstein}},\ }\href {\doibase 10.1063/1.1674185} {\bibfield  {journal}
  {\bibinfo  {journal} {The Journal of Chemical Physics}\ }\textbf {\bibinfo
  {volume} {53}},\ \bibinfo {pages} {1392} (\bibinfo {year}
  {1970})}\BibitemShut {NoStop}%
\bibitem [{\citenamefont {Armour}(1972)}]{Armour_1972_convergence}%
  \BibitemOpen
  \bibfield  {author} {\bibinfo {author} {\bibfnamefont {E.~A.~G.}\
  \bibnamefont {Armour}},\ }\href {\doibase 10.1080/00268977200101321}
  {\bibfield  {journal} {\bibinfo  {journal} {Molecular Physics}\ }\textbf
  {\bibinfo {volume} {24}},\ \bibinfo {pages} {163} (\bibinfo {year}
  {1972})}\BibitemShut {NoStop}%
\bibitem [{\citenamefont {Armour}(1973)}]{Armour_1973}%
  \BibitemOpen
  \bibfield  {author} {\bibinfo {author} {\bibfnamefont {E.~A.~G.}\
  \bibnamefont {Armour}},\ }\href {\doibase 10.1080/00268977300102321}
  {\bibfield  {journal} {\bibinfo  {journal} {Molecular Physics}\ }\textbf
  {\bibinfo {volume} {26}},\ \bibinfo {pages} {1093} (\bibinfo {year}
  {1973})}\BibitemShut {NoStop}%
\bibitem [{\citenamefont {Armour}(1974)}]{Armour_1974}%
  \BibitemOpen
  \bibfield  {author} {\bibinfo {author} {\bibfnamefont {E.~A.~G.}\
  \bibnamefont {Armour}},\ }\href {\doibase
  https://doi.org/10.1016/0009-2614(74)85383-2} {\bibfield  {journal} {\bibinfo
   {journal} {Chemical Physics Letters}\ }\textbf {\bibinfo {volume} {25}},\
  \bibinfo {pages} {614} (\bibinfo {year} {1974})}\BibitemShut {NoStop}%
\bibitem [{\citenamefont {Handy}(1975)}]{Handy_1975}%
  \BibitemOpen
  \bibfield  {author} {\bibinfo {author} {\bibfnamefont {N.}~\bibnamefont
  {Handy}},\ }in\ \href
  {https://link.springer.com/content/pdf/10.1007/978-94-010-1815-9_7.pdf}
  {\emph {\bibinfo {booktitle} {Computational Techniques in Quantum Chemistry
  and Molecular Physics}}}\ (\bibinfo  {publisher} {Springer},\ \bibinfo {year}
  {1975})\ pp.\ \bibinfo {pages} {425--433}\BibitemShut {NoStop}%
\bibitem [{\citenamefont {Huggett}\ and\ \citenamefont
  {Armour}(1976)}]{Huggett_Armour_1976}%
  \BibitemOpen
  \bibfield  {author} {\bibinfo {author} {\bibfnamefont {J.~P.}\ \bibnamefont
  {Huggett}}\ and\ \bibinfo {author} {\bibfnamefont {E.~A.~G.}\ \bibnamefont
  {Armour}},\ }\href {\doibase 10.1088/0022-3700/9/18/017} {\bibfield
  {journal} {\bibinfo  {journal} {Journal of Physics B: Atomic and Molecular
  Physics}\ }\textbf {\bibinfo {volume} {9}},\ \bibinfo {pages} {3263}
  (\bibinfo {year} {1976})}\BibitemShut {NoStop}%
\bibitem [{\citenamefont {Armour}(1980)}]{Armour_1980}%
  \BibitemOpen
  \bibfield  {author} {\bibinfo {author} {\bibfnamefont {E.~A.~G.}\
  \bibnamefont {Armour}},\ }\href {\doibase 10.1088/0022-3719/13/3/009}
  {\bibfield  {journal} {\bibinfo  {journal} {Journal of Physics C: Solid State
  Physics}\ }\textbf {\bibinfo {volume} {13}},\ \bibinfo {pages} {343}
  (\bibinfo {year} {1980})}\BibitemShut {NoStop}%
\bibitem [{\citenamefont {Luo}, \citenamefont {Hackbusch},\ and\ \citenamefont
  {Flad}(2010)}]{Luo_etal_2010}%
  \BibitemOpen
  \bibfield  {author} {\bibinfo {author} {\bibfnamefont {H.}~\bibnamefont
  {Luo}}, \bibinfo {author} {\bibfnamefont {W.}~\bibnamefont {Hackbusch}}, \
  and\ \bibinfo {author} {\bibfnamefont {H.-J.}\ \bibnamefont {Flad}},\ }\href
  {\doibase 10.1080/00268970903521194} {\bibfield  {journal} {\bibinfo
  {journal} {Molecular Physics}\ }\textbf {\bibinfo {volume} {108}},\ \bibinfo
  {pages} {425} (\bibinfo {year} {2010})}\BibitemShut {NoStop}%
\bibitem [{\citenamefont {Bernardi}(1973)}]{Bernardi_1973}%
  \BibitemOpen
  \bibfield  {author} {\bibinfo {author} {\bibfnamefont {F.}~\bibnamefont
  {Bernardi}},\ }\href {https://hal.science/jpa-00207398/document} {\bibfield
  {journal} {\bibinfo  {journal} {Journal de physique}\ }\textbf {\bibinfo
  {volume} {34}},\ \bibinfo {pages} {373} (\bibinfo {year} {1973})}\BibitemShut
  {NoStop}%
\bibitem [{\citenamefont {{des Cloizeaux}}(1960)}]{Cloizeaux_1960}%
  \BibitemOpen
  \bibfield  {author} {\bibinfo {author} {\bibfnamefont {J.}~\bibnamefont {{des
  Cloizeaux}}},\ }\href {\doibase https://doi.org/10.1016/0029-5582(60)90177-2}
  {\bibfield  {journal} {\bibinfo  {journal} {Nuclear Physics}\ }\textbf
  {\bibinfo {volume} {20}},\ \bibinfo {pages} {321} (\bibinfo {year}
  {1960})}\BibitemShut {NoStop}%
\bibitem [{\citenamefont {Brandow}(1967)}]{Brandow_1967}%
  \BibitemOpen
  \bibfield  {author} {\bibinfo {author} {\bibfnamefont {B.~H.}\ \bibnamefont
  {Brandow}},\ }\href {\doibase 10.1103/RevModPhys.39.771} {\bibfield
  {journal} {\bibinfo  {journal} {Rev. Mod. Phys.}\ }\textbf {\bibinfo {volume}
  {39}},\ \bibinfo {pages} {771} (\bibinfo {year} {1967})}\BibitemShut
  {NoStop}%
\bibitem [{\citenamefont {Moshinsky}\ and\ \citenamefont
  {Seligman}(1971)}]{Moshinsky_Seligman_1971}%
  \BibitemOpen
  \bibfield  {author} {\bibinfo {author} {\bibfnamefont {M.}~\bibnamefont
  {Moshinsky}}\ and\ \bibinfo {author} {\bibfnamefont {T.}~\bibnamefont
  {Seligman}},\ }\href {\doibase https://doi.org/10.1016/0003-4916(71)90191-6}
  {\bibfield  {journal} {\bibinfo  {journal} {Annals of Physics}\ }\textbf
  {\bibinfo {volume} {66}},\ \bibinfo {pages} {311} (\bibinfo {year}
  {1971})}\BibitemShut {NoStop}%
\bibitem [{\citenamefont {Gouyet}(1973{\natexlab{a}})}]{Gouyet_1973_champ}%
  \BibitemOpen
  \bibfield  {author} {\bibinfo {author} {\bibfnamefont {J.~F.}\ \bibnamefont
  {Gouyet}},\ }\href {\doibase https://doi.org/10.1002/qua.560070202}
  {\bibfield  {journal} {\bibinfo  {journal} {International Journal of Quantum
  Chemistry}\ }\textbf {\bibinfo {volume} {7}},\ \bibinfo {pages} {139}
  (\bibinfo {year} {1973}{\natexlab{a}})}\BibitemShut {NoStop}%
\bibitem [{\citenamefont {Dahl}(1978)}]{Dahl_1978}%
  \BibitemOpen
  \bibfield  {author} {\bibinfo {author} {\bibfnamefont {J.~P.}\ \bibnamefont
  {Dahl}},\ }\href {\doibase https://doi.org/10.1002/qua.560140207} {\bibfield
  {journal} {\bibinfo  {journal} {International Journal of Quantum Chemistry}\
  }\textbf {\bibinfo {volume} {14}},\ \bibinfo {pages} {191} (\bibinfo {year}
  {1978})}\BibitemShut {NoStop}%
\bibitem [{\citenamefont {Payne}(1982)}]{Payne_1982}%
  \BibitemOpen
  \bibfield  {author} {\bibinfo {author} {\bibfnamefont {P.~W.}\ \bibnamefont
  {Payne}},\ }\href {\doibase 10.1063/1.443769} {\bibfield  {journal} {\bibinfo
   {journal} {The Journal of Chemical Physics}\ }\textbf {\bibinfo {volume}
  {77}},\ \bibinfo {pages} {5630} (\bibinfo {year} {1982})}\BibitemShut
  {NoStop}%
\bibitem [{\citenamefont {Brody}(2013)}]{Brody_2014}%
  \BibitemOpen
  \bibfield  {author} {\bibinfo {author} {\bibfnamefont {D.~C.}\ \bibnamefont
  {Brody}},\ }\href {\doibase 10.1088/1751-8113/47/3/035305} {\bibfield
  {journal} {\bibinfo  {journal} {Journal of Physics A: Mathematical and
  Theoretical}\ }\textbf {\bibinfo {volume} {47}},\ \bibinfo {pages} {035305}
  (\bibinfo {year} {2013})}\BibitemShut {NoStop}%
\bibitem [{\citenamefont {Ashida}, \citenamefont {Gong},\ and\ \citenamefont
  {Ueda}(2020)}]{Ashida_etal_2020}%
  \BibitemOpen
  \bibfield  {author} {\bibinfo {author} {\bibfnamefont {Y.}~\bibnamefont
  {Ashida}}, \bibinfo {author} {\bibfnamefont {Z.}~\bibnamefont {Gong}}, \ and\
  \bibinfo {author} {\bibfnamefont {M.}~\bibnamefont {Ueda}},\ }\href {\doibase
  10.1080/00018732.2021.1876991} {\bibfield  {journal} {\bibinfo  {journal}
  {Advances in Physics}\ }\textbf {\bibinfo {volume} {69}},\ \bibinfo {pages}
  {249} (\bibinfo {year} {2020})}\BibitemShut {NoStop}%
\bibitem [{\citenamefont {Fimple}\ and\ \citenamefont
  {Unwin}(1976)}]{Fimple_Unwin_1976}%
  \BibitemOpen
  \bibfield  {author} {\bibinfo {author} {\bibfnamefont {W.~R.}\ \bibnamefont
  {Fimple}}\ and\ \bibinfo {author} {\bibfnamefont {M.~J.}\ \bibnamefont
  {Unwin}},\ }\href {\doibase https://doi.org/10.1002/qua.560100408} {\bibfield
   {journal} {\bibinfo  {journal} {International Journal of Quantum Chemistry}\
  }\textbf {\bibinfo {volume} {10}},\ \bibinfo {pages} {643} (\bibinfo {year}
  {1976})}\BibitemShut {NoStop}%
\bibitem [{\citenamefont {Lee}\ and\ \citenamefont
  {Thom}(2023)}]{Lee_Thom_2023}%
  \BibitemOpen
  \bibfield  {author} {\bibinfo {author} {\bibfnamefont {N.}~\bibnamefont
  {Lee}}\ and\ \bibinfo {author} {\bibfnamefont {A.~J.}\ \bibnamefont {Thom}},\
  }\href {https://doi.org/10.48550/arXiv.2301.02590} {\bibfield  {journal}
  {\bibinfo  {journal} {arXiv}\ } (\bibinfo {year} {2023})}\BibitemShut
  {NoStop}%
\bibitem [{\citenamefont {Wahlen-Strothman}\ \emph {et~al.}(2015)\citenamefont
  {Wahlen-Strothman}, \citenamefont {Jim\'enez-Hoyos}, \citenamefont
  {Henderson},\ and\ \citenamefont {Scuseria}}]{WahlenStrothman_etal_2015}%
  \BibitemOpen
  \bibfield  {author} {\bibinfo {author} {\bibfnamefont {J.~M.}\ \bibnamefont
  {Wahlen-Strothman}}, \bibinfo {author} {\bibfnamefont {C.~A.}\ \bibnamefont
  {Jim\'enez-Hoyos}}, \bibinfo {author} {\bibfnamefont {T.~M.}\ \bibnamefont
  {Henderson}}, \ and\ \bibinfo {author} {\bibfnamefont {G.~E.}\ \bibnamefont
  {Scuseria}},\ }\href {\doibase 10.1103/PhysRevB.91.041114} {\bibfield
  {journal} {\bibinfo  {journal} {Phys. Rev. B}\ }\textbf {\bibinfo {volume}
  {91}},\ \bibinfo {pages} {041114} (\bibinfo {year} {2015})}\BibitemShut
  {NoStop}%
\bibitem [{\citenamefont {Ochi}, \citenamefont {Arita},\ and\ \citenamefont
  {Tsuneyuki}(2017)}]{Ochi_etal_2017}%
  \BibitemOpen
  \bibfield  {author} {\bibinfo {author} {\bibfnamefont {M.}~\bibnamefont
  {Ochi}}, \bibinfo {author} {\bibfnamefont {R.}~\bibnamefont {Arita}}, \ and\
  \bibinfo {author} {\bibfnamefont {S.}~\bibnamefont {Tsuneyuki}},\ }\href
  {\doibase 10.1103/PhysRevLett.118.026402} {\bibfield  {journal} {\bibinfo
  {journal} {Phys. Rev. Lett.}\ }\textbf {\bibinfo {volume} {118}},\ \bibinfo
  {pages} {026402} (\bibinfo {year} {2017})}\BibitemShut {NoStop}%
\bibitem [{\citenamefont {Ochi}(2023)}]{Ochi-CPC-23}%
  \BibitemOpen
  \bibfield  {author} {\bibinfo {author} {\bibfnamefont {M.}~\bibnamefont
  {Ochi}},\ }\href {\doibase 10.1016/j.cpc.2023.108687} {\bibfield  {journal}
  {\bibinfo  {journal} {Comput Phys Commun}\ ,\ \bibinfo {pages} {108687}}
  (\bibinfo {year} {2023})}\BibitemShut {NoStop}%
\bibitem [{\citenamefont {Umezawa}\ and\ \citenamefont
  {Tsuneyuki}(2003{\natexlab{a}})}]{Umezawa_Tsuneyuki_2003_tc_self}%
  \BibitemOpen
  \bibfield  {author} {\bibinfo {author} {\bibfnamefont {N.}~\bibnamefont
  {Umezawa}}\ and\ \bibinfo {author} {\bibfnamefont {S.}~\bibnamefont
  {Tsuneyuki}},\ }\href {\doibase https://doi.org/10.1002/qua.10455} {\bibfield
   {journal} {\bibinfo  {journal} {International Journal of Quantum Chemistry}\
  }\textbf {\bibinfo {volume} {91}},\ \bibinfo {pages} {184} (\bibinfo {year}
  {2003}{\natexlab{a}})}\BibitemShut {NoStop}%
\bibitem [{\citenamefont {Umezawa}\ and\ \citenamefont
  {Tsuneyuki}(2003{\natexlab{b}})}]{Umezawa_Tsuneyuki_2003_tc_method}%
  \BibitemOpen
  \bibfield  {author} {\bibinfo {author} {\bibfnamefont {N.}~\bibnamefont
  {Umezawa}}\ and\ \bibinfo {author} {\bibfnamefont {S.}~\bibnamefont
  {Tsuneyuki}},\ }\href {\doibase 10.1063/1.1617274} {\bibfield  {journal}
  {\bibinfo  {journal} {The Journal of Chemical Physics}\ }\textbf {\bibinfo
  {volume} {119}},\ \bibinfo {pages} {10015} (\bibinfo {year}
  {2003}{\natexlab{b}})}\BibitemShut {NoStop}%
\bibitem [{\citenamefont {Handy}(1971)}]{Handy_1971_minimization}%
  \BibitemOpen
  \bibfield  {author} {\bibinfo {author} {\bibfnamefont {N.~C.}\ \bibnamefont
  {Handy}},\ }\href {\doibase 10.1080/00268977100101961} {\bibfield  {journal}
  {\bibinfo  {journal} {Molecular Physics}\ }\textbf {\bibinfo {volume} {21}},\
  \bibinfo {pages} {817} (\bibinfo {year} {1971})}\BibitemShut {NoStop}%
\bibitem [{\citenamefont {Handy}(1972)}]{Handy_1972}%
  \BibitemOpen
  \bibfield  {author} {\bibinfo {author} {\bibfnamefont {N.~C.}\ \bibnamefont
  {Handy}},\ }\href {\doibase 10.1080/00268977200100011} {\bibfield  {journal}
  {\bibinfo  {journal} {Molecular Physics}\ }\textbf {\bibinfo {volume} {23}},\
  \bibinfo {pages} {1} (\bibinfo {year} {1972})}\BibitemShut {NoStop}%
\bibitem [{\citenamefont {Handy}(1973)}]{Handy_1973}%
  \BibitemOpen
  \bibfield  {author} {\bibinfo {author} {\bibfnamefont {N.~C.}\ \bibnamefont
  {Handy}},\ }\href {\doibase 10.1063/1.1678918} {\bibfield  {journal}
  {\bibinfo  {journal} {The Journal of Chemical Physics}\ }\textbf {\bibinfo
  {volume} {58}},\ \bibinfo {pages} {279} (\bibinfo {year} {1973})}\BibitemShut
  {NoStop}%
\bibitem [{\citenamefont {Ochi}\ \emph {et~al.}(2012)\citenamefont {Ochi},
  \citenamefont {Sodeyama}, \citenamefont {Sakuma},\ and\ \citenamefont
  {Tsuneyuki}}]{Ochi_etal_2012}%
  \BibitemOpen
  \bibfield  {author} {\bibinfo {author} {\bibfnamefont {M.}~\bibnamefont
  {Ochi}}, \bibinfo {author} {\bibfnamefont {K.}~\bibnamefont {Sodeyama}},
  \bibinfo {author} {\bibfnamefont {R.}~\bibnamefont {Sakuma}}, \ and\ \bibinfo
  {author} {\bibfnamefont {S.}~\bibnamefont {Tsuneyuki}},\ }\href {\doibase
  10.1063/1.3689440} {\bibfield  {journal} {\bibinfo  {journal} {The Journal of
  Chemical Physics}\ }\textbf {\bibinfo {volume} {136}},\ \bibinfo {pages}
  {094108} (\bibinfo {year} {2012})}\BibitemShut {NoStop}%
\bibitem [{\citenamefont {Ochi}\ and\ \citenamefont
  {Tsuneyuki}(2013)}]{Ochi_Tsuneyuki_2013}%
  \BibitemOpen
  \bibfield  {author} {\bibinfo {author} {\bibfnamefont {M.}~\bibnamefont
  {Ochi}}\ and\ \bibinfo {author} {\bibfnamefont {S.}~\bibnamefont
  {Tsuneyuki}},\ }\href {\doibase 10.1088/1742-6596/454/1/012020} {\bibfield
  {journal} {\bibinfo  {journal} {Journal of Physics: Conference Series}\
  }\textbf {\bibinfo {volume} {454}},\ \bibinfo {pages} {012020} (\bibinfo
  {year} {2013})}\BibitemShut {NoStop}%
\bibitem [{\citenamefont {Ochi}, \citenamefont {Sodeyama},\ and\ \citenamefont
  {Tsuneyuki}(2014)}]{Ochi_etal_2014}%
  \BibitemOpen
  \bibfield  {author} {\bibinfo {author} {\bibfnamefont {M.}~\bibnamefont
  {Ochi}}, \bibinfo {author} {\bibfnamefont {K.}~\bibnamefont {Sodeyama}}, \
  and\ \bibinfo {author} {\bibfnamefont {S.}~\bibnamefont {Tsuneyuki}},\ }\href
  {\doibase 10.1063/1.4865500} {\bibfield  {journal} {\bibinfo  {journal} {The
  Journal of Chemical Physics}\ }\textbf {\bibinfo {volume} {140}},\ \bibinfo
  {pages} {074112} (\bibinfo {year} {2014})}\BibitemShut {NoStop}%
\bibitem [{\citenamefont {Umezawa}\ \emph {et~al.}(2006)\citenamefont
  {Umezawa}, \citenamefont {Sarsa}, \citenamefont {Le~Sech},\ and\
  \citenamefont {Chikyow}}]{Umezawa_etal_2006}%
  \BibitemOpen
  \bibfield  {author} {\bibinfo {author} {\bibfnamefont {N.}~\bibnamefont
  {Umezawa}}, \bibinfo {author} {\bibfnamefont {A.}~\bibnamefont {Sarsa}},
  \bibinfo {author} {\bibfnamefont {C.}~\bibnamefont {Le~Sech}}, \ and\
  \bibinfo {author} {\bibfnamefont {T.}~\bibnamefont {Chikyow}},\ }\href
  {\doibase 10.1103/PhysRevA.73.012512} {\bibfield  {journal} {\bibinfo
  {journal} {Phys. Rev. A}\ }\textbf {\bibinfo {volume} {73}},\ \bibinfo
  {pages} {012512} (\bibinfo {year} {2006})}\BibitemShut {NoStop}%
\bibitem [{\citenamefont {Luo}(2010)}]{Luo_2010}%
  \BibitemOpen
  \bibfield  {author} {\bibinfo {author} {\bibfnamefont {H.}~\bibnamefont
  {Luo}},\ }\href {\doibase 10.1063/1.3505037} {\bibfield  {journal} {\bibinfo
  {journal} {The Journal of Chemical Physics}\ }\textbf {\bibinfo {volume}
  {133}},\ \bibinfo {pages} {154109} (\bibinfo {year} {2010})}\BibitemShut
  {NoStop}%
\bibitem [{\citenamefont {Ten-no}(2000)}]{TenNo-CPL-00-a}%
  \BibitemOpen
  \bibfield  {author} {\bibinfo {author} {\bibfnamefont {S.}~\bibnamefont
  {Ten-no}},\ }\href {\doibase https://doi.org/10.1016/S0009-2614(00)01066-6}
  {\bibfield  {journal} {\bibinfo  {journal} {Chemical Physics Letters}\
  }\textbf {\bibinfo {volume} {330}},\ \bibinfo {pages} {169 } (\bibinfo {year}
  {2000})}\BibitemShut {NoStop}%
\bibitem [{\citenamefont {{I. Røeggen and K.
  Aashamar}}(1977)}]{Roeggen_Aashamar_1977}%
  \BibitemOpen
  \bibfield  {author} {\bibinfo {author} {\bibnamefont {{I. Røeggen and K.
  Aashamar}}},\ }\href {\doibase 10.1080/00268977700100391} {\bibfield
  {journal} {\bibinfo  {journal} {Molecular Physics}\ }\textbf {\bibinfo
  {volume} {33}},\ \bibinfo {pages} {453} (\bibinfo {year} {1977})}\BibitemShut
  {NoStop}%
\bibitem [{\citenamefont {Norbeck}\ and\ \citenamefont
  {McWeeny}(1975)}]{Norbeck_McWeeny_1975}%
  \BibitemOpen
  \bibfield  {author} {\bibinfo {author} {\bibfnamefont {J.}~\bibnamefont
  {Norbeck}}\ and\ \bibinfo {author} {\bibfnamefont {R.}~\bibnamefont
  {McWeeny}},\ }\href {\doibase https://doi.org/10.1016/0009-2614(75)85256-0}
  {\bibfield  {journal} {\bibinfo  {journal} {Chemical Physics Letters}\
  }\textbf {\bibinfo {volume} {34}},\ \bibinfo {pages} {206} (\bibinfo {year}
  {1975})}\BibitemShut {NoStop}%
\bibitem [{\citenamefont {McDouall}(1992)}]{McDouall_1992}%
  \BibitemOpen
  \bibfield  {author} {\bibinfo {author} {\bibfnamefont {J.~J.~W.}\
  \bibnamefont {McDouall}},\ }\href {\doibase
  https://doi.org/10.1007/BF01113060} {\bibfield  {journal} {\bibinfo
  {journal} {Theoretica chimica acta}\ }\textbf {\bibinfo {volume} {83}},\
  \bibinfo {pages} {339} (\bibinfo {year} {1992})}\BibitemShut {NoStop}%
\bibitem [{\citenamefont {McDouall}(1993)}]{McDouall_1993}%
  \BibitemOpen
  \bibfield  {author} {\bibinfo {author} {\bibfnamefont {J.~J.}\ \bibnamefont
  {McDouall}},\ }\href {\doibase https://doi.org/10.1007/BF01113433} {\bibfield
   {journal} {\bibinfo  {journal} {Theoretica chimica acta}\ }\textbf {\bibinfo
  {volume} {85}},\ \bibinfo {pages} {395} (\bibinfo {year} {1993})}\BibitemShut
  {NoStop}%
\bibitem [{\citenamefont {Thorsteinsson}\ and\ \citenamefont
  {Cooper}(1998)}]{Thorsteinsson_Cooper_1998}%
  \BibitemOpen
  \bibfield  {author} {\bibinfo {author} {\bibfnamefont {T.}~\bibnamefont
  {Thorsteinsson}}\ and\ \bibinfo {author} {\bibfnamefont {D.~L.}\ \bibnamefont
  {Cooper}},\ }\href {\doibase 10.1080/002689798168998} {\bibfield  {journal}
  {\bibinfo  {journal} {Molecular Physics}\ }\textbf {\bibinfo {volume} {93}},\
  \bibinfo {pages} {663} (\bibinfo {year} {1998})}\BibitemShut {NoStop}%
\bibitem [{\citenamefont {Fuchikami}\ and\ \citenamefont
  {Block}(1982)}]{Fuchikami_Block_1982}%
  \BibitemOpen
  \bibfield  {author} {\bibinfo {author} {\bibfnamefont {N.}~\bibnamefont
  {Fuchikami}}\ and\ \bibinfo {author} {\bibfnamefont {R.}~\bibnamefont
  {Block}},\ }\href {\doibase https://doi.org/10.1016/0378-4363(82)90096-1}
  {\bibfield  {journal} {\bibinfo  {journal} {Physica B+C}\ }\textbf {\bibinfo
  {volume} {112}},\ \bibinfo {pages} {369} (\bibinfo {year}
  {1982})}\BibitemShut {NoStop}%
\bibitem [{\citenamefont {Gouyet}(1973{\natexlab{b}})}]{Gouyet_1973_use}%
  \BibitemOpen
  \bibfield  {author} {\bibinfo {author} {\bibfnamefont {J.~F.}\ \bibnamefont
  {Gouyet}},\ }\href {\doibase 10.1063/1.1680674} {\bibfield  {journal}
  {\bibinfo  {journal} {The Journal of Chemical Physics}\ }\textbf {\bibinfo
  {volume} {59}},\ \bibinfo {pages} {4637} (\bibinfo {year}
  {1973}{\natexlab{b}})}\BibitemShut {NoStop}%
\bibitem [{\citenamefont {Gouyet}(1974)}]{Gouyet_1974}%
  \BibitemOpen
  \bibfield  {author} {\bibinfo {author} {\bibfnamefont {J.~F.}\ \bibnamefont
  {Gouyet}},\ }\href {\doibase 10.1063/1.1681590} {\bibfield  {journal}
  {\bibinfo  {journal} {The Journal of Chemical Physics}\ }\textbf {\bibinfo
  {volume} {60}},\ \bibinfo {pages} {3690} (\bibinfo {year}
  {1974})}\BibitemShut {NoStop}%
\bibitem [{\citenamefont {Mayer}(1983)}]{Mayer_1983}%
  \BibitemOpen
  \bibfield  {author} {\bibinfo {author} {\bibfnamefont {I.}~\bibnamefont
  {Mayer}},\ }\href {\doibase https://doi.org/10.1016/0009-2614(83)80005-0}
  {\bibfield  {journal} {\bibinfo  {journal} {Chemical Physics Letters}\
  }\textbf {\bibinfo {volume} {97}},\ \bibinfo {pages} {270} (\bibinfo {year}
  {1983})}\BibitemShut {NoStop}%
\bibitem [{\citenamefont {{Péter R. Surján and István Mayer and István
  Lukovits}}(1985)}]{Surjan_etal_1985}%
  \BibitemOpen
  \bibfield  {author} {\bibinfo {author} {\bibnamefont {{Péter R. Surján and
  István Mayer and István Lukovits}}},\ }\href {\doibase
  https://doi.org/10.1016/0009-2614(85)85384-7} {\bibfield  {journal} {\bibinfo
   {journal} {Chemical Physics Letters}\ }\textbf {\bibinfo {volume} {119}},\
  \bibinfo {pages} {538} (\bibinfo {year} {1985})}\BibitemShut {NoStop}%
\bibitem [{\citenamefont {{Péter R. Śurján and Raymond A.
  Poirier}}(1986)}]{Surjan_Poirier_1986}%
  \BibitemOpen
  \bibfield  {author} {\bibinfo {author} {\bibnamefont {{Péter R. Śurján and
  Raymond A. Poirier}}},\ }\href {\doibase
  https://doi.org/10.1016/0009-2614(86)80377-3} {\bibfield  {journal} {\bibinfo
   {journal} {Chemical Physics Letters}\ }\textbf {\bibinfo {volume} {128}},\
  \bibinfo {pages} {358} (\bibinfo {year} {1986})}\BibitemShut {NoStop}%
\bibitem [{\citenamefont {{Péter R. Surján and István
  Mayer}}(1991)}]{Surjan_Mayer_1991}%
  \BibitemOpen
  \bibfield  {author} {\bibinfo {author} {\bibnamefont {{Péter R. Surján and
  István Mayer}}},\ }\href {\doibase
  https://doi.org/10.1016/0166-1280(91)80004-R} {\bibfield  {journal} {\bibinfo
   {journal} {Journal of Molecular Structure: THEOCHEM}\ }\textbf {\bibinfo
  {volume} {226}},\ \bibinfo {pages} {47} (\bibinfo {year} {1991})}\BibitemShut
  {NoStop}%
\bibitem [{\citenamefont {Finley}\ and\ \citenamefont
  {Hirao}(2000)}]{Finley_Hirao_2000}%
  \BibitemOpen
  \bibfield  {author} {\bibinfo {author} {\bibfnamefont {J.~P.}\ \bibnamefont
  {Finley}}\ and\ \bibinfo {author} {\bibfnamefont {K.}~\bibnamefont {Hirao}},\
  }\href {\doibase https://doi.org/10.1016/S0009-2614(00)00919-2} {\bibfield
  {journal} {\bibinfo  {journal} {Chemical Physics Letters}\ }\textbf {\bibinfo
  {volume} {328}},\ \bibinfo {pages} {51} (\bibinfo {year} {2000})}\BibitemShut
  {NoStop}%
\bibitem [{\citenamefont {{Thierry Visentin and Christine Cézard and Gaétan
  Weck and Elise Kochanski and Lilyane Padel}}(2001)}]{Visentin_etal_2001}%
  \BibitemOpen
  \bibfield  {author} {\bibinfo {author} {\bibnamefont {{Thierry Visentin and
  Christine Cézard and Gaétan Weck and Elise Kochanski and Lilyane Padel}}},\
  }\href {\doibase https://doi.org/10.1016/S0166-1280(01)00471-7} {\bibfield
  {journal} {\bibinfo  {journal} {Journal of Molecular Structure: THEOCHEM}\
  }\textbf {\bibinfo {volume} {547}},\ \bibinfo {pages} {209} (\bibinfo {year}
  {2001})}\BibitemShut {NoStop}%
\bibitem [{\citenamefont {Cooper}(2002)}]{Cooper_2002}%
  \BibitemOpen
  \bibfield  {author} {\bibinfo {author} {\bibfnamefont {D.}~\bibnamefont
  {Cooper}},\ }\href@noop {} {\emph {\bibinfo {title} {Valence bond theory}}}\
  (\bibinfo  {publisher} {Elsevier},\ \bibinfo {year} {2002})\BibitemShut
  {NoStop}%
\bibitem [{\citenamefont {{Péter R. Surján}}(2011)}]{Surjan_2011}%
  \BibitemOpen
  \bibfield  {author} {\bibinfo {author} {\bibnamefont {{Péter R. Surján}}},\
  }\href {\doibase https://doi.org/10.1007/978-3-642-74755-7} {\emph {\bibinfo
  {title} {Second Quantized Approach to Quantum Chemistry.An Elementary
  Introduction}}}\ (\bibinfo  {publisher} {Springer Berlin, Heidelberg},\
  \bibinfo {year} {2011})\BibitemShut {NoStop}%
\bibitem [{\citenamefont {Takano}(1959)}]{Takano_1959}%
  \BibitemOpen
  \bibfield  {author} {\bibinfo {author} {\bibfnamefont {F.}~\bibnamefont
  {Takano}},\ }\href {\doibase 10.1143/JPSJ.14.348} {\bibfield  {journal}
  {\bibinfo  {journal} {Journal of the Physical Society of Japan}\ }\textbf
  {\bibinfo {volume} {14}},\ \bibinfo {pages} {348} (\bibinfo {year}
  {1959})}\BibitemShut {NoStop}%
\bibitem [{\citenamefont {Cantu}\ \emph {et~al.}(1975)\citenamefont {Cantu},
  \citenamefont {Klein}, \citenamefont {Matsen},\ and\ \citenamefont
  {Seligman}}]{Cantu_etal_1975}%
  \BibitemOpen
  \bibfield  {author} {\bibinfo {author} {\bibfnamefont {A.~A.}\ \bibnamefont
  {Cantu}}, \bibinfo {author} {\bibfnamefont {D.~J.}\ \bibnamefont {Klein}},
  \bibinfo {author} {\bibfnamefont {F.~A.}\ \bibnamefont {Matsen}}, \ and\
  \bibinfo {author} {\bibfnamefont {T.~H.}\ \bibnamefont {Seligman}},\ }\href
  {\doibase 10.1007/BF00963472} {\bibfield  {journal} {\bibinfo  {journal}
  {Theoretica chimica acta}\ }\textbf {\bibinfo {volume} {38}},\ \bibinfo
  {pages} {341} (\bibinfo {year} {1975})}\BibitemShut {NoStop}%
\bibitem [{\citenamefont {{Vladimír Kvasnička}}(1977)}]{Kvasnicka_1977}%
  \BibitemOpen
  \bibfield  {author} {\bibinfo {author} {\bibnamefont {{Vladimír
  Kvasnička}}},\ }\href {\doibase
  https://doi.org/10.1016/0009-2614(77)85377-3} {\bibfield  {journal} {\bibinfo
   {journal} {Chemical Physics Letters}\ }\textbf {\bibinfo {volume} {51}},\
  \bibinfo {pages} {165} (\bibinfo {year} {1977})}\BibitemShut {NoStop}%
\bibitem [{\citenamefont {Kojo}\ and\ \citenamefont
  {Hirose}(2009)}]{Kojo_Hirose_2009}%
  \BibitemOpen
  \bibfield  {author} {\bibinfo {author} {\bibfnamefont {M.}~\bibnamefont
  {Kojo}}\ and\ \bibinfo {author} {\bibfnamefont {K.}~\bibnamefont {Hirose}},\
  }\href@noop {} {\bibfield  {journal} {\bibinfo  {journal} {Journal of
  Computational and Theoretical Nanoscience}\ }\textbf {\bibinfo {volume}
  {6}},\ \bibinfo {pages} {2567} (\bibinfo {year} {2009})}\BibitemShut
  {NoStop}%
\bibitem [{\citenamefont {Saunders}\ and\ \citenamefont
  {Hillier}(1973)}]{saunders1973level}%
  \BibitemOpen
  \bibfield  {author} {\bibinfo {author} {\bibfnamefont {V.}~\bibnamefont
  {Saunders}}\ and\ \bibinfo {author} {\bibfnamefont {I.}~\bibnamefont
  {Hillier}},\ }\href {\doibase https://doi.org/10.1002/qua.560070407}
  {\bibfield  {journal} {\bibinfo  {journal} {International Journal of Quantum
  Chemistry}\ }\textbf {\bibinfo {volume} {7}},\ \bibinfo {pages} {699}
  (\bibinfo {year} {1973})}\BibitemShut {NoStop}%
\bibitem [{\citenamefont {Guest}\ and\ \citenamefont
  {Saunders}(1974)}]{guest1974methods}%
  \BibitemOpen
  \bibfield  {author} {\bibinfo {author} {\bibfnamefont {M.}~\bibnamefont
  {Guest}}\ and\ \bibinfo {author} {\bibfnamefont {V.~R.}\ \bibnamefont
  {Saunders}},\ }\href {\doibase 10.1080/00268977400102171} {\bibfield
  {journal} {\bibinfo  {journal} {Molecular Physics}\ }\textbf {\bibinfo
  {volume} {28}},\ \bibinfo {pages} {819} (\bibinfo {year} {1974})}\BibitemShut
  {NoStop}%
\bibitem [{\citenamefont {Pulay}(1980)}]{pulay1980convergence}%
  \BibitemOpen
  \bibfield  {author} {\bibinfo {author} {\bibfnamefont {P.}~\bibnamefont
  {Pulay}},\ }\href {\doibase https://doi.org/10.1016/0009-2614(80)80396-4}
  {\bibfield  {journal} {\bibinfo  {journal} {Chemical Physics Letters}\
  }\textbf {\bibinfo {volume} {73}},\ \bibinfo {pages} {393} (\bibinfo {year}
  {1980})}\BibitemShut {NoStop}%
\bibitem [{\citenamefont {Salvador}(2009)}]{salvador2009convergence}%
  \BibitemOpen
  \bibfield  {author} {\bibinfo {author} {\bibfnamefont {P.}~\bibnamefont
  {Salvador}},\ }\href {\doibase https://doi.org/10.1002/qua.22103} {\bibfield
  {journal} {\bibinfo  {journal} {International Journal of Quantum Chemistry}\
  }\textbf {\bibinfo {volume} {109}},\ \bibinfo {pages} {2564} (\bibinfo {year}
  {2009})}\BibitemShut {NoStop}%
\bibitem [{\citenamefont {Garniron}\ \emph {et~al.}(2019)\citenamefont
  {Garniron}, \citenamefont {Gasperich}, \citenamefont {Applencourt},
  \citenamefont {Benali}, \citenamefont {Fert{\'e}}, \citenamefont {Paquier},
  \citenamefont {Pradines}, \citenamefont {Assaraf}, \citenamefont {Reinhardt},
  \citenamefont {Toulouse}, \citenamefont {Barbaresco}, \citenamefont {Renon},
  \citenamefont {David}, \citenamefont {Malrieu}, \citenamefont {V{\'e}ril},
  \citenamefont {Caffarel}, \citenamefont {Loos}, \citenamefont {Giner},\ and\
  \citenamefont {Scemama}}]{QP2-JCTC-19}%
  \BibitemOpen
  \bibfield  {author} {\bibinfo {author} {\bibfnamefont {Y.}~\bibnamefont
  {Garniron}}, \bibinfo {author} {\bibfnamefont {K.}~\bibnamefont {Gasperich}},
  \bibinfo {author} {\bibfnamefont {T.}~\bibnamefont {Applencourt}}, \bibinfo
  {author} {\bibfnamefont {A.}~\bibnamefont {Benali}}, \bibinfo {author}
  {\bibfnamefont {A.}~\bibnamefont {Fert{\'e}}}, \bibinfo {author}
  {\bibfnamefont {J.}~\bibnamefont {Paquier}}, \bibinfo {author} {\bibfnamefont
  {B.}~\bibnamefont {Pradines}}, \bibinfo {author} {\bibfnamefont
  {R.}~\bibnamefont {Assaraf}}, \bibinfo {author} {\bibfnamefont
  {P.}~\bibnamefont {Reinhardt}}, \bibinfo {author} {\bibfnamefont
  {J.}~\bibnamefont {Toulouse}}, \bibinfo {author} {\bibfnamefont
  {P.}~\bibnamefont {Barbaresco}}, \bibinfo {author} {\bibfnamefont
  {N.}~\bibnamefont {Renon}}, \bibinfo {author} {\bibfnamefont
  {G.}~\bibnamefont {David}}, \bibinfo {author} {\bibfnamefont {J.~P.}\
  \bibnamefont {Malrieu}}, \bibinfo {author} {\bibfnamefont {M.}~\bibnamefont
  {V{\'e}ril}}, \bibinfo {author} {\bibfnamefont {M.}~\bibnamefont {Caffarel}},
  \bibinfo {author} {\bibfnamefont {P.~F.}\ \bibnamefont {Loos}}, \bibinfo
  {author} {\bibfnamefont {E.}~\bibnamefont {Giner}}, \ and\ \bibinfo {author}
  {\bibfnamefont {A.}~\bibnamefont {Scemama}},\ }\href {\doibase
  10.1021/acs.jctc.9b00176} {\bibfield  {journal} {\bibinfo  {journal} {J.
  Chem. Theory Comput.}\ }\textbf {\bibinfo {volume} {15}},\ \bibinfo {pages}
  {3591} (\bibinfo {year} {2019})}\BibitemShut {NoStop}%
\bibitem [{\citenamefont {Dunning}(1989)}]{Dun-JCP-89}%
  \BibitemOpen
  \bibfield  {author} {\bibinfo {author} {\bibfnamefont {T.~H.}\ \bibnamefont
  {Dunning}},\ }\href {\doibase https://doi.org/10.1063/1.456153} {\bibfield
  {journal} {\bibinfo  {journal} {J. Chem. Phys.}\ }\textbf {\bibinfo {volume}
  {90}},\ \bibinfo {pages} {1007} (\bibinfo {year} {1989})}\BibitemShut
  {NoStop}%
\bibitem [{\citenamefont {Becke}(1988)}]{Bec-JCP-88b}%
  \BibitemOpen
  \bibfield  {author} {\bibinfo {author} {\bibfnamefont {A.~D.}\ \bibnamefont
  {Becke}},\ }\href {\doibase 10.1063/1.454033} {\bibfield  {journal} {\bibinfo
   {journal} {J. Chem. Phys.}\ }\textbf {\bibinfo {volume} {88}},\ \bibinfo
  {pages} {2547} (\bibinfo {year} {1988})}\BibitemShut {NoStop}%
\bibitem [{\citenamefont {Besal{\'u}}\ and\ \citenamefont
  {Carb{\'o}-Dorca}(2011)}]{besalu2011general}%
  \BibitemOpen
  \bibfield  {author} {\bibinfo {author} {\bibfnamefont {E.}~\bibnamefont
  {Besal{\'u}}}\ and\ \bibinfo {author} {\bibfnamefont {R.}~\bibnamefont
  {Carb{\'o}-Dorca}},\ }\href {\doibase 10.1007/s10910-011-9857-9} {\bibfield
  {journal} {\bibinfo  {journal} {Journal of mathematical chemistry}\ }\textbf
  {\bibinfo {volume} {49}},\ \bibinfo {pages} {1769} (\bibinfo {year}
  {2011})}\BibitemShut {NoStop}%
\bibitem [{\citenamefont {Toulouse}, \citenamefont {Colonna},\ and\
  \citenamefont {Savin}(2004)}]{toulouse2004long}%
  \BibitemOpen
  \bibfield  {author} {\bibinfo {author} {\bibfnamefont {J.}~\bibnamefont
  {Toulouse}}, \bibinfo {author} {\bibfnamefont {F.~m.~c.}\ \bibnamefont
  {Colonna}}, \ and\ \bibinfo {author} {\bibfnamefont {A.}~\bibnamefont
  {Savin}},\ }\href {\doibase 10.1103/PhysRevA.70.062505} {\bibfield  {journal}
  {\bibinfo  {journal} {Phys. Rev. A}\ }\textbf {\bibinfo {volume} {70}},\
  \bibinfo {pages} {062505} (\bibinfo {year} {2004})}\BibitemShut {NoStop}%
\end{thebibliography}%

\end{document}